\documentclass[aps,prl,showpacs,notitlepage,singlecolumn,superscriptaddress,11pt,floatfix,
]{revtex4-1}
\usepackage{amsmath}
\usepackage{amssymb}
\usepackage{amsfonts}
\usepackage{stmaryrd}
\usepackage{wasysym}
\usepackage{graphicx}
\usepackage{textcomp}
\usepackage{subfigure}
\usepackage{url}
\usepackage[colorlinks=true,citecolor=blue,urlcolor=blue]{hyperref}
\usepackage{romannum}
\usepackage{mathtools}
\usepackage[utf8]{inputenc}
\usepackage{braket}
\usepackage{amsmath}
\usepackage{xcolor}
\usepackage{color,soul} 
\usepackage{bm}
\usepackage[normalem]{ulem}
\usepackage[utf8]{inputenc}
\usepackage{array}
\usepackage{makecell}

\definecolor{mygreen}{RGB}{0,190,0}
\definecolor{mymagenta}{RGB}{186, 00, 132}

\hypersetup{colorlinks=true, urlcolor=blue, citecolor=cyan, pdfborder={0 0 0}}

\newcommand{\vv}[1]{{\boldsymbol #1}}

\newcolumntype{L}[1]{>{\raggedright\arraybackslash}p{#1}}
\newcolumntype{C}[1]{>{\centering\arraybackslash}p{#1}}
\newcolumntype{R}[1]{>{\raggedleft\arraybackslash}p{#1}}
\usepackage{multirow}
\usepackage{enumitem}

\usepackage{fancyhdr}
\usepackage{pdfpages} 
\usepackage{pgffor} 

\makeatletter
\AtBeginDocument{\let\LS@rot\@undefined}
\makeatother
\def\supplementfilename{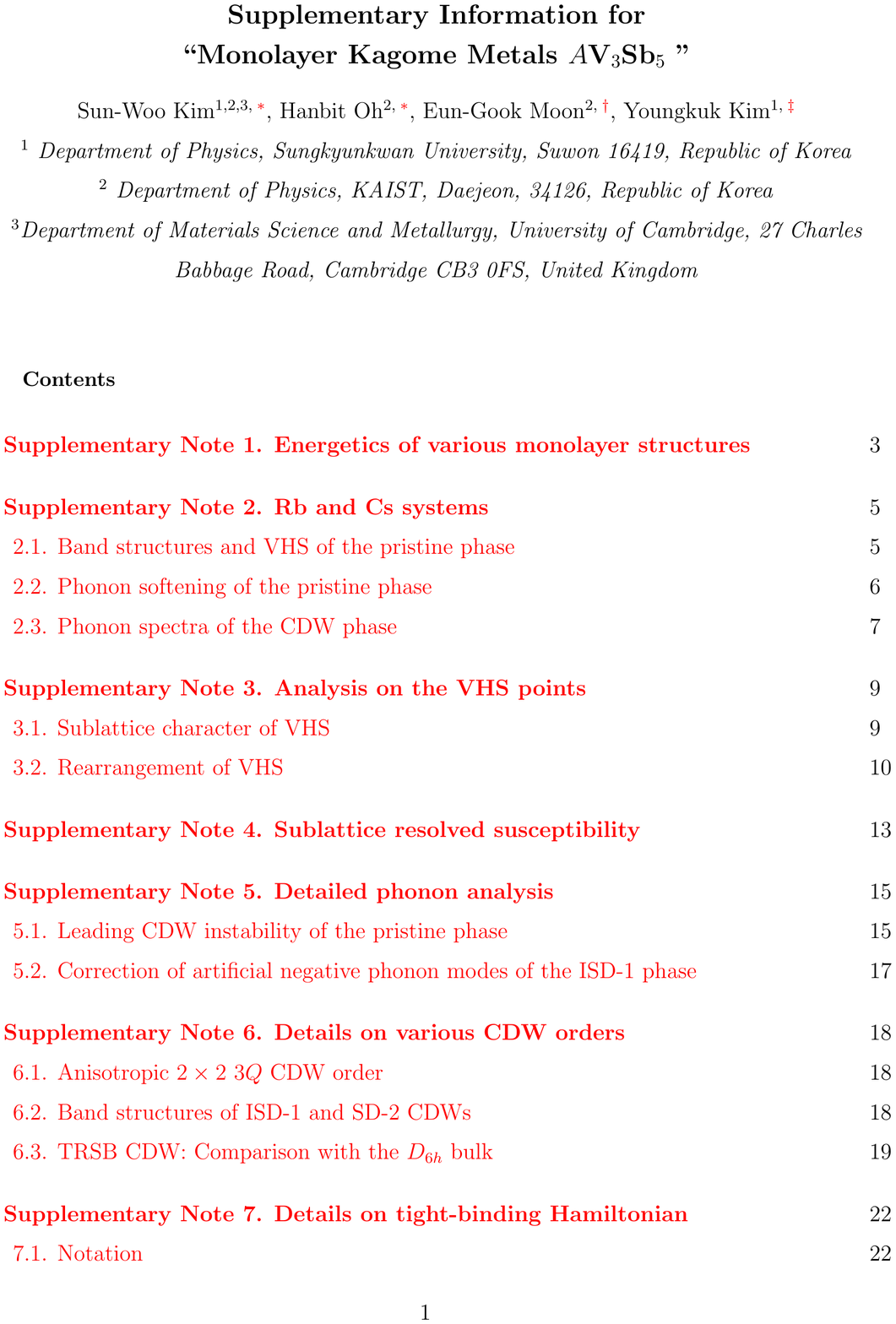}

\pdfximage{\supplementfilename}
\def\numbersupplementpages{\the\pdflastximagepages}

\newif\ifarXiv
\arXivtrue 

\begin{document}
\pagenumbering{arabic}

\title{ Monolayer Kagome Metals $A$V$_3$Sb$_5$}

\author{Sun-Woo Kim $^{1,2,3,\textcolor{red}{*}}$, 
Hanbit Oh $^{2,\textcolor{red}{*}}$, 
Eun-Gook Moon $^{2,\textcolor{red}{\dagger}}$, and 
Youngkuk Kim $^{1,\textcolor{red}{\ddagger}}$ \\
$^{1}$ {\it Department of Physics, Sungkyunkwan University, Suwon 16419, Republic of Korea}\\
$^{2}$ {\it Department of Physics, KAIST, Daejeon, 34126, Republic of Korea}\\
$^{3}$ {\it Department of Materials Science and Metallurgy, University of Cambridge, 27 Charles Babbage Road,
Cambridge CB3 0FS, United Kingdom}\\
$^{\textcolor{red}{*}}$ {\it These authors contributed equally.}\\
$^{\textcolor{red}{\dagger}}$
 {\it \href{mailto:egmoon@kaist.ac.kr}{\textcolor{blue}{$\mathrm{egmoon@kaist.ac.kr}$}}}
\\
$^{\textcolor{red}{\ddagger}}$
 {\it \href{mailto:youngkuk@skku.edu}{\textcolor{blue}{$\mathrm{youngkuk@skku.edu}$}}}
 \\
}

\date{\today}
\maketitle

\section{Abstract}

Recently, layered kagome metals  $A$V$_3$Sb$_5$ ($A$ = K, Rb, and Cs) have emerged as a fertile platform for exploring frustrated geometry, correlations, and topology.
Here, using first-principles and mean-field calculations, we demonstrate that $A$V$_3$Sb$_5$ can crystallize in a mono-layered form, revealing a range of properties that render the system unique.
Most importantly, the two-dimensional monolayer preserves intrinsically different symmetries from the three-dimensional layered bulk, enforced by stoichiometry.
Consequently, the van Hove singularities, logarithmic divergences of the electronic density of states, are enriched, leading to a variety of competing instabilities such as doublets of charge density waves and $s$- and $d$-wave superconductivity.
We show that the competition between orders can be fine-tuned in the monolayer via electron-filling of the van Hove singularities.
Thus, our results suggest the monolayer kagome metal $A$V$_3$Sb$_5$ as a promising platform for designer quantum phases. 

\section{Introduction}

The kagome lattice refers to a two-dimensional (2D) planar crystal composed of corner-sharing triangles.
Unique electronic structures emerge owing to the geometrical frustration of the lattice, featuring a flat band, a pair of Dirac points, and saddle-point van Hove singularities (VHSs). 
A prominent example of candidate kagome metals is the recently discovered vanadium-based kagome metals $A$V$_3$Sb$_5$ ($A$ = K, Rb, and Cs) \cite{ortiz_new_2019,ortiz_CsV3Sb5_2020}. 
A cascade of correlated electronic states have been observed in $A$V$_3$Sb$_5$, associated with charge density waves (CDWs) \cite{ortiz_CsV3Sb5_2020,jiang_unconventional_2021,zhao_cascade_2021,li_observation_2021,denner_analysis_2021,feng_chiral_2021,park_electronic_2021,lin_complex_2021,christensen_theory_2021,ortiz_fermi_2021,kang_twofold_2022,luo_electronic_2022,liang_three-dimensional_2021,chen_roton_2021,xiang_twofold_2021,tan_charge_2021} and superconductivity \cite{ortiz_CsV3Sb5_2020,liang_three-dimensional_2021,chen_roton_2021,xiang_twofold_2021,tan_charge_2021,wang_proximity-induced_2020,zhao_nodal_2021,wu_nature_2021,xu_multiband_2021,mu_s-wave_2021,duan_nodeless_2021}. 
These phases are reported to be accompanied by concomitant unexpected properties such as giant anomalous Hall effects \cite{yang_giant_2020,yu_concurrence_2021} without long-ranged magnetic ordering \cite{kenney_absence_2021}, potential Majorana zero modes \cite{liang_three-dimensional_2021}, and edge supercurrent \cite{wang_proximity-induced_2020}. 
The VHSs in conjugation with the Coulomb interaction is suggested as an impetus of the unconventional properties~\cite{neupert2021charge,jiang2021kagome}.

Although many outstanding materials have been found to comprise a kagome lattice in a layered form \cite{ye2018massive, kang2020dirac,yin2020quantum}, the kagome lattice in genuine two dimensions is rare in nature.
This scarcity leads to prior explorations of the kagome materials based on the assumption that a three-dimensional (3D) layered system can be regarded as decoupled kagome layers.
Similarly, in the case of the vanadium-based kagome metals, current experiments are mainly focused on the 3D layered structures \cite{ortiz_new_2019,ortiz_CsV3Sb5_2020,jiang_unconventional_2021,zhao_cascade_2021,li_observation_2021,ortiz_fermi_2021,kang_twofold_2022,luo_electronic_2022,liang_three-dimensional_2021,chen_roton_2021,xiang_twofold_2021,ortiz_CsV3Sb5_2020,liang_three-dimensional_2021,chen_roton_2021,xiang_twofold_2021,wang_proximity-induced_2020,zhao_nodal_2021,xu_multiband_2021,mu_s-wave_2021,duan_nodeless_2021,yang_giant_2020,yu_concurrence_2021,kenney_absence_2021}, while their theoretical analysis largely relies on an effective kagome model in two dimensions \cite{denner_analysis_2021,feng_chiral_2021,park_electronic_2021,lin_complex_2021,wu_nature_2021}.   
The dimensionality has been tacitly assumed as an irrelevant parameter, but this assumption has been generically refuted in layered systems \cite{gong_2017_discovery,fatemi_electrically_2018}.
A few research groups have made pioneering efforts to tackle this issue by successfully exfoliating thin films of $A$V$_3$Sb$_5$ \cite{song_competition_2021,song_competing_2021,wang_enhancement_2021}. 
However, the importance of dimensionality in this family of kagome metals has remained elusive to date.

In this work, we theoretically demonstrate that the $A$V$_3$Sb$_5$ monolayer is different from the 3D layered bulk by performing density-functional theory (DFT) and mean-field theory (MFT) calculations. 
At the crux of our results is the absence of a dimensional crossover between the monolayer and the layered bulk. 
We argue that symmetry-lowering is inevitable in the monolayer, enforced by the stoichiometry of $A$V$_3$Sb$_5$. 
The reduced symmetries give rise to significant changes in the formation of VHSs. 
Notably, unconventional VHSs appear, referred to as type-II VHSs. 
As a consequence, enhanced electronic instabilities are observed, leading to the emergence of competing orders such as CDW doublets, time-reversal breaking CDWs, $s$- and $d$-wave superconductivity. 
Our calculations predict that the $A$V$_3$Sb$_5$ monolayer can be thermodynamically stable. 
The stable $A$V$_3$Sb$_5$ monolayer becomes a unique platform to study the intriguing interplay between the VHSs and competing order parameters because any monolayer systems are under significantly enhanced fluctuations, as manifested in the celebrated Mermin-Wagner theorem \cite{MM1,MM2,MM3}.
In connection with future experiments, we also calculate the anomalous Hall conductivity that can probe the correlated orders. Possible experimental schemes are discussed to tune the electron-filling based on mechanical and chemical treatment.

\section{Results}

\subsection{Crystal structure and Symmetry}
\begin{figure}[ht]
 \centering
     \includegraphics[width=1\textwidth]{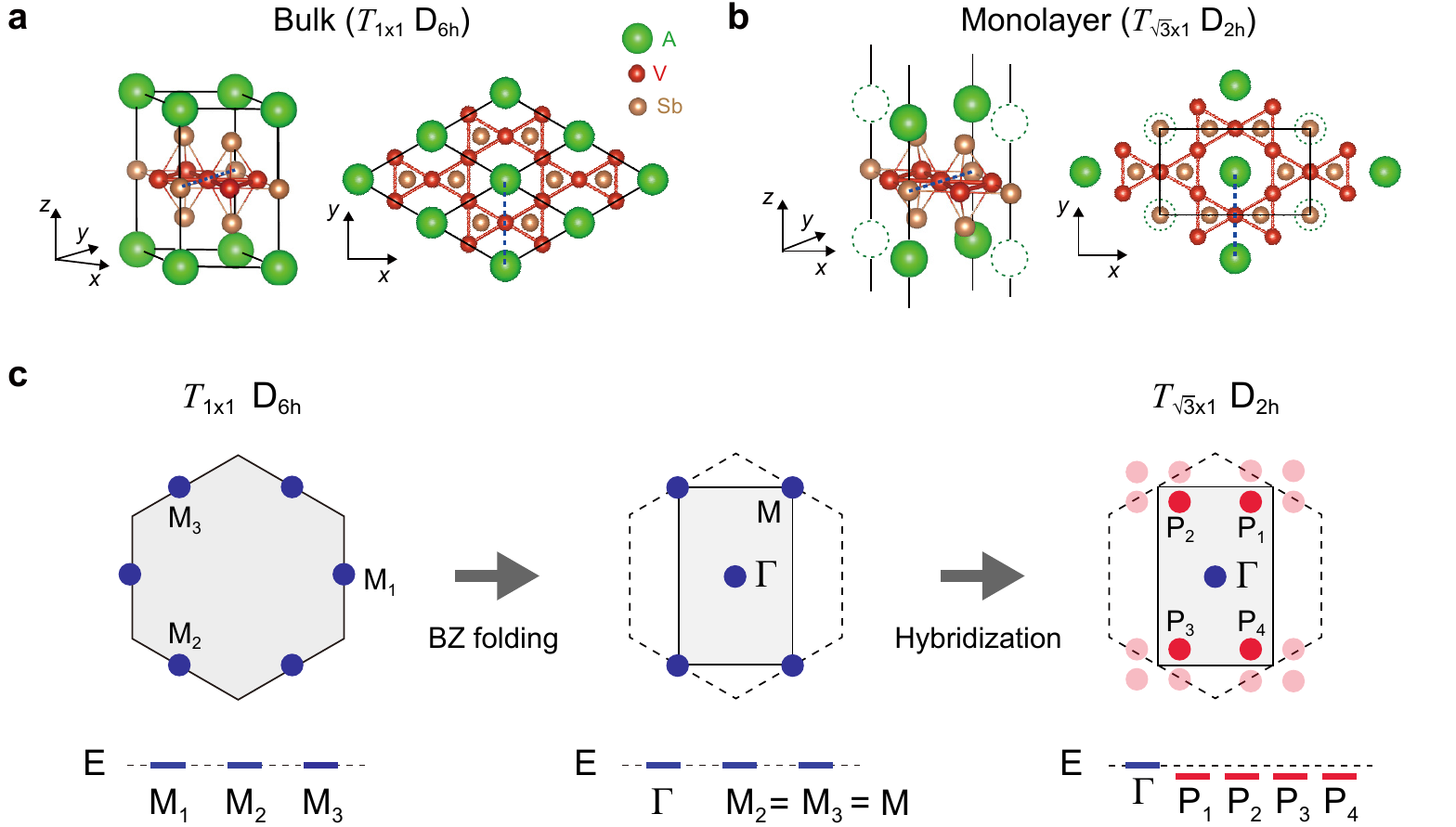} 
  \caption{ \textbf{Symmetry lowering and rearrangement of the van Hove singularity (VHS) in monolayer $A$V$_3$Sb$_5$.}
  \textbf{a, b} Atomic structures of the bulk and monolayer $A$V$_3$Sb$_5$ ($A$ = K, Rb, Cs).
  The monolayer structure preserves $\sqrt{3}\times 1$ translational and $D_{2h}$ point group symmetries. The layered bulk preserves $1\times 1$ translational and $D_{6h}$ point group symmetries.
  Black solid lines indicate the primitive unit cells and blue dashed lines indicate the $y$-axis.
  In \textbf b, dashed open circles represent vacant alkali sites.
  \textbf c Schematic illustration of the VHS points in momentum space rearranged by symmetry lowering.
  The type-I (type-II) VHS points are marked by blue (red) circles.
  In the middle and right panels, the reduced (pristine) BZ of the $\sqrt{3}\times 1$ ($1\times1$) unit cell is indicated by solid (dashed) line. The bottom panels delineate the energy level of the VHS points.
  }
  \label{fig:1}%
\end{figure}

We begin by elucidating the similarities and differences between the crystal structures of the bulk and monolayer $A$V$_3$Sb$_5$ ($A$ = K, Rb, Cs). 
As delineated in Figs.\,\ref{fig:1}\textbf a and \textbf b, both systems comprise multiple sub-layers. 
Most importantly, a 2D kagome sub-layer is formed from V atoms, coexisting with Sb sub-layers. While these are similar in both systems, differences arise from the alkali atoms $A$.
In the monolayer (bulk) system, alkali metals energetically favor to form \textit{rectangular (triangular)} sub-layers shown in Fig.\,\ref{fig:1}\textbf b(\textbf a) (see Supplementary Note 1 for the detailed analysis of the energetics using first-principles calculations).
The different formation of alkali atoms is traceable to the stoichiometry of $A$V$_3$Sb$_5$. 
The kagome layer of the monolayer takes all the electrons donated from the alkali atoms, while those in the bulk system are shared between the adjacent two sub-layers.
Therefore, to preserve the stoichiometry, the number of neighboring alkali atoms is halved by doubling the unit cell, such that they form rectangular sub-layers.

The rectangular sub-layer with the doubled unit cell breaks translational and rotational symmetries of bulk $A$V$_3$Sb$_5$.
The translational symmetry $\mathcal{T}_{1\times 1}$ is reduced to $\mathcal{T}_{\sqrt{3}\times 1}$, and correspondingly, a three-fold rotational symmetry $C_{3z}$ is lifted. This reduces the $D_{6h}$ symmetry of the bulk to $D_{2h}$ in the monolayer.
We stress that the lowered symmetry of the monolayer $A$V$_3$Sb$_5$ is the fundamental symmetry, which can be stable even at room temperature protected by energy barriers from the stoichiometry enforcement, while the same symmetry is only reachable at low temperature from a cascade of phase transitions in bulk $A$V$_3$Sb$_5$ \cite{zhao_cascade_2021,chen_roton_2021}.
We also note that any incommensurate CDW (IC-CDW) orders are prohibited at non-zero temperatures in monolayer systems unless either long-range interactions or substrate effects play a significant role.
Thus, a dimensional crossover is unlikely to occur from the bulk to the monolayer $A$V$_3$Sb$_5$, and unique properties arise as a result.

\subsection{Rearrangement of VHSs}

Most importantly, the lowered symmetry of the monolayer $A$V$_3$Sb$_5$ leads to the rearrangement of the VHS in energy-momentum space.
The mechanism of the rearrangement is illustrated in Fig.\,\ref{fig:1}\textbf c, in which we trace the $\vv k$-points that host the VHS. Hereafter, we refer to these momenta as the VHS points. 
The pristine BZ with the $\mathcal{T}_{1\times 1}$ translational symmetry initially hosts three inequivalent VHS points at $M_i$ ($i$ = 1, 2, 3), as in the case of the bulk $A$V$_3$Sb$_5$ (left panel in Fig.\,\ref{fig:1}\textbf {c}).
Upon the zone folding by lowering $\mathcal{T}_{1\times 1}$ to $\mathcal{T}_{\sqrt{3}\times 1}$, the $M_1$ VHS point is folded to $\Gamma$ and the $M_1$ and $M_2$ VHS points are merged to the $M$ point of the reduced BZ (middle panel in Fig.\,\ref{fig:1}\textbf c). 
The symmetry-lowering further hybridizes the two states at $M$ ($M_2$ and $M_3$), such that it annihilates the VHS at $M$ and creates four new VHS points off $M$, marked as P$_i$ ($i$ = 1, 2, 3, 4) in the right panel of Fig.\,\ref{fig:1}\textbf c.

\begin{figure}[t]
 \centering%
 \includegraphics[width=1\textwidth]{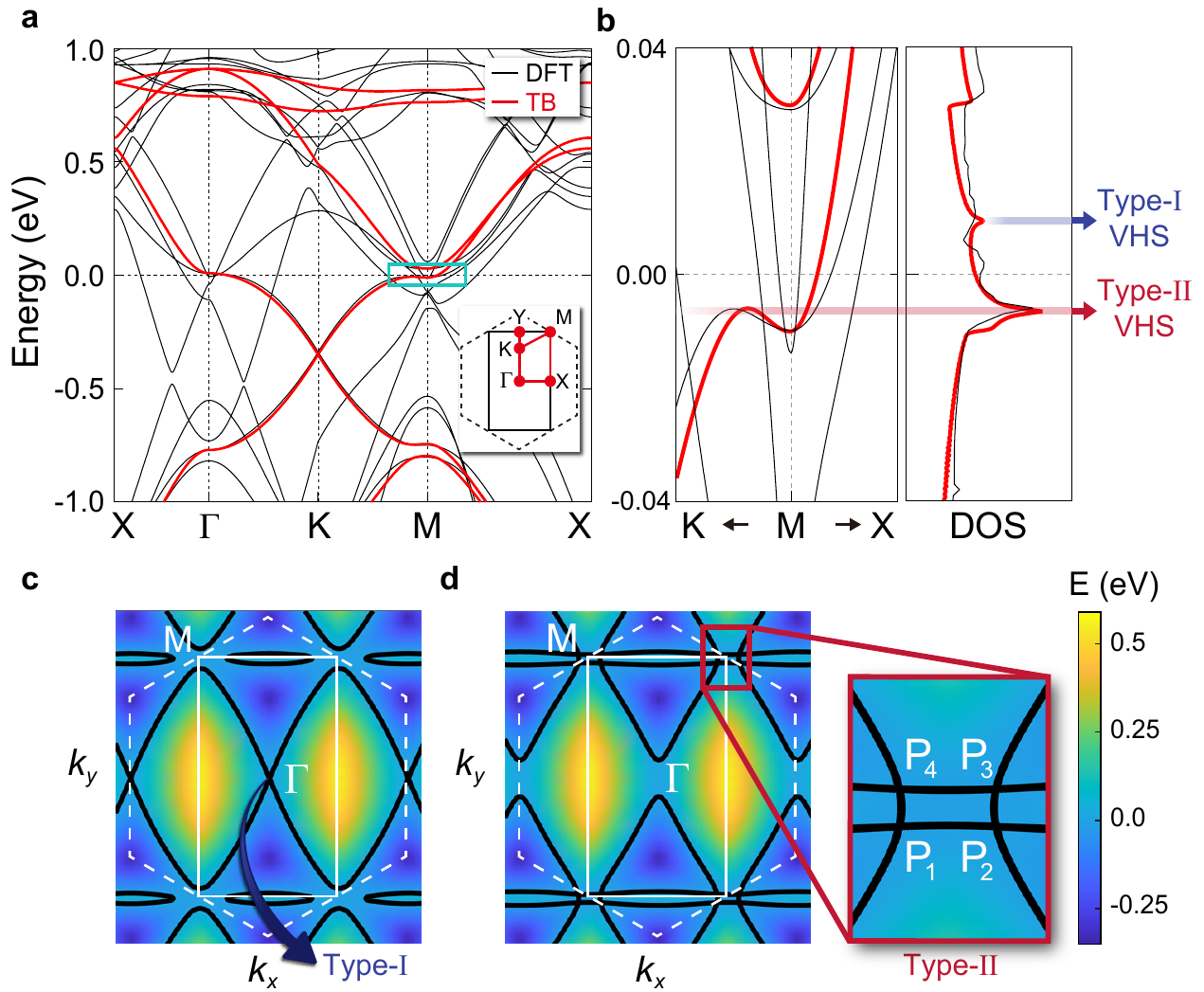} 
  \caption{   \textbf{VHSs in the monolayer $A$V$_3$Sb$_5$.} 
  \textbf a DFT and TB band structures of the KV$_3$Sb$_5$ monolayer. 
  The DFT (TB) bands are colored by black (red). The kagome bands are reproduced by using our TB model (see Methods for the details of the TB model).
  A solid rectangle (dashed hexagon) in the inset depicts the reduced (pristine) BZ of the $\sqrt{3}\times 1$ ($1\times1$) unit cell.
  \textbf b (Left panel) Magnified view of the blue box in \textbf a. 
  (Right panel) Density of states (DOS) of KV$_3$Sb$_5$. The type-I and type-II VHSs are indicated by blue and red arrows, respectively.
  \textbf{c, d} Energy contours of the kagome bands calculated from the TB bands.
  Solid (dashed) white lines show the reduced (pristine) BZ.
  The contour curves in \textbf c and \textbf d correspond to $E = 9$ and $ -6$ meV, respectively.
  The saddle points reside at the crossing points of the contour curves in \textbf c and \textbf d.
  The magnified view of the red box in \textbf d illustrates the position of type-II VHS points off $M$, marked by $P_i$ ($i$ = 1, 2, 3, 4).}
\label{fig:2}%
\end{figure}

The rearrangement of VHS points is observed in our first-principles calculations. Figures\,\ref{fig:2}\textbf a and \textbf b show exemplary DFT bands of KV$_3$Sb$_5$ with archetypal kagome bands distilled by our tight-binding theory (see Methods for the details of the TB model).
In Fig.\,\ref{fig:2}\textbf b, the divergence of the density of states (DOS) is clearly observed at $E = -6 \, (+9)$ meV. 
A close inspection reveals that the diverging DOS at $E = -6$ meV arises from off high-symmetry momenta at $P_i = M + (\pm0.054, \pm0.021)$ \AA$^{-1}$ (Fig.\,\ref{fig:2}\textbf d), while the divergence at $E = 9$ meV arises from the exact high-symmetry $\Gamma$ point (Fig.\,\ref{fig:2}\textbf c).  
In this respect, the VHS points at $E = 9$ meV ($E = -6$ meV) belong to the type-I (type-II) class, where the type-I (II) refers to a class of VHSs that originates from (off) time-reversal invariant momenta \cite{yao2015topological,yuan2019magic,qin2019chiral}.
Our calculations further reveal that the type-II VHSs generically occur in the kagome metals regardless of $A$ = K, Rb, and Cs (see Supplementary Note 2 for the DFT results of the RbV$_3$Sb$_5$ and CsV$_3$Sb$_5$ monolayers). 

The emergence of the type-II VHS is one of the key features of the monolayer $A$V$_3$Sb$_5$. Few remarks are as follows. 
First, the appearance of the type-II VHS off time-reversal invariant momenta is understood as a generic result attributed to the band hybridization allowed by the symmetry lowering.
Since the two VHSs at $M_2$ and $M_3$ before the symmetry lowering are at the exact same energy level, they can be mixed together and contribute to generating the type-II VHSs (See Fig.\,\ref{fig:1}\textbf{c}).
On the other hand, the VHS at $M_1$ still remains a type-I VHS because no states are available to hybridize with.
We find that the number of the generated type-II VHS points (in this case, four) is not universal and depends on the microscopic details of the system (see details in Supplementary Note 3.2). Second, the type-II VHS points $P_i$ ($i$ = 1, 2, 3, 4) (Fig.\,\ref{fig:2}\textbf d) consist of a mixed contribution from both the B and C sublattices (Supplementary Note 3, Fig.S5), referred to as a mixed-type flavor \cite{kiesel2012sublattice,wu_nature_2021,kang_twofold_2022}. 
This is in contrast to the type-I VHS point at $\Gamma$ (Fig.\,\ref{fig:2}\textbf c), which is purely contributed from the A sublattice, referred to as a pure-type flavor.
Third, the increased number of VHS points quantitatively changes the characteristic of the diverging DOS at $-6$ meV. 
Namely, the peak at $-6$ meV is significantly enhanced than that of the type-I VHS at $9$ meV (Fig.\,\ref{fig:2}\textbf b), contributed from the quartet VHS points at $P_i$ ($i$ = 1, 2,3, 4).
Such quantitative changes are of immediate impact on the electronic properties, such as instabilities driven by the type-II VHS, as we will show below.

\subsection{Instability of the monolayer}

\begin{figure}[h]
\centering
 \includegraphics[width=1\textwidth]{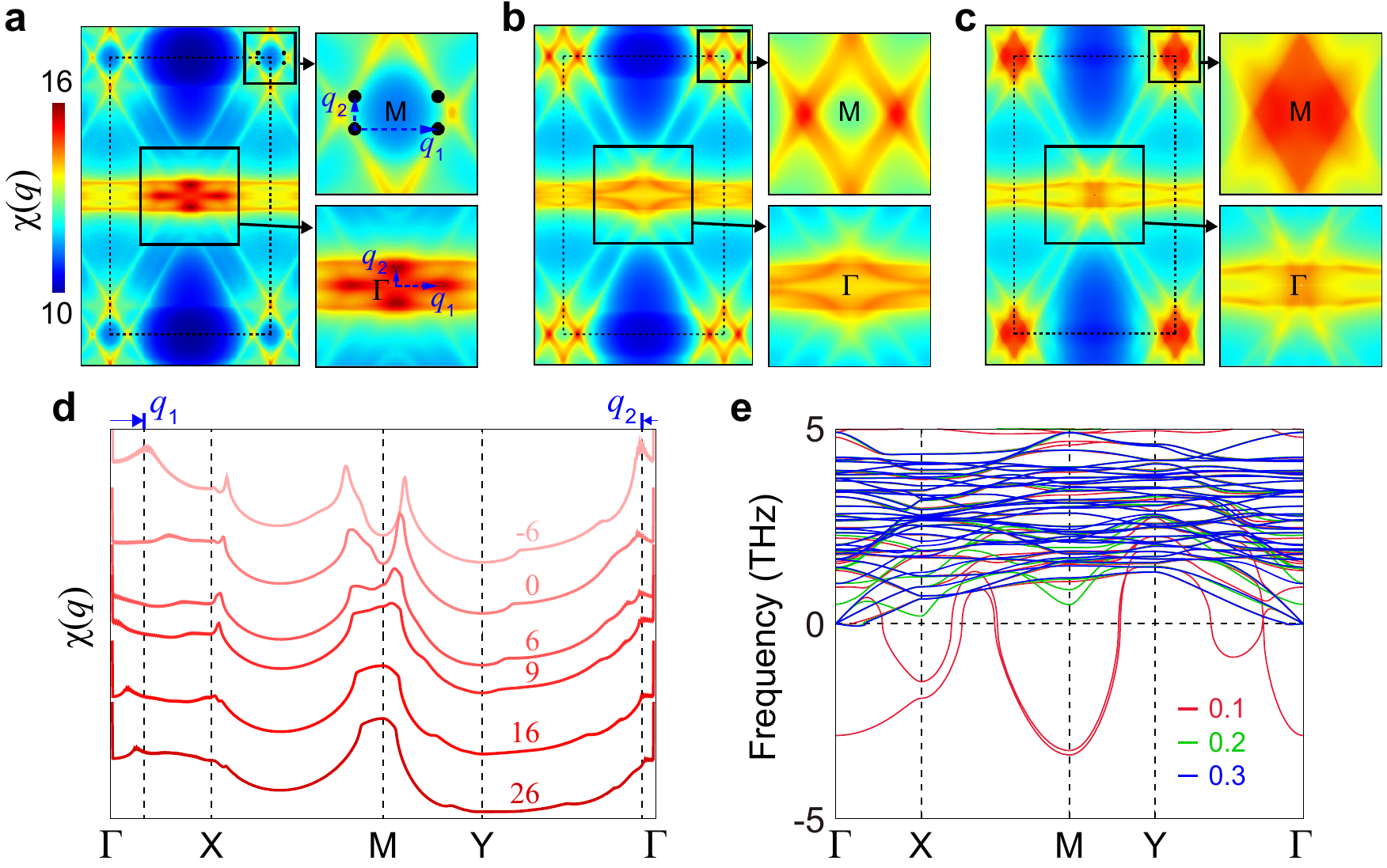}
  \caption{
  \textbf{Electronic susceptibility and phonon bands of the $A$V$_3$Sb$_5$ monolayer.}
  \textbf{a-c} Electronic susceptibility $\chi(\vv{q})$ is calculated in 2D  $\vv q$--space for \textbf{a} the type-II VHS filling ($\mu=-6$ meV), \textbf{b} neutral filling ($\mu=0$ meV), and \textbf{c} type-I VHS filling ($\mu=9$ meV). $\chi(\vv{q})$. The right panels show a magnified view of the boxed areas of the left panel near the high-symmetric $\Gamma$ and $M$ points, respectively. $\vv{q}_1$ and $\vv{q}_2$ in \textbf{a} represent the nesting vectors that connect the type-II VHS points (black circles).
  \textbf{d} Line profiles of $\chi(\bm{q})$ along the high-symmetry lines of the BZ for various chemical potentials. The corresponding chemical potential (in meV) is shown in each profile.  The momenta correspond to $\vv{q}_1$ and $\vv{q}_2$ vectors in \textbf{a} are indicated by arrows with vertical lines. 
  \textbf{e} Phonon bands of the pristine KV$_3$Sb$_5$ monolayer as a function of electronic temperature. The electronic temperature effect is introduced by tuning the smearing factor $\sigma$ (from 0.1 to 0.3 eV) of the Fermi-Dirac distribution function. 
  }
\label{fig:3}
\end{figure}

The pristine monolayer harbors intrinsic instability captured in the electronic susceptibility.
Figures~\ref{fig:3}\textbf{a-d} show the real part of the bare static charge susceptibility $\chi(\bm{q})$ at various fillings (see Methods).
We find that the presence of both type-I and type-II VHSs leads to diversified instability sensitively depending on electron filling.
For example, at the type-II VHS filling (Fig.\,\ref{fig:3}\textbf{a}), a significant contribution arises from the $\vv q$ vectors that connect the van Hove saddle points.
The peaks at $\vv {q}_1$ and $\vv {q}_2$ in $\chi(\bm{q})$ correspond to the nesting vectors mediating distinct $P_i$ points.
The presence of these $\vv q$ vectors that are incommensurate to the reciprocal lattice vector  necessitates the consideration of an incommensurate order parameter, such as IC-CDW phases.
By contrast, when the chemical potential increases, the peaks quickly merge and give rise to the significant enhancement of $\chi(\bm{q})$ arises at the $M$ point (Figs.\,\ref{fig:3}\textbf{b-d}).
This significant peak at $\vv q = M$, as well as at $\vv q = \Gamma$ in the folded BZ, corresponds to the commensurate $2\times2$ CDW vectors. Notably, the large values of $\chi(\bm{q})$ at $M$ survive for a wide range of chemical potentials. These results suggest a significant role of $2\times2$ CDW orders in stabilizing the monolayer.
Moreover, the sublattice character of $\chi(\bm{q})$ reveals that sublattice interference is significant in the monolayer (See Supplementary Note 4 for the calculations of sublattice-resolved susceptibility). This result warrants the consideration of longer-range interactions, which promotes unconventional orders as in the bulk cases~\cite{kiesel2012sublattice,wang_2013_competing,denner_analysis_2021,wu_nature_2021}.

The DFT phonon energy spectra also capture the instability of the pristine monolayer. 
We find that significant phonon softening arises at $M$ and $\Gamma$, associated with the displacement of the V atoms. 
Figure\,\ref{fig:3}\textbf{e} shows a representative example of the DFT phonon spectra for monolayer KV$_3$Sb$_5$.
The $\Gamma$ ($M$) point corresponds to the one (other two) of the $2\times2$ 3$Q$ CDW vectors in the folded BZ.
Thus, these negative branches are a clear indicative of the structural instability associated with the $2 \times 2$ distortion of V atoms, similar to the bulk counterparts~\cite{tan_charge_2021,ye_2022_structural,wu_2022_charge,si_2022_charge}.
We also find that these 2$\times$2 CDW instabilities are generically present in monolayer $A$V$_3$Sb$_5$ for $A$ = K, Rb, and Cs (see Supplementary Fig.\,S3). 
We note that the monolayer hosts the broadened softened phonon modes in the momentum space compared to the bulk~\cite{tan_charge_2021}, which signals various CDW instabilities other than $2\times2$. Nonetheless, we confirm that the 2$\times$2 CDW instability is a leading instability by performing energy profile analysis for the various negative phonon modes (see Supplementary Note 5.1).  
Thus, it is imperative to consider the 2$\times$2 3$Q$ CDWs in the monolayer, such as the star of David (SD) and inverse star of David (ISD) distortions in the monolayer. 

To investigate the role of the electronic instability in $2\times2$ CDW instabilities, we additionally calculated the phonon spectra as a function of electronic temperature, smearing factor $\sigma$, as shown in Fig.\,\ref{fig:3}\textbf{e}.
The softened phonon at $M$ is lifted as we destroy the Fermi surface by increasing the electronic temperature.
These calculations demonstrate that the electronic instability associated with the VHS plays a role in the formation of structural distortion. 
Since the DFT phonon spectra include the effects of electron-phonon coupling, our results suggest that either electronic or phononic contributions cannot be excluded in forming structural instability.
In addition, the occurrence of the $2\times 2$ instability in the phonon spectra is plausible in view of the electronic susceptibility calculations.
Our DFT calculations are performed off the type-II VHS filling, in which large susceptibility values arise near the $M$ point.
While the reduced symmetry featured in the monolayer leads to the rearrangement of the VHSs into  type-I and type-II VHSs at different fillings and correspondingly two types of distinct CDW orders---$2\times2$-commensurate CDWs and IC-CDWs, we believe that the commensurate orders are more likely to occur in the monolayer, especially at non-zero temperature.
The pristine monolayer $A$V$_3$Sb$_5$ is under strong fluctuations as dictated in the celebrated Mermin-Wagner theorem, prohibiting any IC-CDW orders at non-zero temperature. Thus, hereafter we mainly focus on commensurate order parameters in this work and discuss the possibility of incommensurate orders at zero temperature in Supplementary Note 11. 

\subsection{Competing orders}

The rearranged VHSs manifest their properties in competing orders of correlated electronic states. 
In monolayer $A$V$_3$Sb$_5$, we employ the standard mean-field theory with the constructed TB model and uncover phase diagrams with CDWs and superconductivity (SC). 
The onsite and nearest-neighbor Coulomb interactions are introduced,
\begin{eqnarray}
H_{\mathrm{int}}=
U \sum_{\vv{R}}\sum_{\alpha}
n_{\vv{R},\alpha \uparrow}
n_{\vv{R},\alpha \downarrow}
+
V \sum_{\vv{R}}
\sum_{\substack{\langle \alpha,\beta\rangle\\
\sigma,\sigma'
 }}n_{\vv{R},\alpha,\sigma}n_{\vv{R},\beta, \sigma'},
\end{eqnarray}
where $U$ ($V$) describes the on-site (nearest-neighbor) density-density type interaction and $\vv{R}, \alpha$, and $\sigma$ represent the lattice site, sublattice, and spin, respectively. 
We consider two classes of order parameters, CDWs and SC, which can significantly reduce the energy by gapping out the Fermi surface with diverging DOSs.
CDWs and SC are of particular interest as they have been observed in the bulk $A$V$_3$Sb$_5$ in a variety of forms, including SD, ISD, and time-reversal symmetry breaking (TRSB) CDWs \cite{feng_chiral_2021,christensen_theory_2021,tan_charge_2021,denner_analysis_2021}.

\begin{figure}[t]
\centering
 \includegraphics[width=1\textwidth]{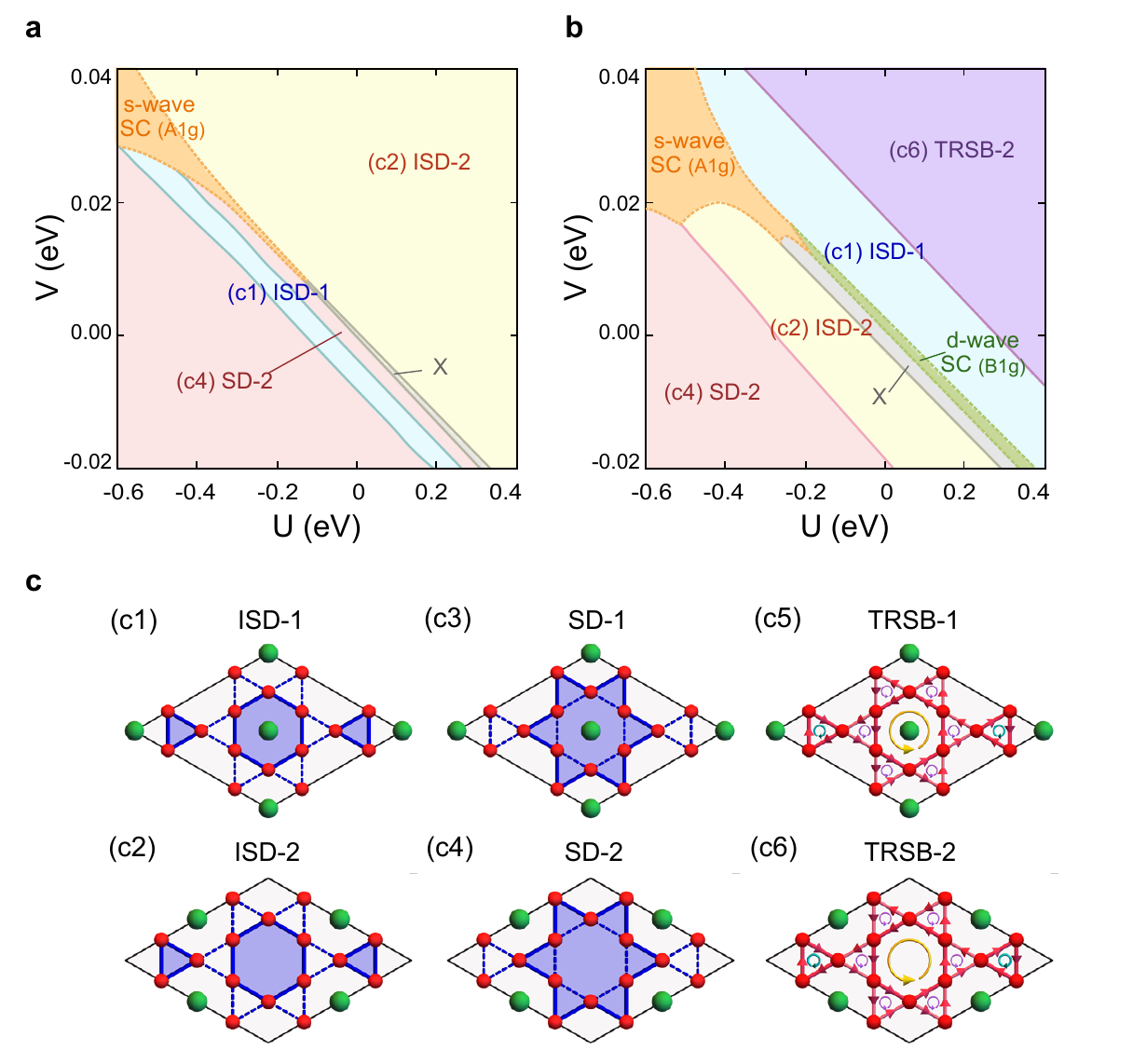} 
  \caption{ 
  \textbf{Phase diagrams at $T \rightarrow 0^+$ and corresponding electronic orders in the  $A$V$_3$Sb$_5$ monolayer.}
  \textbf{a, b} Phase diagrams at the chemical potentials \textbf a $\mu_1 = -6$ meV and \textbf b $\mu_2 = 35$ meV in $U$-$V$ space, where $U$ and $V$ are the onsite and the nearest neighbor density-density type interactions.
  Different CDW and SC orders are indicated by different colors. The gap functions of s-wave and d-wave superconducting states correspond to $A_{g}^{(1)}$ and $B_{1g}^{(1)}$ provided in Table \ref{MT2}. 
  The superconducting phases only exist at zero temperatures, and their phase boundaries are denoted by dashed lines to contrast the boundaries of commensurate CDWs denoted by the solid lines. 
  $X$ denotes a phase unidentified within our mean-field scheme.
  \textbf c Schematic illustration of six distinct CDW configurations.
(c1-c2) Inverse star of David (ISD)-1,2, (c3-c4) star of David (SD)-1,2, (c5-c6) time-reversal symmetry breaking (TRSB)-1,2. 
In (c1-c4), solid and dashed blue lines indicate distinct bonding strengths.
In (c5-c6), red arrows on the bonds indicate the direction of current bond orders.
The oriented circles represent magnetic fluxes 
threading the hexagons and triangles where their magnitudes are proportional to the radius (see Supplementary Note\,6 for the detailed descriptions).
The corresponding CDW and SC order parameters are detailed in Tables\,\ref{MT1} and \ref{MT2}.}
\label{fig:4}
\end{figure}

Remarkably, any CDW order in the monolayer $A$V$_3$Sb$_5$ forms a doublet, as illustrated in Fig.\,\ref{fig:4}\textbf c.
The doublet formation of CDWs is one of the key characteristics of the monolayer $A$V$_3$Sb$_5$, originating from the reduced symmetry. 
The lowered $\mathcal{T}_{\sqrt{3}\times 1}$-translational symmetry of $A$V$_3$Sb$_5$ monolayer plays a crucial role to double the CDW channels of the higher $\mathcal{T}_{1\times 1}$-symmetry, and the number of multiplets is solely determined by their quotient group, $\mathcal{T}_{1\times 1}/\mathcal{T}_{\sqrt{3}\times 1}=\mathbb{Z}_{2}$. 
For example, the two SD-CDW phases, dubbed SD-1 and SD-2, are the $\mathbb{Z}_2$ members, distinguished by the alkali chains hosted on and off the center of SD, respectively, as illustrated in Fig.\,\ref{fig:4}\textbf c.

The corresponding  phenomenological Landau theory of the doublet CDWs becomes exotic. 
Introducing a bosonic real two-component spinor, $\Psi_{{\rm SD}}^{T}
 \equiv (\psi_{{\rm SD-1}}, \psi_{{\rm SD-2}})$ with order parameters of SD-1 ($\psi_{{\rm SD-1}}$) and SD-2 ($\psi_{{\rm SD-2}}$), the Landau functional for the SD-CDW phases is given by
\begin{eqnarray}
\mathcal{F}_{{\rm L}}(\Psi_{{\rm SD}}) = r_{{\rm SD}} (\Psi_{{\rm SD}}^{\dagger}  \Psi_{{\rm SD}} )+s_{{\rm SD}} (\Psi_{{\rm SD}}^{\dagger} \hat{\rho}_z \Psi_{{\rm SD}} )+\cdots,  
\end{eqnarray}
with phenomenological constants $r_{{\rm SD}}$ and $s_{{\rm SD}}$.
Here, the Pauli matrix $\hat{\rho}_z$ describes the spinor space, and higher order terms are omitted for simplicity.
The $s_{{\rm SD}}$-term describes a free-energy difference between SD-1 and SD-2, which is nonzero when the $\mathcal{T}_{1\times 1}$-translational symmetry is broken. 
Depending on $s_{{\rm SD}}$, the system energetically favors one of the doublet CDWs, enriching phase diagrams of the monolayer $A$V$_3$Sb$_5$.

Our mean-field analysis indeed finds enriched phase diagrams of the $A$V$_3$Sb$_5$ monolayer. 
We consider six configurations of CDWs (see Fig.\,\ref{fig:4}\textbf c) and nine spin-singlet channels of SCs (see Table\,\ref{MT2}). 
In Figs.\,\ref{fig:4}\textbf{a, b}, we illustrate representative mean-field phase diagrams of KV$_3$Sb$_5$ in $U$-$V$ space obtained at two different chemical potentials $\mu_1 = - 6$ meV and $\mu_2 = 35$ meV, where
the zero chemical potential is set to the neutral filling. 
We emphasize that our mean-field analysis of $A$V$_3$Sb$_5$ monolayer is reliable at zero and very low temperatures since any monolayer systems suffer from significant thermal fluctuations. Thus, the phase diagrams of Figs.\,\ref{fig:4} need to be understood as the ones in the limit of lowering temperature down to zero, $T\rightarrow0^+$. 
In what follows, we point out key observations made from the phase diagrams.

First, five distinct CDW orders can be accessible by fine-tuning  the chemical potential $\mu$.  
For example, in the vicinity of type-II VHS at $\mu_1 = -6$ meV (Fig.\,\ref{fig:4}\textbf a), SD-2 and ISD-2 dominantly occur with sizable regions of ISD-1 in the energy ranges of -0.6 eV $<$ $U$ $<$ 0.4 eV and -20 meV $<$ $V$ $<$ 40 meV.  
Similarly, a TRSB-2 CDW phase is uncovered under the condition 
$\mu \geq 30$ meV (Fig.\,\ref{fig:4}\textbf b) in a wide range of $U$ and $V$ values.
Moreover, the SD-1 phase appears near the type-I VHS at $\mu_3 =$ 9 meV (see Supplementary Note 9).
A small variation of chemical potential -6 meV $< \mu <$ 40 meV can tune the types and flavors of the VHSs, which should enable an on-demand onset of a variety of CDW phases, ranging from ISD-1/2, SD-1/2, to TRSB-2.
 
Second, competition between CDWs and SC is generically observed. 
A conventional $s$-wave SC phase is observed near $\mu_1 = -6$ meV, which competes with ISD-1/2 and SD-2 at negative $U$ and positive $V$ as shown in Figs.\,\ref{fig:4}\textbf a, \textbf b. 
Similarly, an unconventional $d$-wave SC phase is observed near $\mu_2 = 35$ meV. This competes with ISD-1 at positive $U$ and negative $V$ interactions as shown in Fig.\,\ref{fig:4}\textbf b.
Note that the  chemical potential for the $d$-wave SC is quite higher than the energy of the type-II VHS $\mu_1$ and closer to the edge of the high energy band, which indicates that the VHS itself is not the unique reason to stabilize SC or CDW phases.   
The interplay between the VHS, filling, and interaction strengths should be considered together. 

\begin{figure}[t]
 \centering%
 \includegraphics[width=.9\textwidth]{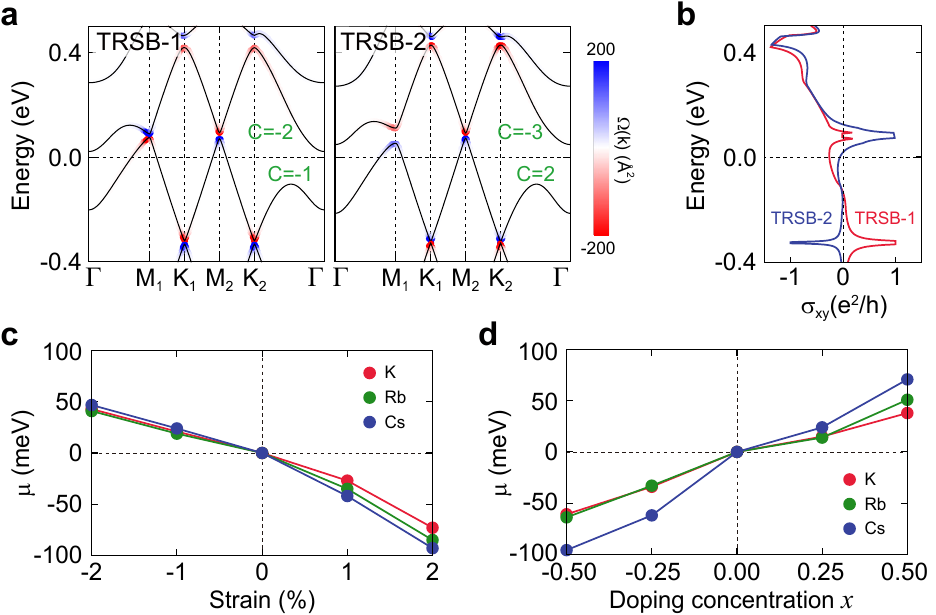} 
  \caption{
   \textbf{Topological charge density waves and engineering of chemical potential.}
  \textbf a TB band structures of the TRSB-1 and TRSB-2 CDW phases.
  The Berry curvature $\Omega_z(\vv{k})$ are overlaid at the corresponding $\vv k$-points in the bands. The Chern numbers $\mathcal C$ induced in the bands are shown in green color.
  \textbf b Anomalous Hall conductivity $\sigma_{xy}$ as a function of energy $E$ for the TRSB-1 and TRSB-2 CDW phases.
  \textbf c DFT calculations of the chemical potential $\mu$ as a function of an uniform biaxial strain in $A$V$_3$Sb$_5$ for $A$ = K, Rb, and Cs.  
  \textbf d DFT calculations of the chemical potential $\mu$ as a function of doping concentration $x$ for $A_{1+x}$V$_3$Sb$_5$.
  The detailed DFT band structures under the biaxial strain and alkali doping are provided in Supplementary Note 12.
  }
  \label{fig:5}%
\end{figure}

The final observation from our mean-field study is the nontrivial topology of the correlated CDW gaps.
Notably, a non-zero Chern number $\mathcal{C}$ is induced  in the energy spectra when gapped by the two time-reversal symmetry broken CDW phases TRSB-1 and TRSB-2.
For example, the lowest unoccupied and highest occupied energy spectra of TRSB-1 (TRSB-2) host the Chern number $\mathcal{C} = -2$ and $\mathcal{C} = -1$ ($\mathcal{C} = -3$ and $\mathcal{C} = 2$), respectively.
As shown in Fig.\,\ref{fig:5}\textbf a, the different Chern numbers between the two phases arise due to the concurrent sign-change of the Berry curvature at high-symmetry momenta $M_1$, $K_1$, and $K_2$.
We note that the monolayer in the  time-reversal symmetry broken CDW phase hosts the Fermi pockets that carry the Berry curvature, referred to as the Fermi Chern pockets~\cite{zhou2021chern} (see Supplementary Fig.\,S13).
However, unlike the bulk case~\cite{zhou2021chern}, asymmetric distribution of the Berry curvature occurs near the $M_1$ and $M_2$ in the monolayer due to the symmetry lowering. This results in distinct experimental observables, such as anomalous Hall conductivity $\sigma_{xy}$ (see detailed comparison between bulk and monolayer in Supplementary Note 6). In Fig\,\ref{fig:5}\textbf{b}, we calculate $\sigma_{xy}$ of the TRSB phases in the monolayer, which demonstrates the enriched structure of the anomalous Hall conductivities, distinct between TRSB-1 and TRSB-2. 
A sign change of $\sigma_{xy}$ is also found as a function of chemical potential near the Fermi energy $E = 0$. 
We believe this nontrivial behavior featured in $\sigma_{xy}$ can be readily observed in the Hall current measurements, leading to the experimental discovery of the TRSB-CDW phases.

The exotic orders should be accessible in monolayer $A$V$_3$Sb$_5$ in a controlled fashion.
Our mean-field phase diagrams (Figs.\,\ref{fig:4}\textbf a and \textbf b)
show that the occurrence of a specific electronic order is highly contingent upon the correct filling of electrons, which can be fine-tuned via mechanical and chemical means.
For example, by applying a uniform biaxial strain in a range of $\pm2\%$ variations, $\mu$ can be tuned from 50 meV to -100 meV (Fig.\,\ref{fig:5}\textbf c).
We also find that  $U$ and $V$ values are functions of applied strains by performing the ab initio calculations with 
constrained random-phase approximation~\cite{cRPA_2004}. This indicates that different phases in the mean-field phase diagrams are accessible using strains (See Supplementary Note 10 for the detailed discussion and computational methods).
Moreover, the light doping of alkali atoms $A$ (see Fig.\,\ref{fig:5}\textbf d) or substituting Sb with Sn \cite{oey2021fermi} can be a fine knob to adjust the chemical potential.
Owing to the 2D geometry of the $A$V$_3$Sb$_5$ monolayer, we believe that there exist more opportunities (such as ionic gating) to tailor the competing orders hosted therein.

\section{Discussion}
We have so far investigated unique features of the $A$V$_3$Sb$_5$ monolayer. 
Our system is unlike the bulk, hosted in a distinct symmetry class. The contrast is even more evident in the phase diagrams that we calculated with the bulk $D_{6h}$ and the monolayer $D_{2h}$ symmetries, respectively
(see Supplementary Fig.\,S19 for the $D_{6h}$ phase diagrams). The lowered $D_{2h}$ symmetry in the monolayer features a tendency to foster the CDW orders. This is in line with the previous experiments, in which a CDW order is observed to suppress SC as the thickness of the $A$V$_3$Sb$_5$ film decreases \cite{song_competing_2021,song_competition_2021}.

\begin{figure}[h]
\centering
 \includegraphics[width=1\textwidth]{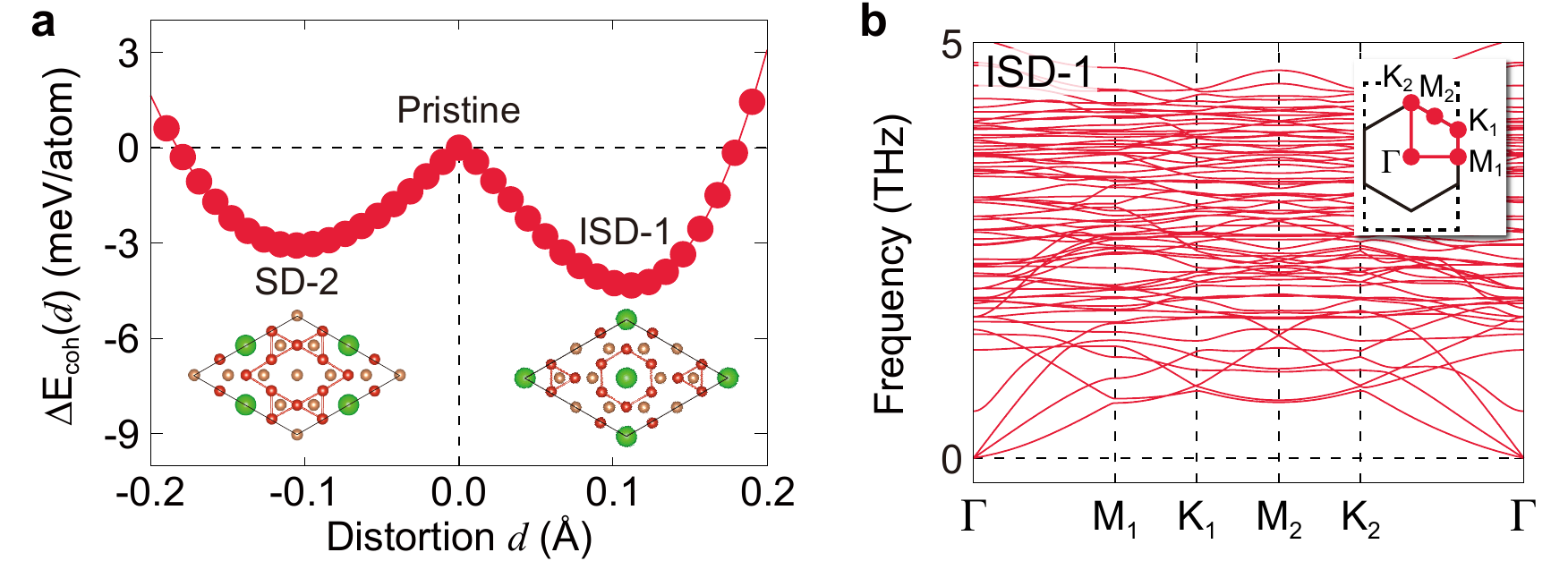}
  \caption{
  \textbf{Stability of the $A$V$_3$Sb$_5$ monolayer.}
   \textbf{a} Cohesive energy difference (per atom) as a function of the CDW distortion $d$ of the KV$_3$Sb$_5$ monolayer.
  The cohesive energy difference is defined as $\Delta E_{\mathrm{coh}} (d) = E_{\mathrm{coh}}(d = 0) - E_{\mathrm{coh}}(d)$, where the positive (negative) value of the distortion $d$ represents the shift of the V atom in the ISD-1 (SD-2) phase with respect to the pristine structure.
  \textbf{b} Phonon bands of monolayer KV$_3$Sb$_5$ in the ISD-1 phase. In the inset, the solid and dashed lines represent the BZs for the $2\times 2$ and $\sqrt{3}\times 1$ structures, respectively.
 For Rb and Cs, see Supplementary Fig.\,S4.
  }
\label{fig:6}
\end{figure}

Our study arguably suggests that monolayer $A$V$_3$Sb$_5$ should be an exciting platform for studying intriguing 2D physics, such as the interplay between thermally suppressed IC-CDW orders and the type-II VHSs.
At zero temperature, an IC-CDW associated with the type-II VHSs appears in the ($U\neq 0,V=0$) phase space. 
We investigate the two different fillings ($\mu=-5, -7$meV) near type-II VHSs and find that the IC-CDW phase is accessible with a negative $U$ in the fine-tuned range of chemical potential (see Supplementary Fig.S21). 
Thus, not only the electron filling but also interaction strengths are important to stabilize the IC-CDW phases. 
At any non-zero temperature, the quasi-long range IC-CDW order can exist, and uncovering the possibility of topological phase transitions such as Kostelitz-Thouless transitions is certainly an important issue in both theoretical and experimental regards.
Therefore, we believe that the $A$V$_3$Sb$_5$ monolayer offers timely new physics in the kagome metal community, calling for future sophisticated theoretical and experimental studies.

Our results also provide further insights into the unconventional CDW orders of the bulk systems. 
Notably, our DFT phonon and electronic susceptibility calculations have shown that the 2$\times 2$ instability of the bulk systems remains a robust feature of the V-based kagome metals against the weakening of the VHS nesting by symmetry-lowering. We have attributed its rationale to significant electronic instability remaining at $\vv q = M$ for a wide range of electron filling. 
Similar mechanisms may explain the robustness of the CDW orders in the bulk systems under a weakened nesting condition due to, for example, out of the exact VHS filling \cite{Isobe}. 
In this regard, further studies in the monolayer should be complementary in pinning down the origin of the CDW orders. The observation of CDW without acoustic phonon anomaly and evolution of CDW amplitude modes may be important to resolve the CDW mechanism of the monolayer as discussed in the bulk case~\cite{Li_2021_observation,liu2022observation}.

We conclude our discussion by arguing that the $A$V$_3$Sb$_5$ monolayer should be possible to synthesize based on the following facts.
First, the cohesive energy generically indicates the thermodynamic stability of the monolayer systems. The cohesive energy is calculated as 3.8 eV/atom for all three alkali atoms (see details in Supplementary Note 1). This value is comparable to the bulk value of $\sim$ 3.9 eV/atom, supporting the thermodynamical stability of the monolayer.
In Fig.\,\ref{fig:6}\textbf{a}, we plot the cohesive energy difference of the ISD-1 and SD-2 phases $\Delta E_{\mathrm{coh}} (d) = E_{\mathrm{coh}}(d = 0) - E_{\mathrm{coh}}(d)$ as a function of the distortion of V atoms $d$. The CDW phases form a local minimum in the configuration space with an energy barrier of around 4 meV/atom.
Second, our DFT phonon calculations indicate that the monolayer should be dynamically stable. Exemplary phonon bands of the ISD-1 phase are shown in Fig.\,\ref{fig:6}\textbf{b}, which are clean of imaginary frequencies, showing the dynamical stability of the monolayer structure.  
Finally, the exfoliation energies of the monolayer are calculated as 42, 45, and 45 meV/$\AA^2$ for $A$ = K, Rb, and Cs, respectively. These values are amount to existing two-dimensional materials, such as graphene ($\sim$ 21 meV/$\AA^2$)~\cite{jung_rigorous_2018}, hBN ($\sim$ 28 meV/$\AA^2$)~\cite{jung_rigorous_2018}, and Ca$_2$N ($\sim$ 68 meV/$\AA^2$)~\cite{zhao_obtaining_2014}. 
We note that the recent experiments have successfully exfoliated thin layers of $A$V$_3$Sb$_5$ up to five layers using the taping methods \cite{song_competing_2021}. Current developments of the chemical solution reaction method could be an appropriate technique to weaken the interlayer interaction of the bulk system and separate the monolayer \cite{song2019creation}.

\begin{center}
\begin{table}[b]
\caption{
\textbf{CDW orders in the mean-field analysis.}
Both time-reversal symmetric and asymmetric types of CDWs are considered.
Translational symmetry-lowering of monolayer $A$V$_{3}$Sb$_{5}$ diversifies the order parameters, leading to a doublet of CDWs (CDW-1/2). 
When time-reversal symmetry is preserved, the sign of $\Phi$ specifies the different CDW patterns resulting in the ISD and SD depending on $\Phi < 0$ and $\Phi > 0$, respectively.
Without time-reversal symmetry, the TRSBs with opposite signs of $\Phi$ are equally classified.
The real space patterns of the CDW orders are illustrated in Fig.\ref{fig:4}\textbf{c}.
Here, the $4\times 4$ bond matrices are introduced with
${O}_{1}(\vv{k})
=(-\phi_1 I_2 +\phi_1^{*}\sigma_{x})\otimes \sigma_{z}
,\ 
{O}_{2}(\vv{k})
=\sigma_z \otimes (-\phi_2 I_2+\phi_2^{*}\sigma_{x})
,\ 
{O}_{3}(\vv{k})
=
\phi_3 \sigma_y \otimes \sigma_y 
+
\phi_3^{*}
\sigma_z \otimes \sigma_z$ and $\phi_{i}\equiv \exp(\vv{k}\cdot \vv{r}_{i})$ (See Eq.\ref{eq_MF_CDW}).
}
\label{MT1}
\renewcommand{\arraystretch}{1.2}
\begin{tabular}{|C{0.15\linewidth}|C{0.24\linewidth}|  C{0.1\linewidth}| C{0.12\linewidth}| }\hline \hline
Label  &\thead{Order parameter\\ $(\Phi_{\mathrm{AB}},\Phi_{\mathrm{AC}},\Phi_{\mathrm{BC}})$}& Pattern &Subgroup\\ \hline
ISD-1  &$\Phi(1,1,1)$,  $\Phi<0$& (c1)&
\multirow{2}{*}{$D_{2h}$}\\
ISD-2  &$\Phi(-1,1,-1)$,  $\Phi<0$& (c2)&\\
\hline
SD-1 & $\Phi(1,1,1)$, $\Phi>0$ & (c3)&
\multirow{2}{*}{$D_{2h}$}\\
SD-2 & $\Phi(-1,1,-1)$,  $\Phi>0$ & (c4)&\\
\hline
 TRSB-1 &$\Phi(i,-i,i)$,  $\Phi\neq0$& (c5)& 
\multirow{2}{*}{$C_{2h} $}\\
TRSB-2 & $\Phi(-i,-i,-i)$,  $\Phi\neq0$ &(c6)& \\
\hline
\hline
\end{tabular}
\end{table}
\end{center}

\begin{center}
\begin{table}[t]
\caption{ 
\textbf{
SC orders in the mean-field analysis.
}
Spin-singlet superconducting pairings are classified by irreducible representations (R) of the point group $D_{2h}$.
The condensation of $\hat{\Gamma}_{R}$ channels can lower the interaction energy, which preserves the  $\mathcal{T}_{1\times1}$-translational symmetry.
The momentum dependence of pairing gap functions is abbreviated by $(c_i,s_i) \equiv (\cos\vv{k} \cdot \vv{R}_i,\sin\vv{k} \cdot \vv{R}_i)$. 
To express pairing functions in both $1 \times 1$ and $2 \times 2$ unit cells, the $3 \times 3$ Gell Mann matrices $\lambda_a$ and the $12 \times 12$ matrices $L_{a}^{\pm, 0}=\lambda_{a}\otimes M_a^{\pm, 0} $ are introduced with $4 \times 4$ matrices $M_a^{\pm,0}$ (see Supplementary Note 7).
The fifth column shows the lowest-order basis functions for the corresponding irreducible representation.}
\label{MT2}
\renewcommand{\arraystretch}{1.2}
\begin{tabular}{|C{0.07\linewidth}|C{0.28\linewidth}|C{0.38\linewidth}|C{0.11\linewidth}|C{0.1\linewidth}| } \hline \hline
R&Pairing in $1\times 1$ unit cell &Pairing in $2\times 2$ unit cell $\hat{\Gamma}_{R}(\vv{k})$  &Label &\thead{Basis\\ function}\\ \hline 
\multirow{3}{*}{ A$_{g}$ }& $\frac{1}{\sqrt{3}}\lambda_{0},\frac{1}{\sqrt{2}}\left[\frac{\sqrt{3}}{2}\lambda_{7}+\frac{1}{2}\lambda_{8}\right],$& $\frac{1}{\sqrt{3}}L_{0}^{0},\frac{1}{\sqrt{2}}\left[\frac{\sqrt{3}}{2}L_{7}^{0}+\frac{1}{2}L_{8}^{0}\right],$&$\!\Gamma_{A_{g}}^{(1)},\!\Gamma_{A_{g}}^{(2)}\!,$& \multirow{3}{*}{$x^2,y^2,z^2$ } \\ 
& $\frac{1}{\sqrt{2}}\left[c_{1}\lambda_{1}+c_{2}\lambda_{2}\right]$, &   $\frac{1}{\sqrt{2}}\left[c_{1}L_{1}^{+}+c_{2}L_{2}^{+}-(s_{1}L_{4}^{-}+s_{2}L_{5}^{-})\right],$& $\Gamma_{A_{g}}^{(3)},$ &\\ 
& $c_{3}\lambda_{3}$ &   $\left[c_{3}L_{3}^{+}-s_{3}L_{6}^{-}\right]$  & $\Gamma_{A_{g}}^{(4)}$&\\   \hline
\multirow{2}{*}{ B$_{1g}$  }& $\frac{1}{\sqrt{2}}\left[-\frac{1}{2}\lambda_{7}+\frac{\sqrt{3}}{2}\lambda_{8}\right],$& $\frac{1}{\sqrt{2}}\left[-\frac{1}{2}L_{7}^{0}+\frac{\sqrt{3}}{2}L_{8}^{0}\right],$ &$\Gamma_{B_{1g}}^{(1)},$&\multirow{2}{*}{$xy$}\\
& $\frac{1}{\sqrt{2}}\left[c_{1}\lambda_{1}-c_{2}\lambda_{2}\right]$ & $\frac{1}{\sqrt{2}}\left[c_{1}L_{1}^{+}-c_{2}L_{2}^{+}-(s_{1}L_{4}^{-}-s_{2}L_{5}^{-})\right]$&$\Gamma_{B_{1g}}^{(2)}$& \\  \hline
\multirow{1}{*}{ B$_{2u}$  }& $\frac{1}{\sqrt{2}}\left[s_{1}\lambda_{4}-s_{2}\lambda_{5}\right]$ &$\frac{1}{\sqrt{2}}\left[s_{1}L_{4}^{+}-s_{2}L_{5}^{+}-(c_{1}L_{1}^{-}-c_{2}L_{2}^{-})\right]$&$\Gamma_{B_{2u}}^{(1)}$ &$y$ \\  \hline 
\multirow{2}{*}{ B$_{3u}$  }& $\frac{1}{\sqrt{2}}\left[s_{1}\lambda_{4}+s_{2}\lambda_{5}\right],$ & $\frac{1}{\sqrt{2}}\left[s_{1}L_{4}^{+}+s_{2}L_{5}^{+}+(c_{1}L_{1}^{-}+c_{2}L_{2}^{-})\right],$ &$\Gamma_{B_{3u}}^{(1)},$&\multirow{2}{*}{$x$}\\
&$s_{3}\lambda_{6}$ & $\left[s_{3}L_{6}^{+}-c_{3}L_{3}^{-}\right]$&$\Gamma_{B_{3u}}^{(2)}$&
\\  
\hline
\hline
\end{tabular}
\end{table}
\end{center}

\section{Methods}

\textit{Tight-binding model. -}
We construct a TB model for the monolayer with $D_{2h}$ symmetry in the $\sqrt{3}\times 1$ unit cell.
Introducing the six-component spinor, 
$\Psi_{\vv{k}}^{T}=(A_{\vv{k}}^{T},B_{\vv{k}}^{T},C_{\vv{k}}^T)$ with $\alpha_{\vv{k}}^{T}=(\alpha_{1,\vv{k}},\alpha_{2,\vv{k}})$,
the Hamiltonian becomes 
$H_{0}=\sum_{\vv{k}}\Psi_{\vv{k}}^{\dagger}\mathcal{H}_{0}(\vv{k})
\Psi_{\vv{k}}$.
The indices for sublattice $\alpha\in\{A,B,C\}$ and site $i\in \{1,2\}$ are used.
The Bloch Hamiltonian $\mathcal{H}_{0}(\vv{k})$ is given by
\begin{eqnarray}
\mathcal{H}_{0}(\vv{k}) &=&
-t
\left[ 
(c_1\lambda_1 -s_1\lambda_4 +c_2\lambda_2 -s_2 \lambda_5+2c_3 \lambda_3)\otimes I_2
+(c_1\lambda_1 +s_1\lambda_4 +c_2\lambda_2 +s_2 \lambda_5)\otimes \sigma_{x} 
\right]\nonumber\\
&&
-t_2
\left[ 
(c_4\lambda_1 -s_4\lambda_4 +c_5\lambda_2 -s_5 \lambda_5)\otimes I_2
+(c_4\lambda_1 +s_4\lambda_4 +c_5\lambda_2 +s_5 \lambda_5+2c_6 \lambda_3)\otimes \sigma_{x} 
\right]\\
&&
+\epsilon I_{6}
+\frac{\delta \epsilon}{6} \left[(2\lambda_{0}+3 \lambda_7 + \sqrt{3}\lambda_8)\otimes \sigma_z \right]
-2\delta t s_{3} \left[ \lambda_6 \otimes \sigma_z \right]
,\nonumber
\end{eqnarray}
where $(c_i,s_i)\equiv(\cos\vv{k} \cdot \vv{r}_i,\sin\vv{k} \cdot \vv{r}_i)$ with $\vv{r}_1=\frac{1}{2}(\sqrt{3},1)$, $\vv{r}_2=\frac{1}{2}(\sqrt{3},-1)$, $\vv{r}_3=(0,1)$, $\vv{r}_4=\vv{r}_2-\vv{r}_3$,
$\vv{r}_5=\vv{r}_1+\vv{r}_3$, and 
$\vv{r}_6=\vv{r}_1+\vv{r}_2$.
Here, $t$, $t_2$, $\epsilon$, $\sigma_i$ and $\lambda_i$ are nearest-neighbor hopping, next nearest-neighbor hopping, onsite energy, $2\times2$ Pauli matrices and $3\times3$ Gell-Mann matrices, respectively.
The sublattice information and the definition of the Gell-Mann matrices are provided in Supplementary Fig.\,S5\textbf a and Supplementary Note\,7, respectively.
The last two terms give rise to onsite energy difference $2\delta \epsilon$ between two A sublattices and staggered hopping $\pm \delta t$ between the B and C sublattices, respectively.
These two terms lead to the symmetry-lowering from $\mathcal{T}_{1\times 1}$ $D_{6h}$ to $\mathcal{T}_{\sqrt{3}\times 1}$ $D_{2h}$.
With parameters $(\epsilon,t,t_2,\delta \epsilon,\delta t)=(0.01,0.42,0.03,-0.033,0.01)$, our TB model well reproduces the DFT bands as well as the irreducible representations of the VHS bands at $\Gamma$ and $M$ ($A_{g}$) for KV$_3$Sb$_5$ (see Figs.\,\ref{fig:2}\textbf a,\textbf b).
For Rb and Cs, we also find good agreement between the TB and DFT (see Supplementary Note 2).
The TB band structures of the TRSB-1/2 phases in Fig.\,\ref{fig:5}\textbf a are obtained by using parameters $(\epsilon, t, t_2, \delta \epsilon, \delta t, \Phi) = (0.035, 0.42, 0.03, -0.033, 0.01, 0.07)$ where $\Phi$ is a CDW magnitude of the TRSB-1/2 phases (see Supplementary Note 8 for detailed order parameters of the TRSB-1/2 phases).

\textit{Electronic susceptibility. -} We calculate the real part of the bare electronic susceptibility in the constant matrix approximation $\chi(\bm{q})$ given by 
\begin{eqnarray}
\chi(\bm{q})
\equiv -\sum_{\bm{k},\mu,\nu}
\left[
\frac{n_{F}(E_{\mu}(\bm{k}))-n_{F}(E_{\nu}(\bm{k}+\bm{q}))}
{E_{\mu}(\bm{k})-E_{\nu}(\bm{k}+\bm{q})}
\right],
\end{eqnarray}
where $n_{F}(\epsilon)$ is the Fermi-Dirac distribution and $(\mu,\nu)$ is band index. For the calculation of $\chi(\bm{q})$ in Fig.\,\ref{fig:3}, TB bands with $200\times348$ $\vv k$-points and $\beta=1000$ are used.

\textit{Mean-field theory. -} 
We perform the conventional zero-temperature mean-field method to investigate the interaction effects.
Six distinct CDW configurations and nine spin-singlet SC orders are chosen to uncover CDW and SC instabilities in our analysis (see Fig.\,\ref{fig:4}\textbf c and Table \ref{MT1},\ref{MT2}). 
We introduce the CDW and SC order parameters $(\Phi,\Delta)$ that break translational and U(1) symmetry.
In the $2\times 2$ unit cell, the mean-field Hamiltonians for CDW and SC orders are written as,
\begin{eqnarray}
\delta H_{\mathrm{MF}}^{\mathrm{CDW}}[\vv{\Phi}]
= \sum_{\vv{k},\sigma}\!
\left[
\Phi_{\mathrm{AB}}\!
\left(
\tilde{A}_{\vv{k}\sigma}^{\dagger}O_{1} (\vv{k})\tilde{B}_{\vv{k}\sigma}
\right)\!
+
\Phi_{\mathrm{AC}}\!
\left(
\tilde{A}_{\vv{k}\sigma}^{\dagger}O_{2} (\vv{k})\tilde{C}_{\vv{k}\sigma}
\right)\!
+
\Phi_{\mathrm{BC}}\!
\left(
\tilde{B}_{\vv{k}\sigma}^{\dagger}O_{3} (\vv{k})\tilde{C}_{\vv{k}\sigma}
\right)
+\mathrm{C.C.}
\right],  \ \ \  \ \ \  \label{eq_MF_CDW}
\end{eqnarray}
and
\begin{eqnarray}
\ 
\delta H_{\mathrm{MF}}[\Delta]
=\Delta \left[\sum_{\vv{k}}
\tilde{\Psi}_{\vv{k}\uparrow}^{\dagger}
\hat{\Gamma}(\vv{k})
\tilde{\Psi}_{\vv{k}\downarrow}^{*}+\mathrm{C.C.}\right], 
\end{eqnarray}
with the twelve-component spinor, 
$\tilde{\Psi}_{\vv{k}}^{T}=(\tilde{A}_{\vv{k}}^{T},\tilde{B}_{\vv{k}}^{T},\tilde{C}_{\vv{k}}^T)$ and $\tilde{\alpha}_{\vv{k}}^{T}=(\alpha_{1,\vv{k}},\alpha_{2,\vv{k}},\alpha_{3,\vv{k}},\alpha_{4,\vv{k}})$.
The indices $i \in \{1,2,3,4\}$ and $\sigma\in \{\uparrow,\downarrow\}$ denote the orbital sites and spins, respectively.
 The specific forms of $4\times 4$ bond matrices $O_{a}(\vv{k})$ and $12\times 12$ pairing gap functions $\hat{\Gamma}(\vv{k})$ are tabulated in Table \ref{MT1}, \ref{MT2}.
 In our mean-field ansatz, the tested CDW order parameters are restricted to the $2\times 2$ CDW orders carrying $l=1$ angular momentum where the hopping strength modulates with the Q-vectors connecting $M_{1},M_{2},M_{3}$ points in the real space,
 \begin{eqnarray}
\Phi_{\mathrm{AB}}&=&\sum_{\vv{R}} \cos(\vv{M}_{3}\cdot \bm{R}) \langle A_{\vv{R}}^{\dagger}B_{\vv{R}}- A_{\vv{R}}^{\dagger}B_{\vv{R}-2\vv{r}_1}
 \rangle \equiv \Phi \sigma_{AB},\\
  \Phi_{\mathrm{AC}}&=&\sum_{\vv{R}} \cos(\vv{M}_{2}\cdot \bm{R}) \langle A_{\vv{R}}^{\dagger}C_{\vv{R}}- A_{\vv{R}}^{\dagger}C_{\vv{R}-2\vv{r}_2}
 \rangle \equiv \Phi \ \sigma_{AC}, \\
\Phi_{\mathrm{BC}}&=&\sum_{\vv{R}} \cos(\vv{M}_{1}\cdot \bm{R}) \langle B_{\vv{R}}^{\dagger}C_{\vv{R}}- B_{\vv{R}}^{\dagger}C_{\vv{R}+2\vv{r}_3}
 \rangle \equiv \Phi\ \  \sigma_{BC},
 \end{eqnarray}
where the overall amplitude $\Phi$ quantifies the modulation strength and the relative phase of $(\sigma_{\mathrm{AB}},\sigma_{\mathrm{AC}},\sigma_{\mathrm{BC}}) \in \{\pm 1,\pm i\}$ determines the pattern of CDW. 
Here, we fix the order parameter $\phi \in (\Phi, \Delta)$ is set as a real value.
After constructing the ground state of the mean-field Hamiltonian $|G,\phi\rangle$ and evaluating the ground state energy $ E[\phi] \equiv \langle G;\phi | H_{0}+H_{\mathrm{int}}|  G;\phi \rangle$, we obtain the phase diagrams presented in Fig.\,\ref{fig:4} (see Supplementary Notes 8 and 9 for details).
161 grids of the ground energy with $\phi_i\in [-0.08,0.08]$ are utilized and their numerical integration at given $\phi_{i}$ are carried out with $80\times 80$ $\vv k$-points.

{\em First-principles calculations. -}
We perform density-functional theory (DFT) calculations using the Vienna $ab$ $initio$ simulation package (VASP)~\cite{VASP1,VASP2} with the projector-augmented wave method~\cite{PAW}.  
For the exchange-correlation energy, the generalized-gradient approximation functional of Perdew-Burke-Ernzerhof~\cite{PBE} is employed. 
The van der Waals correction is included within the zero damping DFT-D3 method of Grimme~\cite{Grimme-D3}.
The kinetic energy cutoff for the plane wave basis is 300 eV. 
The force criteria for optimizing the structures is set to  $0.01$~eV/$\text{\AA}$.
The monolayer $A$V$_3$Sb$_5$ is simulated using a periodic supercell with a vacuum spacing of $\sim 20~\text{\AA}$.
The $\vv k$-space integration is done with $10\times17$ $\vv k$-points for the $\sqrt{3}\times1$ structure. 
For the DOS calculation, we use $46\times69$ $\vv k$-points.
The exfoliation energy is calculated using the Jung-Park-Ihm method \cite{jung_rigorous_2018}.
We employ the finite difference method to obtain phonon dispersions. For the pristine monolayer (Fig. \ref{fig:3}\textbf{e}), we use $4\times4$ supercell with $9\times9$ $\vv k$-points with the Phonopy software~\cite{phonopy}. For the ISD-1 phase (Fig. ~\ref{fig:6}\textbf{b}), we use $8\times8$ supercell with $6\times6$ $\vv k$-points using the nondiagonal supercell method~\cite{monserrat_phonon}.
We evaluate the $(U,V)$ values in monolayer by using the first-principles constrained random phase approximation (cRPA) method~\cite{cRPA_2004} with the weighting approach~\cite{cRPA_weighted}, as implemented in VASP.

\textit{Berry curvature and anomalous Hall conductivity. -}
The berry curvature $\Omega_n(\vv{k})$ associated with the $n$-th energy band of the TB Hamiltonian $H(\vv{k})$ is given by 
\begin{eqnarray}
\Omega_{n}(\vv{k})=-2\sum_{m\neq n}\frac{\text{Im}\braket{u_{n\vv{k}}|\partial_{x} H(\vv{k})
|u_{m\vv{k}}}\braket{u_{m\vv{k}}|
\partial_{y} H(\vv{k})
|u_{n\vv{k}}}}{[E_n(\vv{k})-E_m(\vv{k})]^2}.
\end{eqnarray}
Here, $E_n(\vv{k})$ and $\ket{u_{n\vv{k}}}$ are $n$th eigenvalue and eigenstate of $H(\vv{k})$ and the derivative in the momentum space $\partial_{i}\equiv \frac{\partial}{\partial k_{i}}$ is adopted.
The anomalous Hall conductivity is calculated as a function of energy $E$ by integrating the $n$-band Berry curvatures for $E_n < E$ over the BZ
\begin{eqnarray}
\sigma_{xy}(E)=\frac{1}{2\pi}\frac{e^2}{h}\int d^2k \sum_{E_n<E}\Omega_{n}(\vv{k}).
\end{eqnarray}
The numerical integration is performed by using $500\times500$ $\vv k$ points.

\section{Data availability}
The data that support the findings of this study are available within the paper and Supplementary Information. Additional relevant data are available from the corresponding authors upon request.

\section{Code availability}
The code that supports the findings of this study is available from the corresponding authors upon request.

\section{Acknowledgments}

\begin{acknowledgments}
This work was supported by the Korean National Research Foundation (NRF) Basic Research Laboratory (NRF-2020R1A4A3079707). 
Y.K.  acknowledges the support from the NRF Grant numbers (NRF-2021R1A2C1013871, NRF-2021M3H3A1038085). E.-G. M acknowledges the support from the NRF funded by the Ministry of Science and ICT (No. 2019M3E4A1080411, No. 2021R1A2C4001847, No. 2022M3H4A1A04074153), National Measurement Standard Services and Technical Services for SME funded by
Korea Research Institute of Standards and Science (KRISS –2022 – GP2022-0014). 
The computational resource was provided by the Korea Institute of Science and Technology Information (KISTI) (KSC-2020-CRE-0108) and the Cambridge Tier-2 system operated by the University of Cambridge Research Computing Service and funded by EPSRC [EP/P020259/1].
\end{acknowledgments}

\section{Author contributions}
E.G.M. and Y.K. designed and organized the research. 
S.W.K. and H.O. performed all calculations.
E.G.M. and Y.K. supervised the research. 
All authors discussed the results and contributed to writing the manuscript.

\section{Competing interests}
The authors declare no competing interests.

\ifarXiv
    

\section*{Supplementary material}

**Supplementary Note 1. Energetics of various monolayer structures**

In this section, we provide details about the constraint of stoichiometry on the monolayer structure and about the energetics of the monolayer structure satisfying the stoichiometry of $A$V$_3$Sb$_5$. We first present the effect of stoichiometry on the stability of monolayer structures by comparing the cohesive energy between monolayer structures with different stoichiometry. The cohesive energy per atom $E_{coh}$ is given by

$$E_{coh} = [n_A E_A + n_V E_V + n_{Sb} E_{Sb} - E_{tot}]/(n_A + n_V + n_{Sb}), \qquad (1)$$

where $E_A$, $E_V$, $E_{Sb}$, and $E_{tot}$ are the total energies of a single $A$ atom, a single V atom, a single Sb atom, and monolayer, respectively. The cohesive energy $E_{coh}$ is calculated as 3.8 (3.6) eV for the monolayer structures having the stoichiometry of $A$V$_3$Sb$_5$ ($A_2$V$_3$Sb$_5$), which implies that the monolayer structure with the stoichiometry $A_2$V$_3$Sb$_5$ is less likely realized in the experiment. This is in line with the experimental observations [1, 2] that show the $A$-terminated surface is unstable for exfoliated samples because $A$ atoms are prone to form a cluster. Thus, we only consider the monolayer structure satisfying the stoichiometry of

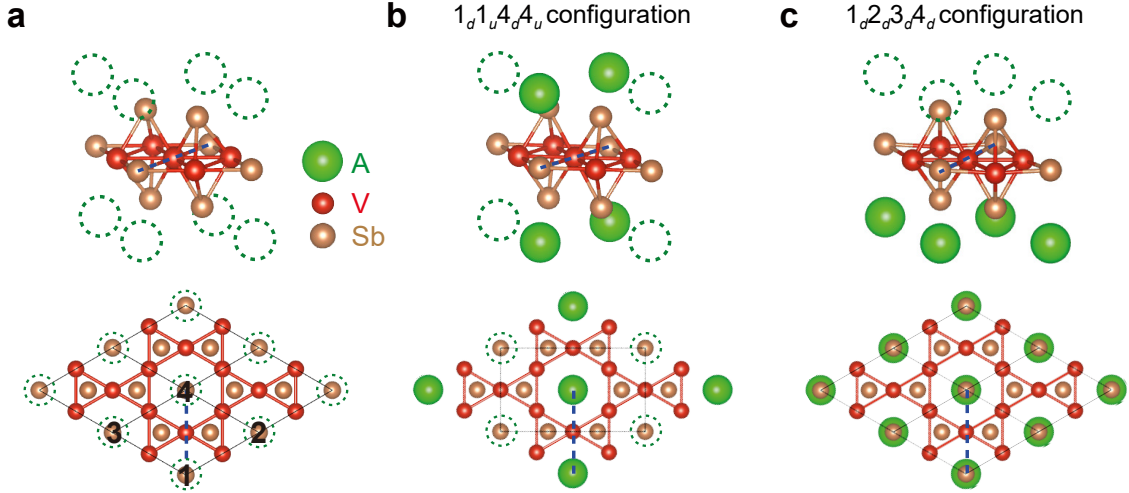

FIG. S1: **Various monolayer structures satisfying the stoichiometry of $A$V$_3$Sb$_5$.** **a** Possible atomic sites (green dashed open circles; labeled as 1 to 4 with up and down positions relative to V-Sb triple layers) of alkali metal atoms $A$ in the monolayer structure. Black lines indicate the $1 \times 1$ unit cell. Dashed blue lines are drawn to show the direction in the geometries. **b,c** Two exemplary configurations $1_d 1_u 4_d 4_u$ and $1_d 2_d 3_d 4_d$. The $1_d 1_u 4_d 4_u$ configuration is the lowest energy structure of the monolayer $A$V$_3$Sb$_5$ which is studied in the main text.

3

$A$V$_3$Sb$_5$.

Next, keeping the stoichiometry of $A$V$_3$Sb$_5$, we obtain energetics of various monolayer structures by considering all the possible configurations of the alkali atoms (Fig. S1 and Table S1). There are six symmetrically distinct monolayer structures depending on the configurations of the alkali atoms: $1_d1_u4_d4_u$, $1_d1_u2_d3_d$, $1_d1_u2_d3_u$, $1_d2_d3_d4_d$, $1_d2_d3_d4_u$, and $1_d2_u3_d4_u$ where $1_d$ refers the position of a alkali atom located at site 1 (see site indices in Fig. S1**a**) and below the V-Sb triple layers ("down" position; $d$). The configurations of the alkali atoms determine the lattice type and symmetry of monolayer structures (see Table S1). We find that $1_d1_u4_d4_u$ structure having rectangular lattice with $\mathcal{T}_{\sqrt{3}\times1}$ translational and $D_{2h}$ point group symmetries (Fig. S1**b**) is the most energetically favorable for all three alkali atoms. Note that the rectangular lattice is commonly observed in the surface of exfoliated bulk [3–5], which supports our calculation results.

TABLE S1: **Lattice type, symmetry, and energetics for various pristine monolayer structures satisfying the chemical stoichiometry of $A$V$_3$Sb$_5$.** Six symmetrically distinct configurations of $A$ atoms are given. The configurations are labelled by the site index $i$ ($i = 1, 2, 3, 4$) and height index $u$ or $d$ (up and down positions relative to the V-Sb triple layers), as shown in Fig. S1. For energetics, zero energy is set to the lowest total energy in each column.

| Configuration | Lattice | Symmetry | Energetics (Unit: meV/f.u.) | | |
|---|---|---|---|---|---|
| | | | K | Rb | Cs |
| $1_d1_u4_d4_u$ | Rectangular | $\mathcal{T}_{\sqrt{3}\times1}$ $D_{2h}$ | 0.0 | 0.0 | 0.0 |
| $1_d1_u2_d3_d$ | Triangular | $\mathcal{T}_{2\times2}$ $C_{2v}$ | 29.4 | 39.3 | 49.2 |
| $1_d1_u2_d3_u$ | Triangular | $\mathcal{T}_{2\times2}$ $D_2$ | 0.3 | 0.7 | 1.1 |
| $1_d2_d3_d4_d$ | Triangular | $\mathcal{T}_{1\times1}$ $C_{6v}$ | 125.9 | 166.4 | 206.5 |
| $1_d2_d3_d4_u$ | Triangular | $\mathcal{T}_{2\times2}$ $C_{2v}$ | 30.9 | 41.1 | 51.1 |
| $1_d2_u3_d4_u$ | Triangular | $\mathcal{T}_{2\times2}$ $S_2$ | 1.7 | 2.7 | 3.2 |

**Supplementary Note 2. Rb and Cs systems**

**2.1. Band structures and VHS of the pristine phase**

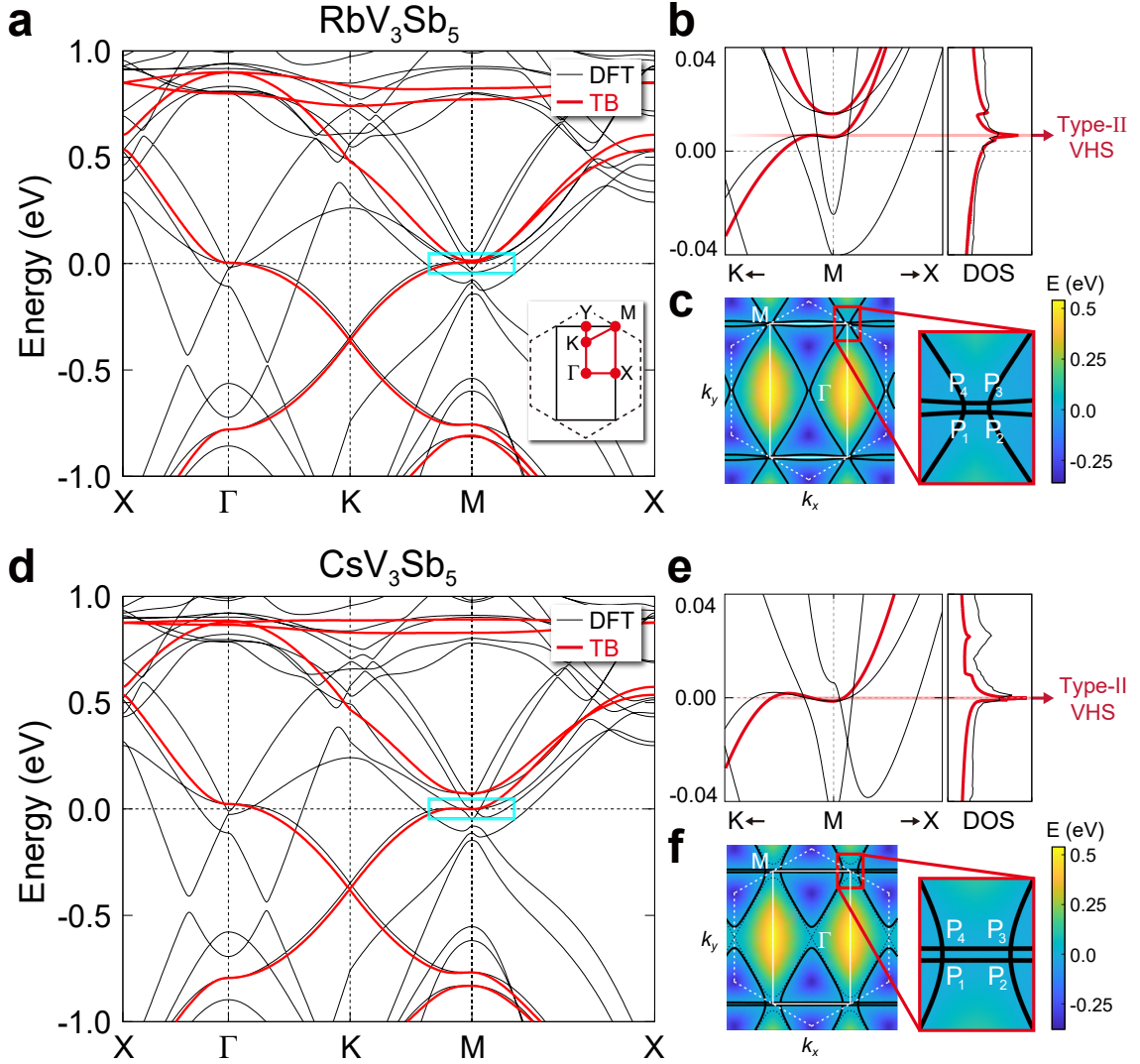

FIG. S2: **Electronic structures and type-II VHS points in the $RbV_3Sb_5$ and $CsV_3Sb_5$ monolayer.** **a** DFT and TB band structures of the $RbV_3Sb_5$ monolayer. The black and red color schemes are used to indicate the DFT and TB bands. Solid (dashed) line in inset of **a** depicts the BZ of the $\sqrt{3} \times 1$ ($1 \times 1$) unit cell. **b** Magnified view of the boxed region in **a**. The density of states (DOS) is also shown where the energy of type-II VHSs is indicated by the red arrow. **c** Energy contours containing saddle points of the TB bands. White dashed (solid) lines show the pristine (reduced) BZ. The black solid and dashed contour curves correspond to $E = 6$ and $4$ meV, respectively. **d-f** Same data set as **a-c** but for the $CsV_3Sb_5$ monolayer. In **f**, the black solid and dashed contour curves correspond to $E = 0$ and $24$ meV, respectively.

5

Figure S2 shows the band structures and type-II VHSs of the RbV$_3$Sb$_5$ and CsV$_3$Sb$_5$ monolayer. We find that the TB band structures well reproduce the VHS bands of the DFT results for the RbV$_3$Sb$_5$ and CsV$_3$Sb$_5$ monolayer. The TB parameters used for the RbV$_3$Sb$_5$ and CsV$_3$Sb$_5$ monolayer are $(\epsilon, t, t_2, \delta\epsilon, \delta t, \delta\epsilon_2) = (0.01, 0.42, 0.025, -0.0485, 0.0023, -0.0044)$ and $(0.036, 0.42, 0.005, -0.026, 0.0185, -0.012)$, respectively. Here, we introduce the additional onsite term $\frac{\delta\epsilon_2}{6}\left[(2\lambda_0 + 3\lambda_7 + \sqrt{3}\lambda_8) \otimes \sigma_0\right]$ to reproduce the exact energy of type-I VHSs in DFT results, which does not alter the symmetry. We also find that the type-II VHSs appear in the RbV$_3$Sb$_5$ and CsV$_3$Sb$_5$ monolayer, as shown in the DOS and TB energy contours in Fig. S2.

## 2.2. Phonon softening of the pristine phase

We perform a systematic study of the DFT phonon spectra of the monolayer. We find that there exists phonon softening in the pristine monolayer, as in the case of the bulk [6]. The DFT phonon spectra of the pristine monolayer for all three alkali atoms (Fig. S3) consistently show the phonon softening at the $\Gamma$ and $M$ points, associated with the distortion of V atoms. Here, the $\Gamma$ ($M$) point corresponds to the one (other two) of the $2 \times 2$ $3Q$ CDW vectors in the folded BZ. These phonon spectra of the pristine monolayer clearly indicate structural instability, which justifies the consideration of the CDW in the monolayer.

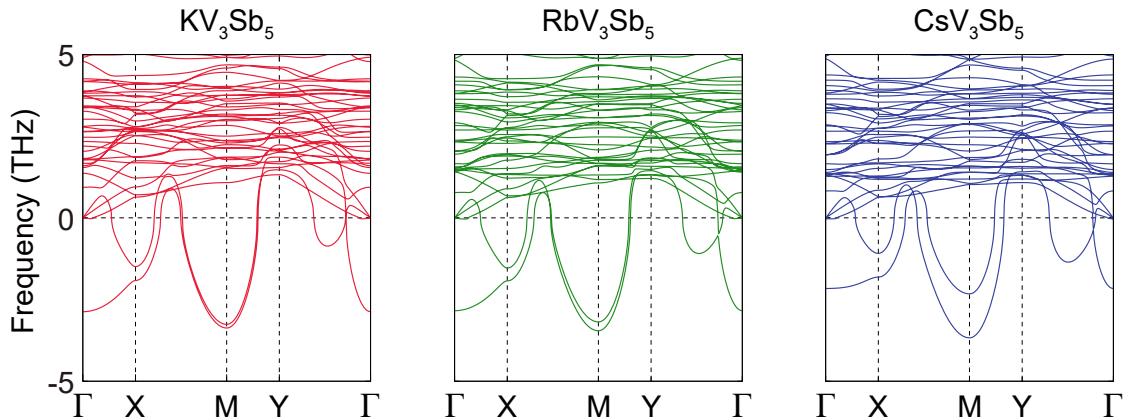

FIG. S3: **Phonon bands of the pristine phases for the $A$V$_3$Sb$_5$ monolayer family.** Here, the smearing factor $\sigma = 0.01$ eV is used.

## 2.3. Phonon spectra of the CDW phase

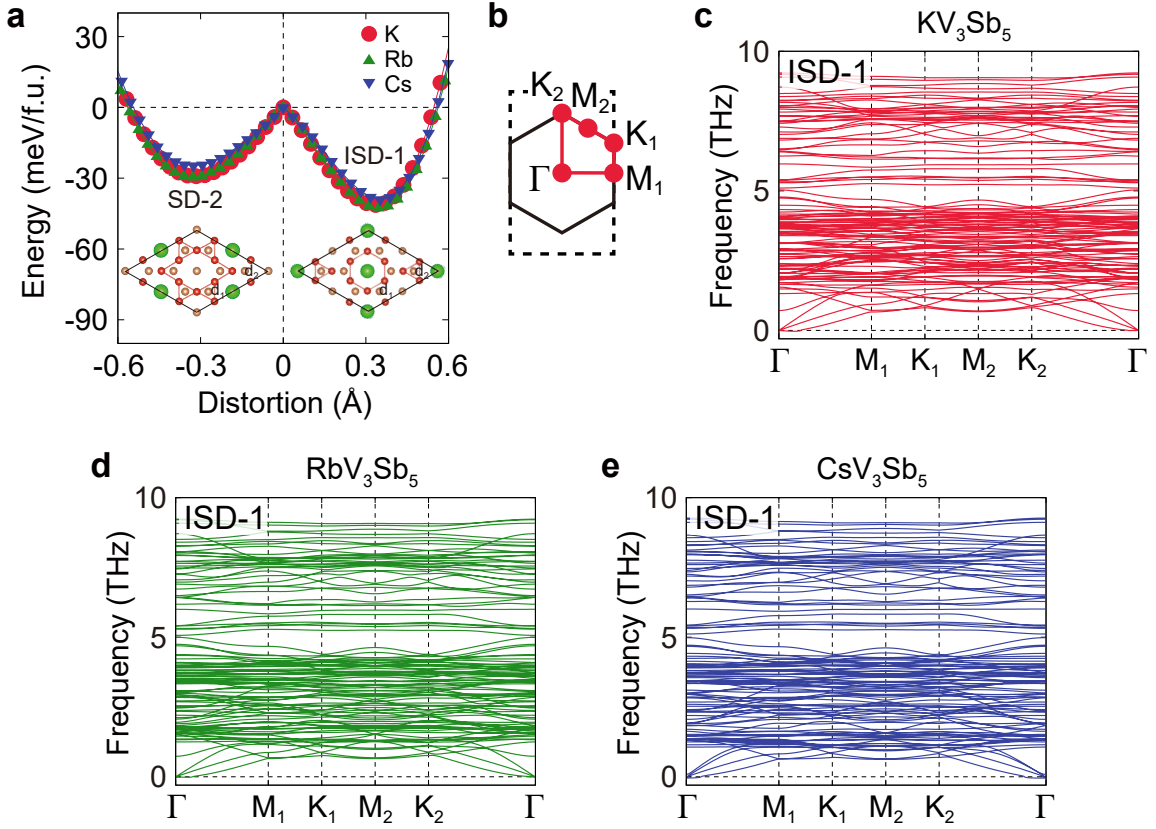

FIG. S4: **Energy profile and phonon band structures of the DFT ground state for the $A$V$_3$Sb$_5$ monolayer.** **a** Energy profile as a function of the distortion for the $A$V$_3$Sb$_5$ monolayer. The distortion is defined as $d \equiv d_1 - d_2$ where positive (negative) distortion corresponds to the $2 \times 2$ ISD-1 (SD-2) CDW phase whose geometry is displayed in the inset. **b** BZ and high symmetry points for the $2 \times 2$ unit cell. BZ for the $\sqrt{3} \times 1$ unit cell is drawn in a dashed line. **c-e** Phonon band structures for the ISD-1 phase of the $A$V$_3$Sb$_5$ monolayer. The $4 \times 4$ supercell with $9 \times 9$ $k$-points is used for the phonon band calculations.

This section provides the details of the DFT ground state and its stability in the $A$V$_3$Sb$_5$ monolayer. By extensively inspecting diverse CDW distortions, we find a stable structure in two dimensions. The $2 \times 2$ ISD-1 CDW phase is the DFT ground state and is dynamically stable for all three alkali atoms $A$ = K, Rb, and Cs (Fig. S4). The $2 \times 2$ ISD-1 phase is energetically favored over the pristine counterpart by 41, 43, and 39 meV per formula unit for $A$ = K, Rb, and Cs, respectively (Fig. S4a). It is noteworthy that the energy gain under the ISD distortion in the monolayer is much higher than that of the bulk structure (2, 8, and 15 meV per formula unit for $A$ = K, Rb, and Cs, respectively), which implies the

7

enhanced instability of the 2D system. Moreover, nearly identical energy gains and CDW distortion magnitudes, irrespective of the type of alkali atoms $A$, for the ISD-1 phase reflect the strong 2D nature of V atoms in the $A$V$_3$Sb$_5$ monolayer. On the other hand, the 2 × 2 SD-2 CDW phase is metastable in the configuration space, as shown in the energy profile (Fig. S4**a**). We also find that the SD-1 and ISD-2 phases are unstable against their pristine counterparts, which indicates that the energetically preferred positions of alkali atoms are different between the ISD and SD phases.

**Supplementary Note 3. Analysis on the VHS points**

### 3.1. Sublattice character of VHS

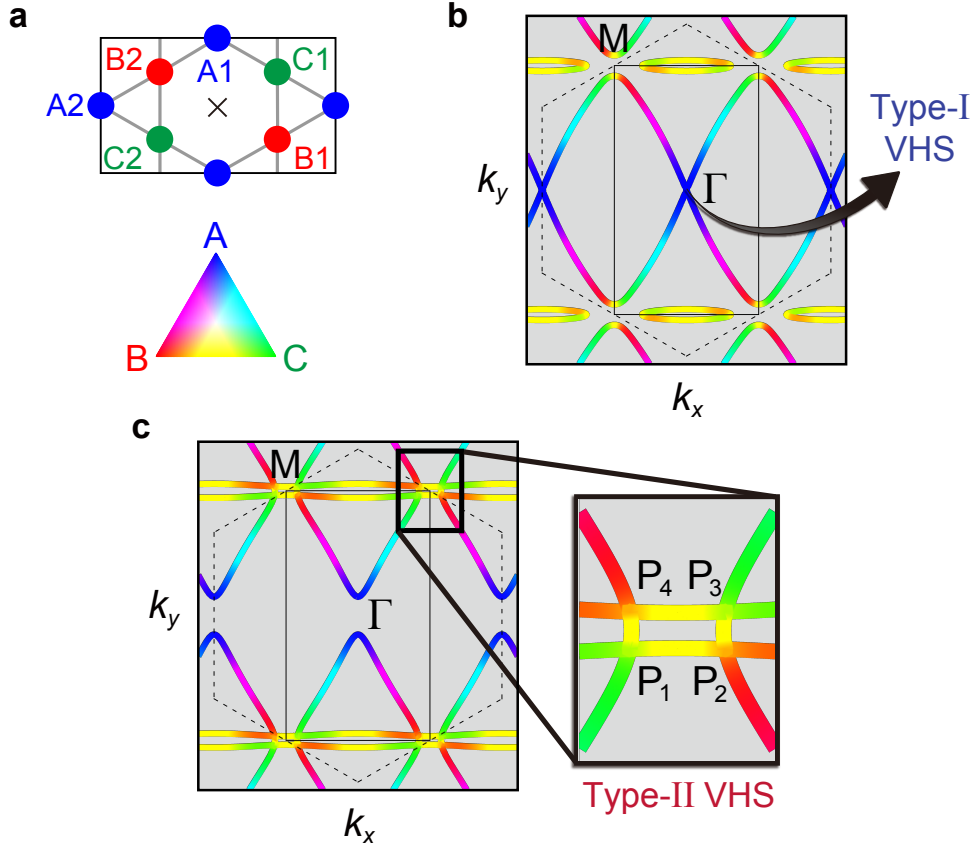

FIG. S5: **Sublattice character of the monolayer $A$V$_3$Sb$_5$.** **a** Six sublattices in the $\sqrt{3} \times 1$ unit cell for the TB model. The three different sublattices A, B, and C are colored in blue, red, and green, respectively, as shown in the color triangle. The position of alkali atoms is symbolically marked by × in the center of the unit cell. **b,c** Sublattice-projected energy contours of the TB bands for the KV$_3$Sb$_5$ monolayer. The contour curves in **b** and **c** correspond to $E = 9$ meV (type-I VHS) and $-6$ meV (type-II VHS), respectively. Thin black dashed (solid) lines show the BZ of the $1 \times 1$ ($\sqrt{3} \times 1$) unit cell.

Figure S5 shows the sublattice character of the VHS points in the monolayer. The sublattice character of the four type-II VHS points at $P_i$ shows the mixing of B and C sublattices due to the symmetry lowering. Specifically, the ratio between two sublattices is B : C = 23% : 77% for $P_1$ and $P_3$ points and B : C = 77% : 23% for $P_2$ and $P_4$ points. Such sublattice-mixed type-II VHS points are in contrast to the sublattice-pure type-I VHS point

9

at Γ which consists of the single A sublattice.

### 3.2. Rearrangement of VHS

The appearance of type-I and type-II VHS points in the monolayer $AV_3Sb_5$ can be understood with the analogy of hybridization. Starting from the bulk system with $(D_{6h}, T_{1\times 1})$ symmetries, the monolayer system is achieved by breaking both point group and translation symmetries, where the two states $\bm{k}$ and $\bm{k} + \bm{M}_1$ are mixed as a results of zone folding. Then, the type-II VHS is associated with the hybridization between two different $M$ points at the Brillouin zone boundary (say, $\bm{k} = \bm{M}_2$ or $\bm{M}_3$) while the type-I VHS at $\Gamma$ is just a shift of one M point (say, $\bm{k} = \bm{M}_1$), as illustrated in Fig.S6. The symmetry-lowering of the monolayer system hybridizes the two bands with different sublattice constitutions at the zone boundary, represented by different colors (red and green in Fig.S6) and produces the four type-II VHSs.

One of our observations is that the number of the resultant VHS points (four in our tight-binding model, Fig.S6) is not universal but crucially depends on the microscopic details of the system such as a Fermi-surface nesting condition parameter. To show these non-universal behaviors, it is useful to apply the effective two-band model, focusing on the bands which include the saddle points. One can perform the Taylor expansion near the saddle point $\bm{k} = \bm{M}_2 + \bm{p}$, and obtain the effective $2 \times 2$ Hamiltonian,

$$\mathcal{H}_{\text{two}}^0(\bm{p}) = \begin{pmatrix} \epsilon_2(\bm{p}) & 0 \\ 0 & \epsilon_3(\bm{p}) \end{pmatrix}, \quad \delta\mathcal{H}_{\text{two}}(\bm{p}) = \begin{pmatrix} 0 & \Delta \\ \Delta & 0 \end{pmatrix}, \quad (2)$$

where $\epsilon_a(\bm{p})$ is expanded energy dispersion near $M_a$ and $\Delta$ is an exemplified symmetry breaking term from $(T_{1\times 1}, D_{6h})$ to $(T_{\sqrt{3}\times 1}, D_{2h})$. We rotate the axes to diagonalize the Hessian matrix of the energy near the VHS of $(T_{1\times 1}, D_{6h})$ system which gives

$$\epsilon_2(\bm{p}) = \frac{\alpha}{2} \begin{pmatrix} p_x & p_y \end{pmatrix} \begin{pmatrix} 1 & 0 \\ 0 & -\tan(\frac{\pi}{3} + \beta)^2 \end{pmatrix} \begin{pmatrix} p_x \\ p_y \end{pmatrix} + \cdots, \quad (3)$$

where the higher order terms, $O(\bm{p}^3)$, are omitted in $\cdots$. Here, the two coefficients $(\alpha > 0, -\frac{\pi}{6} < \beta < \frac{\pi}{6})$ determine the behavior of VHS and Fermi-surface contours. For

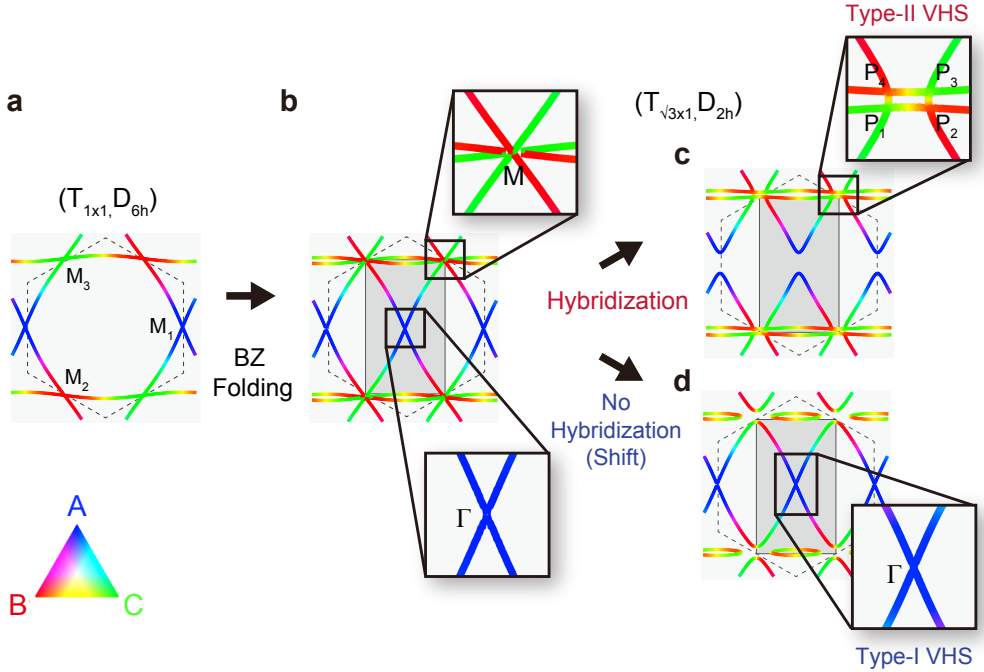

FIG. S6: **Schematic illustration of the type-I and -II VHS points in momentum space rearranged by symmetry lowering. a,b**. The VHS point distribution in $(T_{1\times 1}, D_{6h})$ symmetry illustrated at $\sqrt{3}\times 1$ $(1\times 1)$ unit cell which is indicated by solid (dashed) line, respectively. **c,d**. The hybridization due to symmetry-lowering into $(T_{\sqrt{3}\times 1}, D_{2h})$ importantly affects to the type of VHS near $M$ and $\Gamma$. More specifically, the two VHSs at $M$ point become mixed and split into four type-II VHS points consisting of a mixed sublattice character, while the VHS at $\Gamma$ still survives as a type-I VHS with a single sublattice character.

example, the parameter $\beta$ encodes a Fermi-surface nesting condition, and its absolute value $|\beta|$ quantifies how far away from the perfect nesting condition ($\beta = 0$), which depends on the next nearest hopping parameter $t_2$. The original Hamiltonian $\mathcal{H}^0_{\text{two}}$ enjoys a $D_{6h}$ symmetry, and $\epsilon_3(\boldsymbol{p})$ is constructed by $\frac{2\pi}{3}$ rotation of $\epsilon_2(\boldsymbol{p})$,

$$\epsilon_2(\boldsymbol{p}) = \frac{\alpha}{2} \begin{pmatrix} p'_x & p'_y \end{pmatrix} \begin{pmatrix} 1 & 0 \\ 0 & -\tan(\frac{\pi}{3}+\beta)^2 \end{pmatrix} \begin{pmatrix} p'_x \\ p'_y \end{pmatrix} + \cdots, \quad (4)$$

with an unitary transformation,

$$\begin{pmatrix} p'_x \\ p'_y \end{pmatrix} = \begin{pmatrix} \cos(\frac{2\pi}{3}) & -\sin(\frac{2\pi}{3}) \\ \sin(\frac{2\pi}{3}) & \cos(\frac{2\pi}{3}) \end{pmatrix} = \begin{pmatrix} p_x \\ p_y \end{pmatrix}. \quad (5)$$

By diagonalizing the total effective two band Hamiltonian, $\mathcal{H}^0_{\text{two}}(\boldsymbol{p}) + \delta\mathcal{H}_{\text{two}}(\boldsymbol{p})$, we find

that the number of shifted VHS points becomes either 4 or 2, depending on the sign of the parameter $\beta > 0$ or $\beta < 0$, as illustrated in Fig. S7.

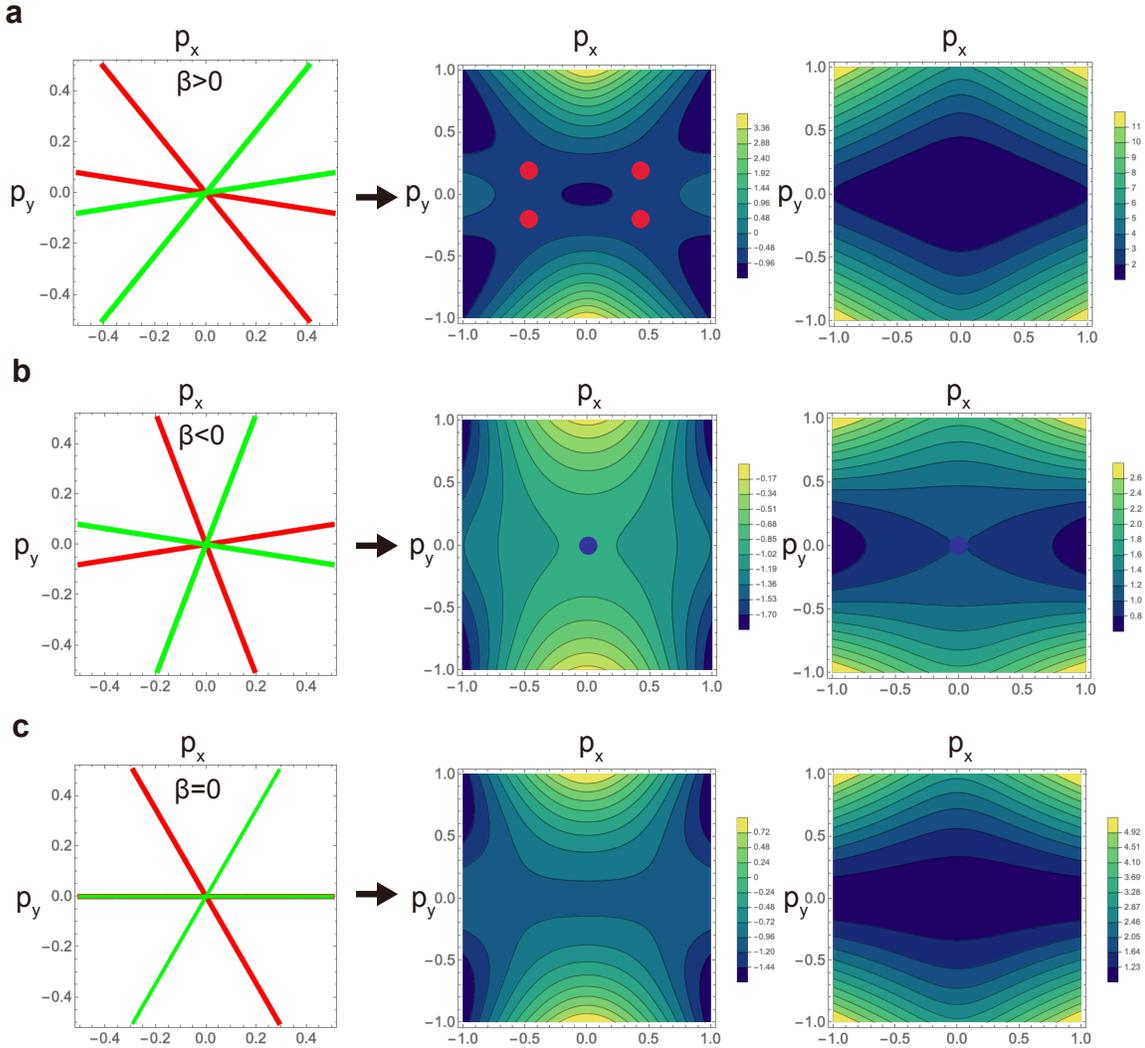

FIG. S7: **Number of VHSs depending on Fermi-surface nesting parameter $\beta$.** Based on the effective two-band model, Eq.4, we find that the number of VHSs is not universal but crucially depends on the microscopic details of VHSs. Figures **a-c** are for the different Fermi-surface nesting parameters, $(\beta = \frac{\pi}{20}, -\frac{\pi}{20}, 0)$. In leftmost panels of **a-c**, the the contours of the Fermi-surface before the hybridization ($\Delta = 0$) are illustrated in the momentum space, $(p_x, p_y)$, where $\boldsymbol{p} = \boldsymbol{k} - M$ is the momentum deviation from the original VHS points (M). The other two panels of **a-c** show the rearranged VHSs and Fermi contour of the energy dispersion of two bands after the hybridization ($\Delta = 0.1$), where the plot legend is drawn with a unit of $\alpha/2$. For $\beta > 0$, the four type-II VHS points (red circles) newly appear at the shifted point, while the two type-I VHSs (blue circle) survive at $M$ in each band for $\beta < 0$. In the fine-tuned case, $\beta = 0$, both derivative and Hessian matrix of energy are zero at $k_y = 0$ line, disappearing all VHS points and the higher order expansion may be essential to examine the possibility of the high-order VHSs.

**Supplementary Note 4. Sublattice resolved susceptibility**

We demonstrate the sublattice interference in the monolayer system by calculating the sublattice resolved susceptibility, following the previous literature [7]. The explicit form of the susceptibility is given by

$$\chi^{l_1,l_2}(\boldsymbol{q}) \equiv -\sum_{\boldsymbol{k},\mu,\nu} a_\nu^{l_1}(\boldsymbol{k}+\boldsymbol{q}) a_\mu^{l_1}(\boldsymbol{k})^* a_\nu^{l_2}(\boldsymbol{k}+\boldsymbol{q})^* a_\mu^{l_2}(\boldsymbol{k}) \left[ \frac{n_F(E_\mu(\boldsymbol{k})) - n_F(E_\nu(\boldsymbol{k}+\boldsymbol{q}))}{E_\mu(\boldsymbol{k}) - E_\nu(\boldsymbol{k}+\boldsymbol{q})} \right] \quad (6)$$

with sublattice index, $(l_1, l_2)$, and band index, $(\mu, \nu)$. Here, the Fermi distribution, $n_F(\epsilon)$, and the $l_i$-th component of the eigenvector of the $\mu$ band, $a_\mu^{l_i}(\boldsymbol{k})$, are introduced. The eigenvector of the largest eigenvalue of $\chi^{l_1,l_2}(\boldsymbol{q})$ around $M$ point encode the sublattice interference effects. For example, in the two opposite limits, the sublattice weights of (A,B,C) sublattice are given by either (0,0.5,0.5), (0.5,0,0.5) or (0.5,0.5,0) in the case displaying a perfect sublattice interference, while the values become either (1,0,0), (0,1,0) or (0,0,1) in the perfectly sublattice interference absent limit, as illustrated in Fig. S8**a**.

Based on the tight-binding model of monolayer $AV_3Sb_5$ introduced in the main text, we found that the sublattice interference is suppressed, but it is not absent. The largest eigenvalues of the susceptibility at the $M$ point have the sublattice weights, (0.482,0.258,0.258) and (0.476,0.262,0.262), for the type-I, type-II VHS filling, which are significantly modified from the perfectly sublattice interference absent case (Fig. S8**b,c**). Hence, we can argue that $V$ still plays a crucial role in offering new physics in monolayer $AV_3Sb_5$, as in the bulk system.

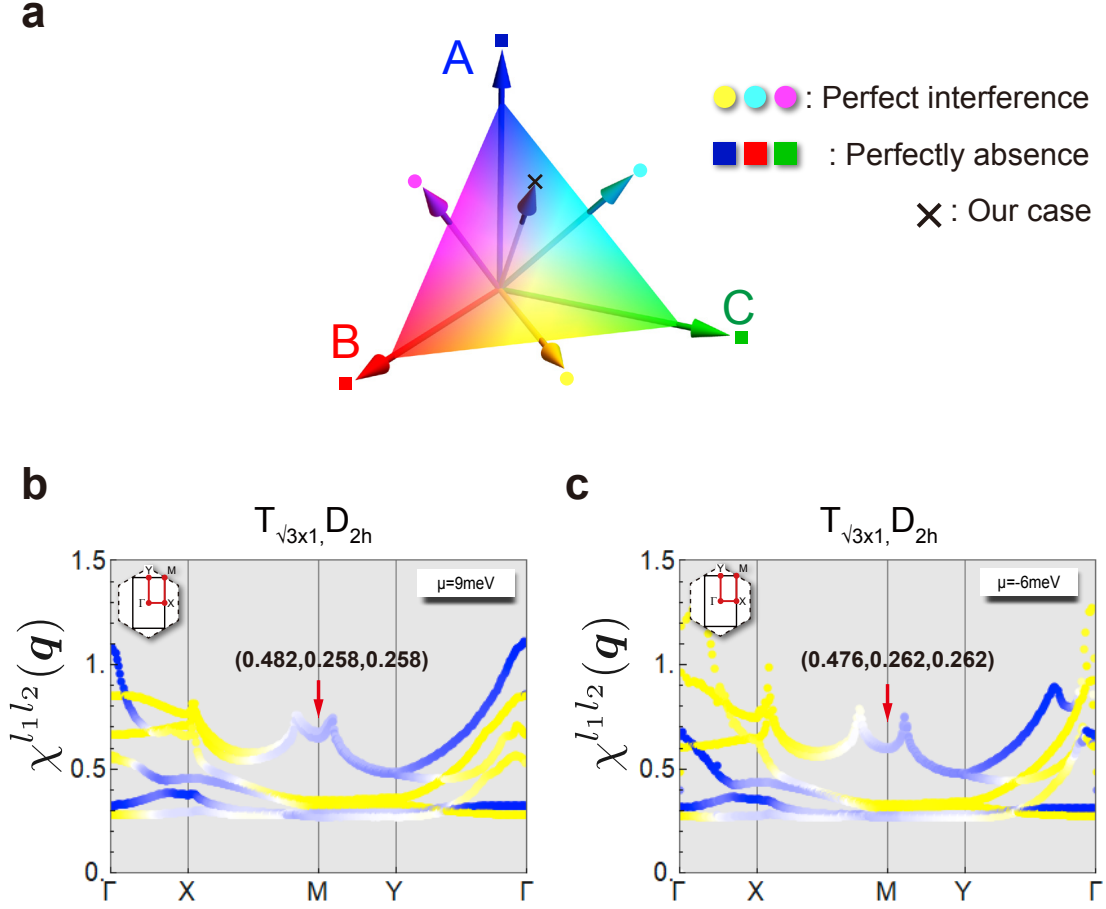

FIG. S8: **Sublattice-resolved bare susceptibility of $AV_3Sb_5$ systems.** **a.** The three different sublattices A, B, and C are colored in blue, red, and green, respectively. It is useful to utilize the eigenvector associated with the largest eigenvalues of $\chi^{l_1,l_2}(\bm{q})$ to quantify the sublattice interference. For example, the sublattice weight would be either (0,0.5,0.5), (0.5,0,0.5) or (0.5,0.5,0) in the limit displaying a perfect sublattice interference, while the values become either (1,0,0), (0,1,0) or (0,0,1) in the perfectly sublattice interference absent limit. The sublattice character of the monolayer $AV_3Sb_5$ system corresponds to the intermediate case between two limits. **b-c.** The eigenvalues of bare susceptibility matrix $\chi^{l_1,l_2}(\bm{q})$ and their sublattice characters are carried out based on our tight-binding Hamiltonian at $\beta = 1000$. (b) is for the $AV_3Sb_5$ systems at type-I VHS filling, while (c) is at type-II VHS filling. Focusing on the sublattice character of largest eigenvalues of $\chi^{l_1,l_2}(\bm{q}=M)$, we find that (0.482,0.258,0.258), and (0.476,0.262,0.262) for b,c, respectively.

14

**Supplementary Note 5. Detailed phonon analysis**

### 5.1. Leading CDW instability of the pristine phase

In this subsection, we show that the 2×2 CDW instability is a leading instability by performing energy profile analysis along the various negative phonon modes. Fig. S9**a** shows the phonon band structure of the pristine monolayer. The negative modes appear at a broad region in the momentum space. A close inspection reveals that they appear delocalized as they are localized at multiple $k$-point, including high symmetry points $\Gamma$, $M$, and $X$. Among these, the negative energies at the $\Gamma$ point arise from the folding of the $M$ point of the $1 \times 1$ unit cell to the $\sqrt{3} \times 1$ unit cell, responsible for the 2×2 instability. We evaluated the total energy profiles for the reconstructions driven by the negative phonon modes at high symmetry $X$ and $M$ points as well as the nesting vectors $q_1$ and $q_2$ mediating the type-II VHS points (Fig. S9**b**), in which the latter is to consider the incommensurate CDWs. Here, we choose $q_1$ and $q_2$ to be commensurate, namely $(\frac{1}{6}, 0, 0)$ and $(0, \frac{1}{10}, 0)$, respectively, which are the closest rational numbers to the incommensurate nesting vectors mediating the type-II VHS points (see also Fig. 3**a** in the main text). The figure clearly shows that the reconstructions driven by $X$, $q_1$, and $q_2$ are less stable than that of $M$. Thus, the $2 \times 2$ reconstruction is energetically favored over the other reconstructions captured in the negative phonon energies of the monolayer.

15

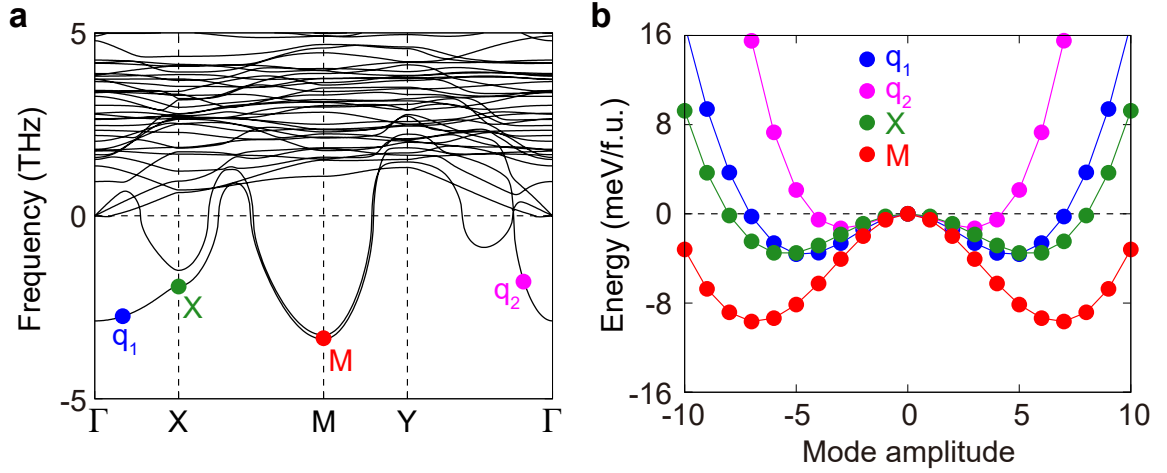

FIG. S9: **Negative phonon mode analysis of the $A\mathbf{V}_3\mathbf{Sb}_5$ monolayer. a** Phonon bands of the pristine KV$_3$Sb$_5$ monolayer at the smearing factor $\sigma = 0.1$ eV. Here, $q_1 = \frac{1}{6}\mathbf{b}'_1$ and $q_2 = \frac{1}{10}\mathbf{b}'_2$ ($\mathbf{b}'_1$ and $\mathbf{b}'_2$ are the reciprocal lattice vectors of the $\sqrt{3} \times 1$ unit cell) are the closest commensurate vectors to the incommensurate nesting vectors mediating the type-II VHS points (see also Fig. 3a in the main text). **b** Energy profiles (per formula unit) as a function of the amplitude of the phonon mode for various negative energy states. The corresponding energy and momenta are indicated in **a**. We stress that the minimum of total energy appears at $M$ ($\sim -9.7$ meV/f.u), which corresponds to the 2×2 CDW reconstruction.

## 5.2. Correction of artificial negative phonon modes of the ISD-1 phase

In this subsection, we demonstrate the correction of the artificial negative phonon modes of the ISD-1 phase. We calculate the phonon bands by increasing the supercell size from 4×4 to 8×8 (Fig. S10), which allows us to get more accurate results without the negative branches. Thus, we conclude that the phonon bands show the stability of the 2×2 reconstruction in the $D_{2h}$ monolayer.

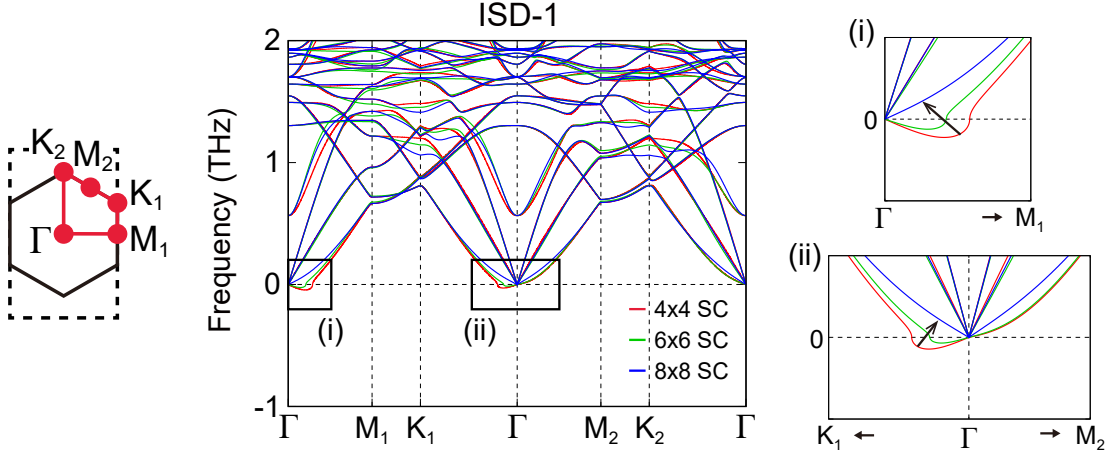

FIG. S10: **Convergence test of the negative phonon energies of the ISD-1 structure.** The phonon bands of the ISD-1 structure are calculated using three different supercells (SCs): 4×4, 6×6, and 8×8. The erroneous negative energies, which appears in the 4×4 and 6×6 SC calculations, converge to positive ones as we increase the computational accuracy using a large SC. The insets (i) and (ii) display the magnified views of the boxed regions near $\Gamma$. The arrows indicate the converging behavior of the negative branches (from $4 \times 4$ to $8 \times 8$ SC). The left panel depicts the Brillouin zone (solid line) and the corresponding high-symmetry points of the $2 \times 2$ unit cell.

**Supplementary Note 6. Details on various CDW orders**

**6.1. Anisotropic $2 \times 2$ $3Q$ CDW order**

The lowered symmetries of the monolayer can favor the $3Q$ SD and ISD phases as they respect the lower symmetries. A close look at the SD and ISD distortion of the monolayer reveals that they adapt themselves into the lowered symmetries by losing $C_3$ symmetries and being split into the doublets, which we refer to as SD-1 and SD-2 and ISD-1 and ISD-2, respectively. As delineated in Fig. S11, the adjacent two vanadium atoms have different bonding lengths between the AB and BC sublattices. These patterns are unlike the $D_{6h}$ bulk counterpart, in which the bonding lengths are the same between the adjacent V atoms. Since these types of distortion respect the $\sqrt{3} \times 1$ symmetries, it is likely that the monolayer symmetry favors them.

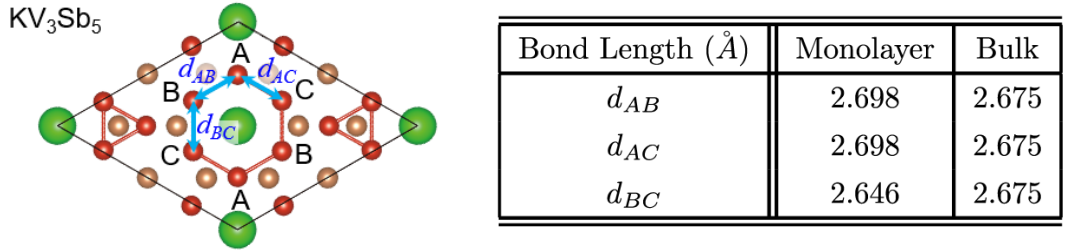

| Bond Length (Å) | Monolayer | Bulk |
|---|---|---|
| $d_{AB}$ | 2.698 | 2.675 |
| $d_{AC}$ | 2.698 | 2.675 |
| $d_{BC}$ | 2.646 | 2.675 |

FIG. S11: **Bond lengths of the ISD-1 phase in the monolayer.** Sublattice indices A, B, and C (for hexagons) are depicted on left. The calculated bond lengths between the sublattices are given in Table (in Å). In contrast to the isotropic CDW order of the bulk counterpart, the CDW in the monolayer has anisotropic bond lengths, such that $d_{BC} < d_{AB} = d_{AC}$.

**6.2. Band structures of ISD-1 and SD-2 CDWs**

We calculate the band structure of the $2 \times 2$ ISD-1 and SD-2 CDWs of the $AV_3Sb_5$ monolayer for $A =$ K, Rb, Cs, as shown in Fig. S12. Due to the symmetry breaking, $M_1$ and $M_2$ (or $K_1$ and $K_2$) have no longer the same energy.

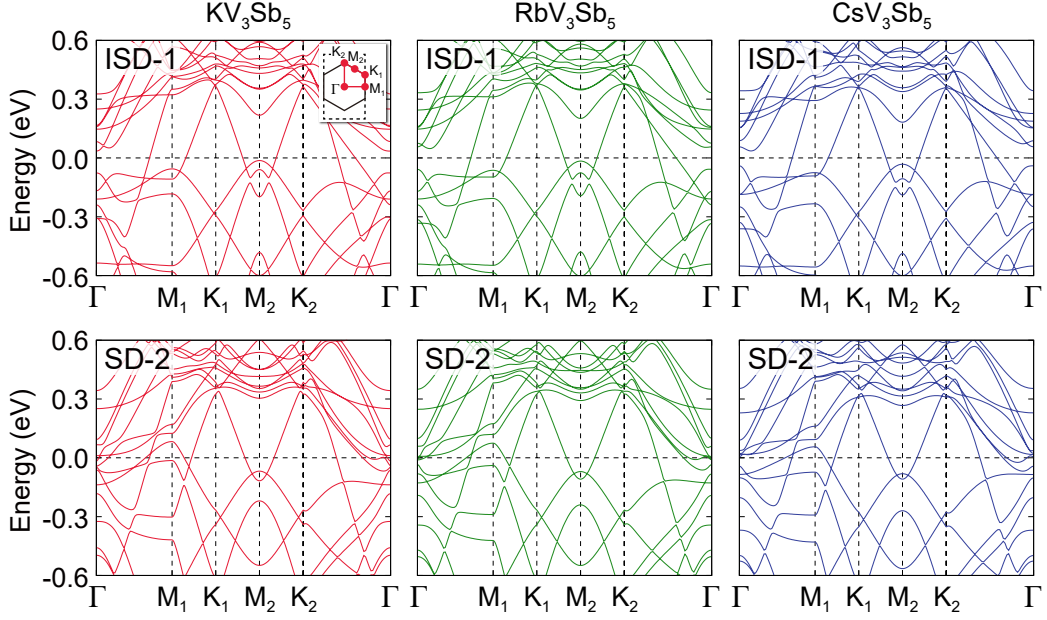

FIG. S12: **The band structures of symmetry-breaking phases in the monolayer $AV_3Sb_5$.** Left, middle, and right panels indicate the band structures of ISD-1 and SD-2 phases for K, Rb, and Cs systems, respectively.

### 6.3. TRSB CDW: Comparison with the $D_{6h}$ bulk

In this section, we compare the TRSB-1 and TRSB-2 CDWs of the $D_{2h}$ monolayer with their $D_{6h}$ bulk counterpart (Fig. S13). The complex CDW introduced in arXiv:2110.06266 (Ref. [8]) can be considered as a bulk counterpart of the TRSB-1 and TRSB-2 CDWs that we studied in the monolayer. As compared in Fig. S13, the current patterns of our TRSB-1 and TRSB-2 are the same as those of the complex CDW order of Ref. [8] with the magnetic fluxes threading in all the hexagons and triangles. Notably, the Chern Fermi pockets are featured in our system as well, realizing the phase referred to as a doped orbital Chern insulator by Zhou and Wang.

We explicitly show that the Chern numbers are distributed differently over the bands near the Fermi level between the $D_{2h}$ monolayer and the $D_{6h}$ bulk, as delineated in Fig. S13. In particular, the different distributions of the Berry curvatures on the Fermi surface are found near the $M1$ and $M2$ (as well as $K1$ and $K2$) points, which can be mainly attributed to distinct parameters that are chosen such that they depict the absence of the $C_{3z}$ rotational symmetry of the monolayer. In detail, the threading flux values of the TRSB phase in Ref. [8] are different from our TRSB-1 and TRSB-2 phases (Table. S2). The complex CDW

19

phase in Ref. [8] has flux values as $\phi_{1,...,4} = (0.840, -0.113, -0.291, 0.014)\pi$ while our TRSB-1 and TRSB-2 phases have the flux values as $\phi_{1,...,4} = (0.318, -0.103, -0.159, 0.051)\pi$ and $\phi_{1,...,4} = (0.313, -0.108, -0.157, 0.054)\pi$, respectively.

Correspondingly, distinct anomalous Hall conductivities were calculated between the bulk and monolayer systems. The bulk system has the anomalous Hall conductivity of around $-1.1 \frac{e^2}{h}$, whereas the monolayer in TRSB-1 and TRSB-2 phases lead to $-0.26$ and $-0.06$ $\frac{e^2}{h}$, respectively, Here, $e$ and $h$ are electronic charge and the Plank constant, respectively. This difference in the anomalous Hall conductivities should be of crucial importance for distinguishing the bulk and monolayer systems.

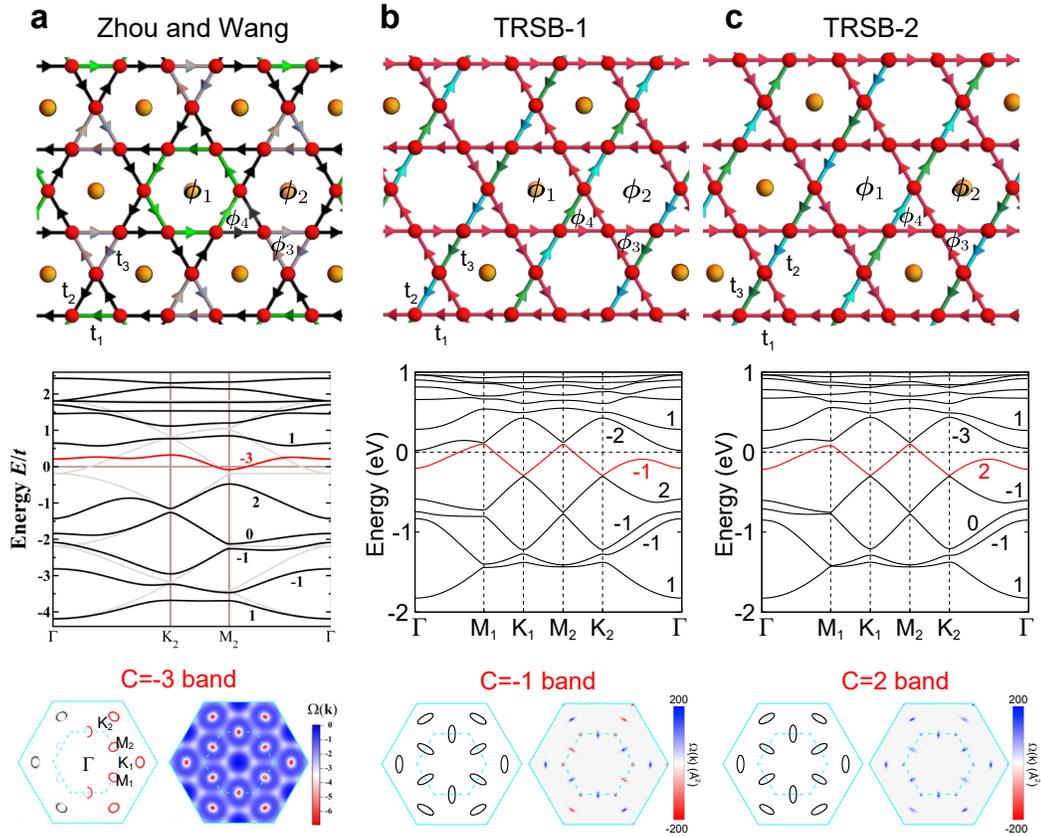

FIG. S13: **Current pattern, band structure, Fermi surface, and Berry curvature of various TRSB phases. a** Current pattern, band structure, Fermi surface, and Berry curvature of the TRSB phase studied in Zhou and Wang [8]. **b,c** Current pattern, band structure, Fermi surface, and Berry curvature of the TRSB-1 and TRSB-2 phases were studied in this work. The magnitudes of threading fluxes $\phi_1, \phi_2, \phi_3$, and $\phi_4$ as well as the hopping parameters are given in Table S2. The different hopping parameters are displayed in different colors. Chern numbers of each band are displayed in the band structures. Energy contours and Berry curvature distribution of the band lying at the Fermi level (colored in red in the band structures) are drawn in the bottom panel.

TABLE S2: **Magnetic fluxes and complex hopping parameters of the TRSB phases in Fig. S13.** The unit of hopping parameters is the electron volt (eV). The ± sign of the imaginary hopping constants denotes the direction of the bond current depicted in Fig S13.

| Model | $\phi_1$ | $\phi_2$ | $\phi_3$ | $\phi_4$ | $t_1$ | $t_2$ | $t_3$ |
|---|---|---|---|---|---|---|---|
| Ref. [8] | $0.840\pi$ | $-0.113\pi$ | $-0.291\pi$ | $0.014\pi$ | $0.345 \pm 0.162i$ | $0.499 \pm 0.123i$ | $0.349 \pm 0.110i$ |
| TRSB-1 | $0.318\pi$ | $-0.103\pi$ | $-0.159\pi$ | $0.051\pi$ | $0.42 \pm 0.07i$ | $0.41 \pm 0.07i$ | $0.43 \pm 0.07i$ |
| TRSB-2 | $0.313\pi$ | $-0.108\pi$ | $\pm 0.157\pi$ | $0.054\pi$ | $0.42 \pm 0.07i$ | $0.41 \pm 0.07i$ | $0.43 \pm 0.07i$ |

**Supplementary Note 7. Details on tight-binding Hamiltonian**

**7.1. Notation**

In this section, we provide detailed information on notations about our fermionic systems. We introduce the three-dimensional Gamma matrices ($\lambda_a$),

$$\lambda_1 = \begin{pmatrix} 0 & 1 & 0 \\ 1 & 0 & 0 \\ 0 & 0 & 0 \end{pmatrix}, \quad \lambda_2 = \begin{pmatrix} 0 & 0 & 1 \\ 0 & 0 & 0 \\ 1 & 0 & 0 \end{pmatrix}, \quad \lambda_3 = \begin{pmatrix} 0 & 0 & 0 \\ 0 & 0 & 1 \\ 0 & 1 & 0 \end{pmatrix}, \quad \lambda_4 = \begin{pmatrix} 0 & -i & 0 \\ i & 0 & 0 \\ 0 & 0 & 0 \end{pmatrix},$$

$$\lambda_5 = \begin{pmatrix} 0 & 0 & -i \\ 0 & 0 & 0 \\ i & 0 & 0 \end{pmatrix}, \quad \lambda_6 = \begin{pmatrix} 0 & 0 & 0 \\ 0 & 0 & -i \\ 0 & i & 0 \end{pmatrix}, \quad \lambda_7 = \begin{pmatrix} 1 & 0 & 0 \\ 0 & -1 & 0 \\ 0 & 0 & 0 \end{pmatrix}, \quad \lambda_8 = \frac{1}{\sqrt{3}}\begin{pmatrix} 1 & 0 & 0 \\ 0 & 1 & 0 \\ 0 & 0 & -2 \end{pmatrix},$$

with $\lambda_0 = I_3$. It is convenient to introduce the four dimension matrices ($M_a$) to express the sublattice subspace in $2 \times 2$ unit cell,

$$M_1^\pm = \frac{1}{2}\left[(I_2 \pm \sigma_x) \otimes I_2\right], \quad M_2^\pm = \frac{1}{2}\left[I_2 \otimes (I_2 \pm \sigma_x)\right], \quad M_3^\pm = \frac{1}{2}\left[\pm I_2 \otimes I_2 + \sigma_x \otimes \sigma_x\right],$$

with $M_0 = I_4$ and define twelve dimensional matrices ($L_a^0, L_a^\pm$),

$$L_a^0 = \lambda_a \otimes M_0, \quad L_a^\pm = \begin{cases} \lambda_a \otimes M_1^\pm & \text{for } a = 1, 4, \\ \lambda_a \otimes M_2^\pm & \text{for } a = 2, 5, \\ \lambda_a \otimes M_3^\pm & \text{for } a = 3, 6. \end{cases} \quad (7)$$

**7.2. Tight-binding Hamiltonian in $2 \times 2$ unit cell**

This section provides the details on tight-binding (TB) Hamiltonian in $2 \times 2$ unit cell. Introducing the twelve-component spinor, the Hamiltonian becomes $H_0 = \sum_{\boldsymbol{k},\sigma} \tilde{\Psi}^\dagger_{\boldsymbol{k}\sigma} \mathcal{H}_0(\boldsymbol{k}) \tilde{\Psi}_{\boldsymbol{k}\sigma}$ with $\tilde{\Psi}^T_{\boldsymbol{k}} = (\tilde{A}^T_{\boldsymbol{k}}, \tilde{B}^T_{\boldsymbol{k}}, \tilde{C}^T_{\boldsymbol{k}})$ and $\tilde{\alpha}^T_{\boldsymbol{k}} = (\alpha_{1,\boldsymbol{k}}, \alpha_{2,\boldsymbol{k}}, \alpha_{3,\boldsymbol{k}}, \alpha_{4,\boldsymbol{k}})$. The site of the sublattice index of the $2 \times 2$ unit cell and the corresponding Brillouin zone are illustrated in Fig. S14. Our

minimal TB Hamiltonian may be decomposed as four different terms,

$$\mathcal{H}_0(\boldsymbol{k}) \;=\; \mathcal{H}_{NN}(\boldsymbol{k}) + \mathcal{H}_{NNN}(\boldsymbol{k}) + \mathcal{H}_1(\boldsymbol{k}) + \mathcal{H}_2(\boldsymbol{k}), \qquad (8)$$

with

$$\begin{aligned}
\mathcal{H}_{NN}(\boldsymbol{k}) \;=\;& -t\left[c_1\lambda_1 \otimes (I_2+\sigma_x) \otimes I_2 + s_1\lambda_4 \otimes (-I_2+\sigma_x) \otimes I_2 \right.\\
& + c_2\lambda_2 \otimes I_2 \otimes (I_2+\sigma_x) \otimes s_2\lambda_5 \otimes I_2 \otimes (-I_2+\sigma_x)\\
& \left. + c_3\lambda_3(I_2\otimes I_2 + \sigma_x\otimes\sigma_x) + s_3\lambda_6(I_2\otimes I_2 - \sigma_x\otimes\sigma_x)\right] + \epsilon I_{12},
\end{aligned}$$

$$\begin{aligned}
\mathcal{H}_{NNN}(\boldsymbol{k}) \;=\;& -t_2\left[c_4\lambda_1 \otimes (I_2+\sigma_x) \otimes \sigma_x + s_4\lambda_4 \otimes (-I_2+\sigma_x) \otimes \sigma_x \right.\\
& + c_5\lambda_2 \otimes \sigma_x \otimes (I_2+\sigma_x) \otimes s_5\lambda_5 \otimes \sigma_x \otimes (I_2-\sigma_x)\\
& \left. + c_6\lambda_3(I_2\otimes\sigma_x + \sigma_x\otimes I_2) + s_6\lambda_6(I_2\otimes\sigma_x - \sigma_x\otimes I_2)\right],
\end{aligned}$$

$$\mathcal{H}_1(\boldsymbol{k}) \;=\; \frac{\delta\epsilon}{6}\left[(2\lambda_0 + 3\lambda_7 + \sqrt{3}\lambda_8)\otimes\sigma_z\otimes\sigma_z\right],$$

$$\mathcal{H}_2(\boldsymbol{k}) \;=\; -\delta t\left[c_3\lambda_3(I_2\otimes I_2 + \sigma_x\otimes\sigma_x) + s_3\lambda_6(I_2\otimes I_2 - \sigma_x\otimes\sigma_x)\right].$$

The notations $(c_i, s_i) = (\cos\boldsymbol{k}\cdot\boldsymbol{r}_i, \sin\boldsymbol{k}\cdot\boldsymbol{r}_i)$ with $\boldsymbol{r}_1 = \frac{1}{2}(\sqrt{3},1)$, $\boldsymbol{r}_2 = \frac{1}{2}(\sqrt{3},-1)$, $\boldsymbol{r}_3 = (0,1)$, $\boldsymbol{r}_4 = \boldsymbol{r}_2 - \boldsymbol{r}_3$, $\boldsymbol{r}_5 = \boldsymbol{r}_1 + \boldsymbol{r}_3$, and $\boldsymbol{r}_6 = \boldsymbol{r}_1 + \boldsymbol{r}_2$ is used. Here, $t$, $t_2$, $\epsilon$ are nearest-neighbor hopping, next nearest-neighbor hopping, onsite energy and $\delta t$, $\delta\epsilon$ are anisotropic hopping,

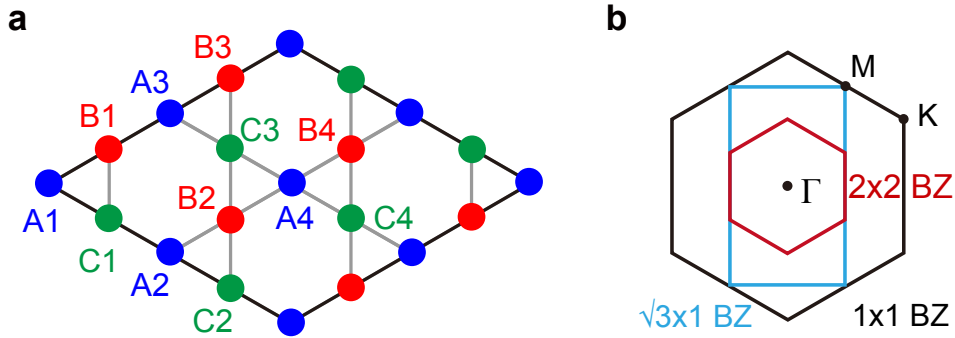

FIG. S14: **$2\times 2$ unit cell and its Brillouin zone for TB model of the monolayer $A\mathbf{V}_3\mathbf{Sb}_5$.** **a** Twelve sublattices in the $2\times 2$ unit cell for the TB model. The three different sublattices A, B, and C are colored in blue, red, and green, respectively. **b** The BZs of the $2\times 2$, $\sqrt{3}\times 1$, and $1\times 1$ unit cell are illustrated by red, cyan, and black lines, respectively.

23

anisotropic onsite energy which break symmetries from $\mathcal{T}_{1\times1}\ D_{6h}$ to $\mathcal{T}_{\sqrt{3}\times1}\ D_{2h}$. The tight-binding parameter set $(\epsilon, t, t_2, \delta\epsilon, \delta t) = (0.01, 0.42, 0.03, -0.033, 0.01)$ is consistent with DFT band structures of $KV_3Sb_5$ and used in most of our discussion unless otherwise stated.

### 7.3. Justification of the tight-binding model

We augmented the justification of our minimal model in terms of the relevant orbitals contributing to bands around the Fermi level. The orbital-projected bands shown in Fig. S15 reveal that the VHS band described by our model mainly comprises three $d$-orbitals of V atoms: $d_{xy}$, $d_{3z^2-r^2}$, and $d_{x^2-y^2}$. Our group-theoretical analysis of the DFT Bloch state at $M$ shows that the VHS state is in the $A_g$ irreducible representation of the little group of $M$. This indicates that the three $d$ orbitals of V are hybridized to form a $A_g$ orbital, based on which our minimal model is constructed. Therefore, our minimal model was able to reproduce the DFT VHS band successfully.

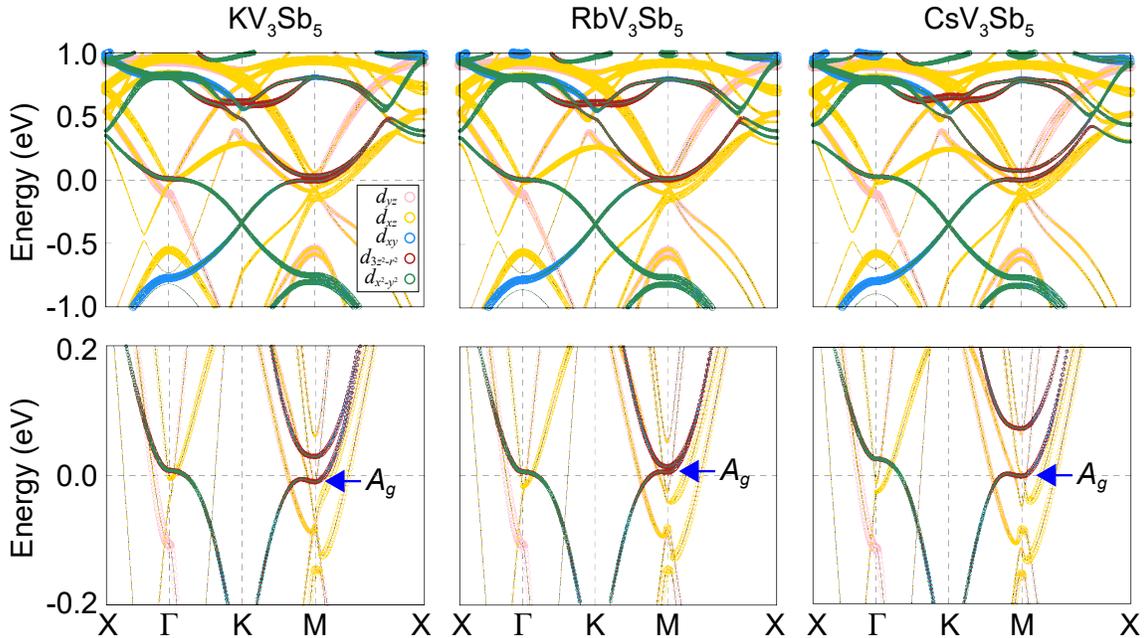

FIG. S15: **Orbital-projected band structures of the $AV_3Sb_5$ monolayer.** $d_{yz}$, $d_{xz}$, $d_{xy}$, $d_{3z^2-r^2}$ and $d_{x^2-y^2}$ orbitals of V atoms are projected on the band structure where the magnitude of open circles is proportional to the projected weight.

Our minimal model only considers the $A_g$ VHS and thus other two VHSs are excluded because they are relatively away from the Fermi level. Thus, they should be less important

in describing the low-energy electronic property of the $A$V$_3$Sb$_5$ monolayer. This is, in fact, a unique feature of the monolayer. Unlike the bulk materials [9], the monolayer always hosts the $A_g$ VHS nearest to the Fermi level irrespective of the type of alkali metal $A$ (Fig. S15). Thus, our minimal model generically captures the low-energy electronic structure of the $A$V$_3$Sb$_5$ monolayer without the inclusion of the other VHS bands.

# Supplementary Note 8. Details on mean-field analysis

## 8.1. Model and strategy

We carry out the zero temperature mean-field phase diagram following the steps.

(i) We consider the non-interacting Hamiltonian $H_0$ incorporating with density-density interactions $H_{\text{int}}$ whose form may be expressed as

$$H_{\text{int}} = U \sum_{\boldsymbol{R}} \sum_{\alpha_i} n_{\boldsymbol{R},\alpha_i\uparrow} n_{\boldsymbol{R},\alpha_i\downarrow} + V \sum_{\boldsymbol{R}} \sum_{\substack{\langle \alpha_i,\beta_j \rangle \\ \sigma,\sigma'}} n_{\boldsymbol{R},\alpha_i,\sigma} n_{\boldsymbol{R},\beta_j,\sigma'} \equiv U_{\text{int}} + V_{\text{int}}, \qquad (9)$$

where ($\alpha \in \{A, B, C\}, i \in \{1, 2, 3, 4\}$) and $\sigma \in \{\uparrow, \downarrow\}$ are the indices for sublattice and spin, respectively.

(ii) We first check the CDW instability under the specific choice of CDW configurations (see Table S3 and Fig. 3c). We construct the CDW mean-field Hamiltonian in $2 \times 2$ unit cell which is convenient to see the physics after translational-symmetry breaking. Introducing a CDW order ($\Phi$), the mean-field Hamiltonian is given by

$$\delta H_{\text{MF}}[\Phi] = \Phi \sum_{\boldsymbol{k},\sigma} \tilde{\Psi}^{\dagger}_{\boldsymbol{k}\sigma} \begin{pmatrix} 0 & O_1(\boldsymbol{k}) & O_2(\boldsymbol{k}) \\ O_1(\boldsymbol{k})^{\dagger} & 0 & O_3(\boldsymbol{k}) \\ O_2(\boldsymbol{k})^{\dagger} & O_3(\boldsymbol{k})^{\dagger} & 0 \end{pmatrix} \tilde{\Psi}_{\boldsymbol{k}\sigma}, \qquad (10)$$

with the twelve-component spinor, $\tilde{\Psi}^T_{\boldsymbol{k}}$. After evaluating the ground state of the mean-field Hamiltonian $|G; \Phi\rangle$ and its energy expectation value $E[\Phi; U, V]$, the phase diagram associated with CDWs can be obtained as a function of interaction strengths.

(iii) We further consider the SC instability. The superconducting pairing gap functions are classified based on the point group and fermionic anticommuting relation (see Table S4). Among them, we consider the spin-singlet pairings whose mean-field Hamiltonian is given by

$$\delta H_{\text{MF}}[\Delta] = \Delta \left[ \sum_{\boldsymbol{k}} \tilde{\Psi}^{\dagger}_{\boldsymbol{k}\uparrow} \hat{\Delta}(\boldsymbol{k}) \tilde{\Psi}^{*}_{\boldsymbol{k}\downarrow} + \text{C.C.} \right], \qquad (11)$$

where the overall pairing amplitude ($\Delta$) is chosen as a real number. Performing the same process, we obtain the full phase diagram including SC phases.

We obtain 161 data points of the ground state energy $E[\phi_i]$ where the order parameter range is $\phi_i \in [-0.08, 0.08]$. The integrations at given $\phi_i$ are evaluated in the momentum space with $80 \times 80$ $k$-points. In the following two sections, we provide detailed information on the ground states of the mean-field Hamiltonians of CDW ($\phi = \Phi$) and SC ($\phi = \Delta$).

### 8.2. Charge density wave (CDW)

This section provides details on the CDW order parameters and their mean-field theories. We start with the mean-field Hamiltonian density in $2 \times 2$ unit cell,

$$\delta \mathcal{H}_{\mathrm{MF}}[\Phi; \boldsymbol{k}] = \Phi \begin{pmatrix} 0 & O_1 & O_2 \\ O_1^\dagger & 0 & O_3 \\ O_2^\dagger & O_3^\dagger & 0 \end{pmatrix}, \quad (12)$$

where $\Phi$ is a CDW order parameter. The order parameter couples with the charge distribution of the V ions whose pattern is encoded in the bond matrix $O_a(\boldsymbol{k})$, where the index $a = 1, 2, 3$ labels AB, AC, BC bond. Strikingly, the number of CDW modes of the monolayer system ($\mathcal{T}_{\sqrt{3} \times 1}$ $D_{2h}$) is doubled compared to the bulk system ($\mathcal{T}_{1 \times 1}$ $D_{6h}$). It is understood by the quotient group of two translational symmetries $\mathbb{Z}_2 = \mathcal{T}_{1 \times 1}/\mathcal{T}_{\sqrt{3} \times 1}$ and this $\mathbb{Z}_2$ factor specifies the doublet CDW. Hereafter, the two states in the CDW doublet are denoted as "CDW-1/2", in which they are connected to each other by $\mathcal{T}_{\sqrt{3} \times 1}$ translation of the alkali atoms (see Fig. 3c). In our analysis, the four different CDW bond order patterns $\{(\mathcal{O}_1, \mathcal{O}_2, \mathcal{O}_3), -(\mathcal{O}_1, -\mathcal{O}_2, \mathcal{O}_3), i(\mathcal{O}_1, -\mathcal{O}_2, \mathcal{O}_3), -i(\mathcal{O}_1, \mathcal{O}_2, \mathcal{O}_3)\}$ are tested where

$$\mathcal{O}_1 = (-\phi_1 I_2 + \phi_1^* \sigma_x) \otimes \sigma_z, \ \mathcal{O}_2 = \sigma_z \otimes (-\phi_2 I_2 + \phi_2^* \sigma_x), \ \mathcal{O}_3 = \phi_3 \sigma_y \otimes \sigma_y + \phi_3^* \sigma_z \otimes \sigma_z. \ (13)$$

The former/latter two correspond to the TRS/TRSB cases. The sign of order parameter $\Phi$ distinguishes two different phases, star of David phase (SD; $\Phi > 0$) and inverse star of David phase (ISD; $\Phi < 0$), for the TRS case. However, the sign is irrelevant for the TRSB case since reversing the sign just connects a given state and its time-reversal partner. To sum up, six different CDW configurations (SD-1/2, ISD-1/2, and TRSB-1/2) are possible and

their symmetry properties are summarized in Table S3.

We now evaluate the ground state of the mean-field Hamiltonian, $H_{\mathrm{MF}}[\Phi] = H_0 + \delta H_{\mathrm{MF}}[\Phi]$. Introducing the unitary transformation, $\tilde{\Psi}_{\bm{k}} = \mathcal{U}(\bm{k})\overline{\Psi}_{\bm{k}}$, the Hamiltonian is

$$H_{\mathrm{MF}}[\Phi] = \sum_{\bm{k},\alpha_i,\sigma} E_{\mathrm{MF}}^{\alpha_i}[\Phi;\bm{k}]\, \overline{\Psi}^{\dagger}_{\bm{k},\alpha_i,\sigma} \overline{\Psi}_{\bm{k},\alpha_i,\sigma}, \tag{14}$$

where $E_{\mathrm{MF}}^{\alpha_i}[\Phi;\bm{k}]$ is an $\alpha_i$-th eigenenergy. Note that the state with tilde/overline is for the basis before/after the Hamiltonian diagonalization. Applying the quasiparticle creation operators $\overline{\Psi}^{\dagger}_{\bm{k},\alpha_i,\sigma}$ to the electron vacuum gives the ground state energy,

$$|G;\Phi\rangle = \prod_{\alpha_i} \prod_{\bm{k} \in M_{\alpha_i}[\Phi]} \overline{\Psi}^{\dagger}_{\bm{k},\alpha_i,\uparrow} \overline{\Psi}^{\dagger}_{\bm{k},\alpha_i,\downarrow}|0\rangle, \tag{15}$$

where the momentum product is done over the subspace, $M_{\alpha_i}[\Phi] = \{\forall \bm{k} \in \mathrm{B.Z.} \mid E_{\mathrm{MF}}^{\alpha_i}[\Phi;\bm{k}] < 0\}$, of the Brillouin zone (BZ) of $2 \times 2$ unit cells (see Fig. S14).

We obtain the energy expectation value by sandwiching the ground state to the Hamiltonian, $E[\Phi] \equiv \langle G;\Phi|H_0 + H_{\mathrm{int}}|G;\Phi\rangle$. The explicit forms of each part are given by

$$\langle G;\Phi|H_0|G;\Phi\rangle = 2\sum_{\bm{k}} \mathrm{Tr}\left[\mathcal{U}^{\dagger}(\bm{k})\mathcal{H}_0(\bm{k})\mathcal{U}(\bm{k})\Xi(\bm{k})\right], \tag{16}$$

$$\langle G;\Phi|U_{\mathrm{int}}|G;\Phi\rangle = \frac{U}{N}\sum_{\alpha_i}\left[\sum_{\bm{k}}[\mathcal{U}(\bm{k})\Xi(\bm{k})\mathcal{U}^{\dagger}(\bm{k})]_{\alpha_i,\alpha_i}\right]^2, \tag{17}$$

$$\langle G;\Phi|V_{\mathrm{int}}|G;\Phi\rangle = \frac{4}{N}\sum_{\alpha_i,\beta_j}\mathcal{V}_{\alpha_i,\beta_j}(\bm{0})\left[\sum_{\bm{k}}[\mathcal{U}(\bm{k})\Xi(\bm{k})\mathcal{U}^{\dagger}(\bm{k})]_{\alpha_i,\alpha_i}\right]\left[\sum_{\bm{k}'}[\mathcal{U}(\bm{k}')\Xi(\bm{k}')\mathcal{U}^{\dagger}(\bm{k}')]_{\beta_j,\beta_j}\right]$$

$$- \frac{2}{N}\sum_{\alpha_i,\beta_j}\mathcal{V}_{\alpha_i,\beta_j}(\bm{0})\left|\sum_{\bm{k}} e^{-i\bm{k}\cdot\bm{\delta}_{\alpha_i\beta_j}}[\mathcal{U}(\bm{k})\Xi(\bm{k})\mathcal{U}^{\dagger}(\bm{k})]_{\alpha_i,\beta_j}\right|^2, \tag{18}$$

where the direct and exchange terms are contributed as illustrated in Fig. S16. Here we define $12 \times 12$ diagonal matrix, $\Xi(\bm{k})_{\alpha_i,\alpha_i} \equiv \theta(-E_{\mathrm{MF}}^{\alpha_i}[\Phi;\bm{k}])$, with the Heaviside step function $\theta(x)$ and the $12 \times 12$ matrix component of the nearest-neighbour interaction after the Fourier

28

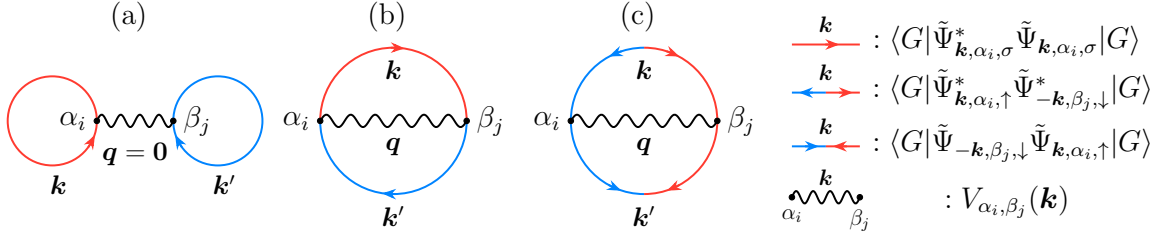

FIG. S16: **Three possible processes contributing to $\langle G|H_{\text{int}}|G\rangle$.** The first two processes (a-b) are the conventional direct and exchange process of the four-fermion interactions [10], while the final pairing process (c) is only included for superconducting ground states. Different color denotes the sublattice index ($\alpha_i$) and detailed diagrammatic rules are explained in the right panel.

transformation $\mathcal{V}_{\alpha_i,\beta_j}$,

$$\begin{aligned}\mathcal{V}(\boldsymbol{q}) = {} & V\left[c_1\lambda_1 \otimes (I_2 + \sigma_x) \otimes I_2 + s_1\lambda_4 \otimes (-I_2 + \sigma_x) \otimes I_2 \right.\\ & +c_2\lambda_2 \otimes I_2 \otimes (I_2 + \sigma_x) \otimes s_2\lambda_5 \otimes I_2 \otimes (-I_2 + \sigma_x) \\ & \left.+c_3\lambda_3(I_2 \otimes I_2 + \sigma_x \otimes \sigma_x) + s_3\lambda_6(I_2 \otimes I_2 - \sigma_x \otimes \sigma_x)\right].\end{aligned} \qquad (19)$$

The matrix satisfies $\mathcal{V}_{\alpha_i,\beta_j}(\boldsymbol{q}) = Ve^{i\boldsymbol{q}\cdot(\boldsymbol{r}_{\beta_j}-\boldsymbol{r}_{\alpha_i})}[\mathcal{V}_{\alpha_i,\beta_j}(\mathbf{0})]$ with $\mathcal{V}(\mathbf{0}) = 2[L_1^+ + L_2^+ + L_3^+]$.

TABLE S3: **CDW orders of monolayer $A$V$_3$Sb$_5$ considered in the mean-field analysis.** Both time-reversal symmetric and broken types of CDWs are considered. Translational symmetry-lowering effect in monolayer $A$V$_3$Sb$_5$ plays a crucial role in classifications of order parameters, splitting out a degenerate doublet, CDW-1/2, for each CDW order parameter. The CDW-1 and CDW-2 are connected to each other by a $\sqrt{3} \times 1$ translation of the position of alkali metals.

| Time-reversal symmetry | Bond matrix $O_a(\boldsymbol{k})$ | Label | Subgroup |
|---|---|---|---|
| TRS | $(\mathcal{O}_1, \mathcal{O}_2, \mathcal{O}_3)$ | ISD-1 ($\Phi < 0$) or SD-1 ($\Phi > 0$) | $D_{2h}$ |
| | $(-\mathcal{O}_1, \mathcal{O}_2, -\mathcal{O}_3)$ | ISD-2 ($\Phi < 0$) or SD-2 ($\Phi > 0$) | |
| TRSB | $i(\mathcal{O}_1, -\mathcal{O}_2, \mathcal{O}_3)$ | TRSB-1 ($\Phi \neq 0$) | $C_{2h}$ |
| | $-i(\mathcal{O}_1, \mathcal{O}_2, \mathcal{O}_3)$ | TRSB-2 ($\Phi \neq 0$) | |

### 8.3. Superconductivity (SC)

This section provides details on the SC order parameters and their mean-field theories. Introducing the Nambu basis $\xi_{\bm{k}}^T = (\tilde{\Psi}_{\bm{k}\uparrow}^T, \tilde{\Psi}_{-\bm{k}\downarrow}^\dagger)$, the mean-field Hamiltonian, or equivalently BdG Hamiltonian, may be written in the $2 \times 2$ unit cell as

$$H_{\mathrm{MF}}[\Delta] = \sum_{\bm{k}} \xi_{\bm{k}}^\dagger \mathcal{H}_{\mathrm{BdG}}(\bm{k}) \xi_{\bm{k}}, \quad \mathcal{H}_{\mathrm{BdG}}[\Delta; \bm{k}] = \begin{pmatrix} \mathcal{H}_0(\bm{k}) & \Gamma(\Delta; \bm{k}) \\ \Gamma^\dagger(\Delta; \bm{k}) & -\mathcal{H}_0^T(-\bm{k}) \end{pmatrix}, \quad (20)$$

where only the spin-singlet pairing channels $\Gamma(\bm{k})$ with pairing amplitude $\Delta$ are considered.

If the pairing gap functions are pinned down, evaluating the ground state of the mean-field Hamiltonian is straightforward. Upon the Bogoliubov transformation with a $24 \times 24$ unitary matrix $\mathcal{U}(\bm{k})$,

$$\xi_{\bm{k}} = \mathcal{U}(\bm{k}) \overline{\xi}_{\bm{k}}, \quad \mathcal{U}(\bm{k}) = \begin{pmatrix} \mathcal{U}_{pp} & \mathcal{U}_{ph} \\ \mathcal{U}_{hp} & \mathcal{U}_{hh} \end{pmatrix}, \quad (21)$$

we arrive at the mean-field Hamiltonian, $H_{\mathrm{MF}}[\Delta]$,

$$H_{\mathrm{MF}}[\Delta] = \sum_{\bm{k}, \alpha_i, \sigma} E_{\mathrm{MF}}^{\alpha_i}[\Delta; \bm{k}] \left[ \overline{\Psi}_{\alpha_i, \bm{k}, \uparrow}^\dagger \overline{\Psi}_{\alpha_i, \bm{k}, \uparrow} - \overline{\Psi}_{\alpha_i, -\bm{k}, \downarrow} \overline{\Psi}_{\alpha_i, -\bm{k}, \downarrow}^\dagger \right], \quad (22)$$

where $E_{\mathrm{MF}}^{\alpha_i}[\Phi; \bm{k}]$ is $\alpha_i$-th eigenenergy of Bogoliubon. The overline notation is used for denoting the basis after the Bogoliubov transformation. Applying the annihilation operator of Bogoliubov quasiparticles $\overline{\Psi}_{\bm{k}, \alpha_i, \sigma}$ to the electron vacuum, the superconducting ground state is constructed as

$$|G; \Delta\rangle = \prod_{\alpha_i} \prod_{\bm{k}} \overline{\Psi}_{-\bm{k}, \alpha_i \downarrow} \overline{\Psi}_{\bm{k}, \alpha_i \uparrow} |0\rangle, \quad (23)$$

where the momentum product is performed over the BZ of the $2 \times 2$ unit cells.

We now determine the possible pairing channels originating from the interactions Eq.(9). Utilizing the Fierz identity, one find that the nine spin-singlet pairing gap functions labeled

by different irreducible representations may be stabilized under the four-fermion interactions,

$$\Gamma[\{\Delta_R\}; \boldsymbol{k}] = \sum_{n=1}^{4}\Delta_{A_g}^{(n)}\hat{\Gamma}_{A_g}^{(n)} + \sum_{n=1}^{2}\Delta_{B_{1g}}^{(n)}\hat{\Gamma}_{B_{1g}}^{(n)} + \Delta_{B_{2u}}^{(1)}\hat{\Gamma}_{B_{2u}}^{(1)} + \sum_{n=1}^{2}\Delta_{B_{3u}}^{(n)}\hat{\Gamma}_{B_{3u}}^{(n)}, \quad (24)$$

where the real order parameter $\Delta_R^{(n)}$ is associated with an irreducible representation $R$. In Table S4, we tabulate all pairing gap functions $\hat{\Gamma}_R^{(n)}$ written in the $1 \times 1$ and $2 \times 2$ unit cells.

We now evaluate the energy expectation value $E[\Delta] \equiv \langle G; \Delta | H_0 + H_{\text{int}} | G; \Delta \rangle$ from the ground state which yields the final results,

$$\langle G; \Delta | H_0 | G; \Delta \rangle = \sum_{\boldsymbol{k}} \text{Tr}\Big(\mathcal{U}_{ph}^\dagger(\boldsymbol{k})\mathcal{H}_0(\boldsymbol{k})\mathcal{U}_{ph}(\boldsymbol{k}) - \mathcal{U}_{hh}^\dagger(\boldsymbol{k})\mathcal{H}_0^T(-\boldsymbol{k})\mathcal{U}_{hh}(\boldsymbol{k}) + \mathcal{H}_0(\boldsymbol{k})\Big), \quad (25)$$

$$\langle G; \Delta | U_{\text{int}} | G; \Delta \rangle = \frac{U}{N} \sum_{\alpha_i} \left| \sum_{\boldsymbol{k}} \left[\mathcal{U}_{hh}(\boldsymbol{k})\mathcal{U}_{ph}^\dagger(\boldsymbol{k})\right]_{\alpha_i, \alpha_i} \right|^2 + \left[\sum_{\boldsymbol{k}} \left[\mathcal{U}_{ph}(\boldsymbol{k})\mathcal{U}_{ph}^\dagger(\boldsymbol{k})\right]_{\alpha_i, \alpha_i}\right]^2, \quad (26)$$

$$\begin{aligned}
\langle G; \Delta | V_{\text{int}} | G; \Delta \rangle &= \frac{2}{N} \sum_{\alpha_i, \beta_j} \mathcal{V}_{\alpha_i, \beta_j}(\boldsymbol{0}) \Big| \sum_{\boldsymbol{k}} e^{-i\boldsymbol{k} \cdot \boldsymbol{\delta}_{\alpha_i \beta_j}} [\mathcal{U}_{hh}(\boldsymbol{k})\mathcal{U}_{ph}(\boldsymbol{k})^\dagger]_{\alpha_i \beta_j} \Big|^2 \\
&+ \frac{4}{N} \sum_{\alpha_i, \beta_j} \mathcal{V}_{\alpha_i, \beta_j}(\boldsymbol{0}) \Big[\sum_{\boldsymbol{k}} \mathcal{U}_{ph}(\boldsymbol{k})\mathcal{U}_{ph}(\boldsymbol{k})^\dagger\Big]_{\alpha_i \alpha_i} \Big[\sum_{\boldsymbol{k}'} \mathcal{U}_{ph}(\boldsymbol{k}')\mathcal{U}_{ph}(\boldsymbol{k}')^\dagger\Big]_{\beta_j \beta_j} \\
&- \frac{2}{N} \sum_{\alpha_i, \beta_j} \mathcal{V}_{\alpha_i, \beta_j}(\boldsymbol{0}) \Big| \sum_{\boldsymbol{k}} e^{-i\boldsymbol{k} \cdot \boldsymbol{\delta}_{\alpha_i \beta_j}} [\mathcal{U}_{ph}(\boldsymbol{k})\mathcal{U}_{ph}(\boldsymbol{k})^\dagger]_{\alpha_i \beta_j} \Big|^2. \quad (27)
\end{aligned}$$

Note that three possible fully contracted terms contribute to the ground state energy of the superconducting state and their diagrammatic interpretation is expressed in Fig. S16.

TABLE S4: **Nine SC orders of monolayer AV$_3$Sb$_5$ are considered in the mean-field analysis.** Spin-singlet superconducting pairings are classified by irreducible representations (R) of the point group $D_{2h}$. Condensation of $\hat{\Gamma}_R$ channels can lower the interaction Hamiltonian energy, which preserves $\mathcal{T}_{1\times 1}$ symmetry. The momentum dependence of pairing gap function is encoded in $(c_i, s_i) \equiv (\cos \boldsymbol{k} \cdot \boldsymbol{R}_i, \sin \boldsymbol{k} \cdot \boldsymbol{R}_i)$. To express pairing functions in both $1\times 1$ and $2 \times 2$ unit cells, the $3 \times 3$ Gell Mann matrice $\lambda_a$ and the $12 \times 12$ matrices $L_a^{\pm,0} = \lambda_a \otimes M_a^{\pm,0}$ are introduced with $4 \times 4$ matrices $M_a^{\pm,0}$. The lowest order of basis functions associated with representation $R$ are provided in the fifth column.

| R | Pairing in $1 \times 1$ unit cell | Pairing in $2 \times 2$ unit cell $\hat{\Gamma}_R(\boldsymbol{k})$ | Label | Basis function |
|---|---|---|---|---|
| $A_g$ | $\frac{1}{\sqrt{3}}\lambda_0, \frac{1}{\sqrt{2}}\left[\frac{\sqrt{3}}{2}\lambda_7 + \frac{1}{2}\lambda_8\right]$, $\frac{1}{\sqrt{2}}[c_1\lambda_1 + c_2\lambda_2]$, $c_3\lambda_3$ | $\frac{1}{\sqrt{3}}L_0^0, \frac{1}{\sqrt{2}}\left[\frac{\sqrt{3}}{2}L_7^0 + \frac{1}{2}L_8^0\right]$, $\frac{1}{\sqrt{2}}\left[c_1 L_1^+ + c_2 L_2^+ - (s_1 L_4^- + s_2 L_5^-)\right]$, $\left[c_3 L_3^+ - s_3 L_6^-\right]$ | $\Gamma_{A_g}^{(1)}, \Gamma_{A_g}^{(2)}$, $\Gamma_{A_g}^{(3)}$, $\Gamma_{A_g}^{(4)}$ | $x^2, y^2, z^2$ |
| $B_{1g}$ | $\frac{1}{\sqrt{2}}\left[-\frac{1}{2}\lambda_7 + \frac{\sqrt{3}}{2}\lambda_8\right]$, $\frac{1}{\sqrt{2}}[c_1\lambda_1 - c_2\lambda_2]$ | $\frac{1}{\sqrt{2}}\left[-\frac{1}{2}L_7^0 + \frac{\sqrt{3}}{2}L_8^0\right]$, $\frac{1}{\sqrt{2}}\left[c_1 L_1^+ - c_2 L_2^+ - (s_1 L_4^- - s_2 L_5^-)\right]$ | $\Gamma_{B_{1g}}^{(1)}$, $\Gamma_{B_{1g}}^{(2)}$ | $xy$ |
| $B_{2u}$ | $\frac{1}{\sqrt{2}}[s_1\lambda_4 - s_2\lambda_5]$ | $\frac{1}{\sqrt{2}}\left[s_1 L_4^+ - s_2 L_5^+ - (c_1 L_1^- - c_2 L_2^-)\right]$ | $\Gamma_{B_{2u}}^{(1)}$ | $y$ |
| $B_{3u}$ | $\frac{1}{\sqrt{2}}[s_1\lambda_4 + s_2\lambda_5]$, $s_3\lambda_6$ | $\frac{1}{\sqrt{2}}\left[s_1 L_4^+ + s_2 L_5^+ + (c_1 L_1^- + c_2 L_2^-)\right]$, $\left[s_3 L_6^+ - c_3 L_3^-\right]$ | $\Gamma_{B_{3u}}^{(1)}$, $\Gamma_{B_{3u}}^{(2)}$ | $x$ |

### 8.4. Details of variational method

In this section, we show that our variational method is essentially equivalent to solving a self-consistency equation at zero temperature. The self-consistency equation is obtained by differentiating free energy with respect to the mean-field order parameter to find a stationary point. On the other hand, the variational method directly determines the minimum of the free energy by varying the order parameter of the wave function. Therefore, the equivalence between the two methods is well established, unless the solution is located at local minimum points. To check the equivalence, we plot the magnitudes of the order parameter carried out through the two methods in Fig. S17, where the perfect matching between the two methods is demonstrated.

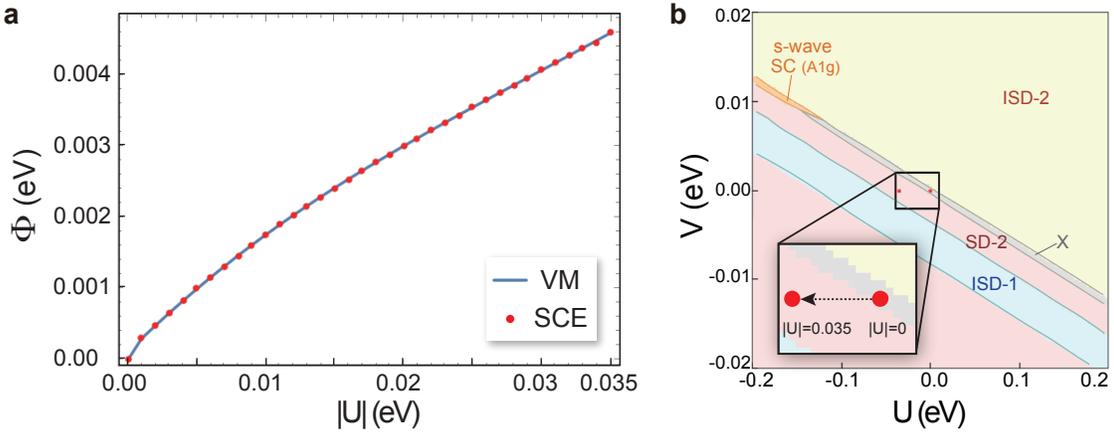

FIG. S17: **Order parameter (SD-2) amplitude of phase as a function of $U$ and the corresponding phase diagram at the chemical potential $\mu = -6\text{meV}$. a.** Order parameter amplitude is obtained by solving the self-consistency equation (SCE) and variational method (VM) at zero temperature, which gives equivalent results $(U < 0, V = 0)$ **b.** The mean-field phase diagram at the chemical potential $\mu = -6\text{meV}$. In particular interest phase space considered in (a), the SD-2 CDW order is stabilized (Inset).

### 8.5. Comment on the attractive interaction

In this section, we would like to add a few words on the effect of the phonons in the mean-field phase diagram. As well established in the famous BCS theory, the phonon-electron coupling induces an effective attractive interaction after the renormalization. For

example, when the phonon is integrated out, the electron-electron interaction is written as,

$$W_{\text{eff}}(\boldsymbol{q},\omega) = W_0(\boldsymbol{q})\frac{\omega^2}{\omega^2 - \omega_D^2}, \tag{28}$$

with bare interaction $V_0(\boldsymbol{q})$, and Debye frequency $\omega_D$. It is clear that the sign can be reversed in the low-frequency limit, $\omega < \omega_D$. In this analogy, the attractive values of $(U,V)$ are taken into account for the mean-field calculations.

**Supplementary Note 9. Details on mean-field phase diagram**

We here show the phase diagrams in the range of chemical potential $\mu = -10 \sim 30$meV to elucidate the effect of fillings on competing orders of the monolayer $AV_3Sb_5$, while only the two representative cases are provided in the main text. Our zero-temperature mean-field theory is performed in the same manner by adopting the same tight-binding (TB) and extended Hubbard model explained in the previous section.

Figure S18**a** illustrates DFT and TB band structures where the TB parameters are set as $(\epsilon, t, t_2, \delta\epsilon, \delta t) = (0.01, 0.42, 0.03, -0.033, 0.01)$. Figure S18**b** shows density of states indicating the two different types of VHSs. Figures S18**c-g** display the mean-field phase diagrams in $(U, V)$ space at $\mu = -10, -6, 9, 20$, and 30 meV, where the six CDW configurations and nine SC orders in Table S3 and S4 are considered. A few remarks are as follows. First, the phase diagrams are significantly changed and a variety of exotic phases can be achieved as a function of the chemical potential. Second, the electronic instabilities associated with the type-I and II VHSs are quite distinct. While the SD phases dominate at type-I VHS filling (Fig. S18**e**), both ISD and SD phases are stabilized in the wide range of $(U, V)$ at type-II VHS filling (Fig. S18**d**). The s-wave SC ($\Gamma^{(1)}_{A_g}$) appears for negative $U$ at both filings, but d-wave SC ($\Gamma^{(1)}_{B_{1g}}$) is only shown at type-I VHS filling. Third, the TRSB CDW phase appears when the chemical potential is shifted above both VHS points (Fig. S18**g**). The topological property of the TRSB CDW phase manifests in nonzero Chern number and subsequent anomalous Hall conductivity (see Fig. 5**a,b**).

We next pin down the stoichiometry-enforced symmetry lowering effects of the monolayer system ($\mathcal{T}_{\sqrt{3}\times 1}, D_{2h}$) on the phase diagrams, by comparing with the phase diagrams obtained for a structure having $\mathcal{T}_{1\times 1}$ and $D_{6h}$ symmetries. Note that $\mathcal{T}_{1\times 1}$ and $D_{6h}$ symmetries are the same as the symmetries of the bulk system (Fig. 1**a**). For the $\mathcal{T}_{1\times 1}$ and $D_{6h}$ symmetries, CDW and SC order parameters are newly classified in Table S5 and S6. The key difference of ($\mathcal{T}_{1\times 1}, D_{6h}$) symmetric system is that the two distinct states CDW-1 and CDW-2 in the presence of ($\mathcal{T}_{\sqrt{3}\times 1}, D_{2h}$) symmetry now become equivalent due to the restored translational symmetry. Figures S19**a-b** show DFT and TB band structures and density of states of the $D_{6h}$ system where the TB parameters are set as $(\epsilon, t, t_2, \delta\epsilon, \delta t) = (0.01, 0.42, 0.03, 0, 0)$. Figures S19**c-g** display the mean-field phase diagrams at $\mu = -10, 0, 10, 20$, and 30 meV. Our analysis makes the following observation. First, the superconducting phase in the

$D_{2h}$ system is less favored compared to the $D_{6h}$ system, which is clearly seen in the phase diagrams at the type-I VHS filling of the two systems (Figs. S18e and S19e). Interestingly, a spin-singlet $f$-wave SC ($\Gamma_{B_{2u}}$) is found below the VHS filling for the $D_{6h}$ system ($\mu = -10$ meV). We note that the emergence of the spin-triplet $f$-wave SC is reported in the previous theoretical study for the bulk system [7]. Second, the TRSB CDW has stabilized in a wide range of chemical potential for the $D_{6h}$ system compared to the $D_{2h}$ system. The occurrence of the TRSB CDW is consistent with the various experiments [11–14] observing TRSB CDW.

TABLE S5: **CDW orders in the $D_{6h}$ symmetry considered in the mean-field analysis.** Both time-reversal symmetric and broken types of CDWs are considered. Two distinct states in the CDW doublet, (CDW-1/2), in the $\mathcal{T}_{\sqrt{3}\times 1}$ and $D_{2h}$ symmetries are now equivalent in the $\mathcal{T}_{1\times 1}$ and $D_{6h}$ symmetries.

| Time-reversal symmetry | Bond matrix $O_a(\boldsymbol{k})$ | Label | Subgroup |
|---|---|---|---|
| TRS | $(\mathcal{O}_1, \mathcal{O}_2, \mathcal{O}_3)$ | ISD ($\Phi < 0$) or SD ($\Phi > 0$) | $D_{6h}$ |
| TRSB | $-i(\mathcal{O}_1, \mathcal{O}_2, \mathcal{O}_3)$ | TRSB ($\Phi \neq 0$) | $C_{6h}$ |

TABLE S6: **Three SC orders in the $D_{6h}$ symmetry are considered in the mean-field analysis.** One-dimensional spin-singlet superconducting pairings are classified by the irreducible representations (R) of the point group $D_{6h}$. The momentum dependence of pairing gap function is encoded in $(c_i, s_i) \equiv (\cos \boldsymbol{k} \cdot \boldsymbol{R}_i, \sin \boldsymbol{k} \cdot \boldsymbol{R}_i)$ with $\omega \equiv e^{\frac{2\pi}{3}i}$. The lowest order of basis functions associated with representation $R$ is provided in the fifth column.

| R | Pairing in $1 \times 1$ unit cell | Pairing in $2 \times 2$ unit cell $\Gamma_R(\boldsymbol{k})$ | Label | Basis function |
|---|---|---|---|---|
| $A_{1g}$ | $\frac{1}{\sqrt{3}}\lambda_0$, $\frac{1}{\sqrt{3}}[c_1\lambda_1 + c_2\lambda_2 + c_3\lambda_3]$ | $\frac{1}{\sqrt{3}}L_0^0$, $\frac{1}{\sqrt{3}}[c_1L_1^+ + c_2L_2^+ + c_3L_3^+ - (s_1L_4^- + s_2L_5^- + s_3L_6^-)]$ | $\Gamma_{A_{1g}}^{(1)}$, $\Gamma_{A_{1g}}^{(2)}$ | $x^2 + y^2, z^2$ |
| $B_{2u}$ | $\frac{1}{\sqrt{3}}[s_1\lambda_4 + s_2\lambda_5 - s_3\lambda_6]$ | $\frac{1}{\sqrt{3}}[s_1L_4^+ + s_2L_5^+ - s_3L_6^+ - (c_1L_4^- + c_2L_5^- - c_3L_6^-)]$ | $\Gamma_{B_{2u}}^{(1)}$ | $x(x^2 - 3y^2)$ |

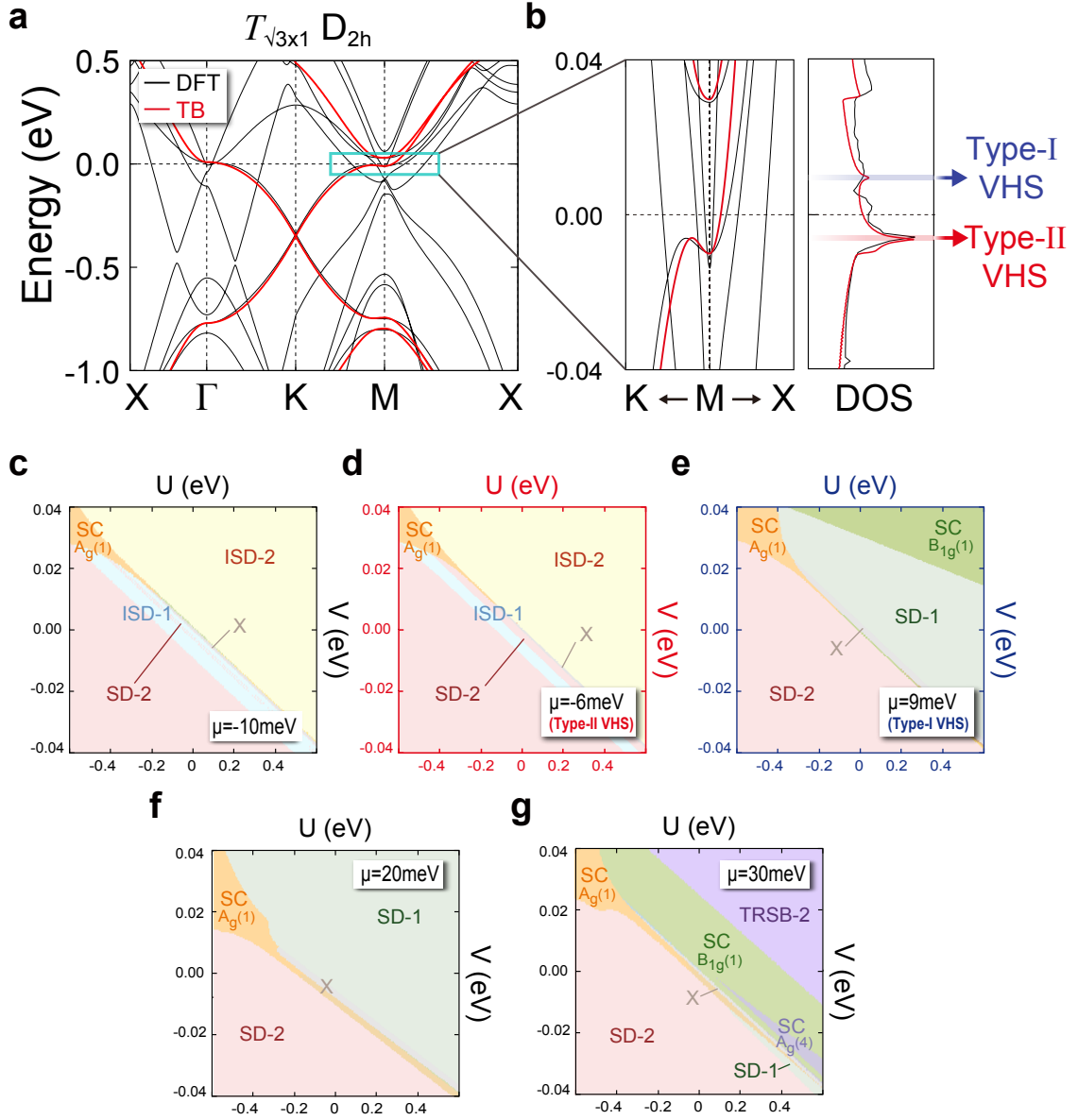

FIG. S18: **Electronic structures and chemical potential dependence of phase diagram of the monolayer $AV_3Sb_5$.** **a-b** DFT and TB band structures of monolayer $KV_3Sb_5$, denoted by black and red color. Density of states and the position of type-I/II VHSs are illustrated. **c-g** Mean-field phase diagram at the chemical potentials $\mu = -10, -6, 9, 20,$ and 30 meV. For numerical evaluations, we set the tight-binding parameters $(\epsilon, t, t_2, \delta\epsilon, \delta t) = (0.01, 0.42, 0.03, -0.033, 0.01)$. Emergence and competition of various CDW and superconducting phases are highly contingent upon the fillings. The CDW and superconducting orders considered in the mean-field theory are tabulated in Tables S3 and S4. For numerical evaluations, we obtain 161 data points $\phi_i \in [-0.08, 0.08]$ of the ground state energy, with $80 \times 80$ $k$-points.

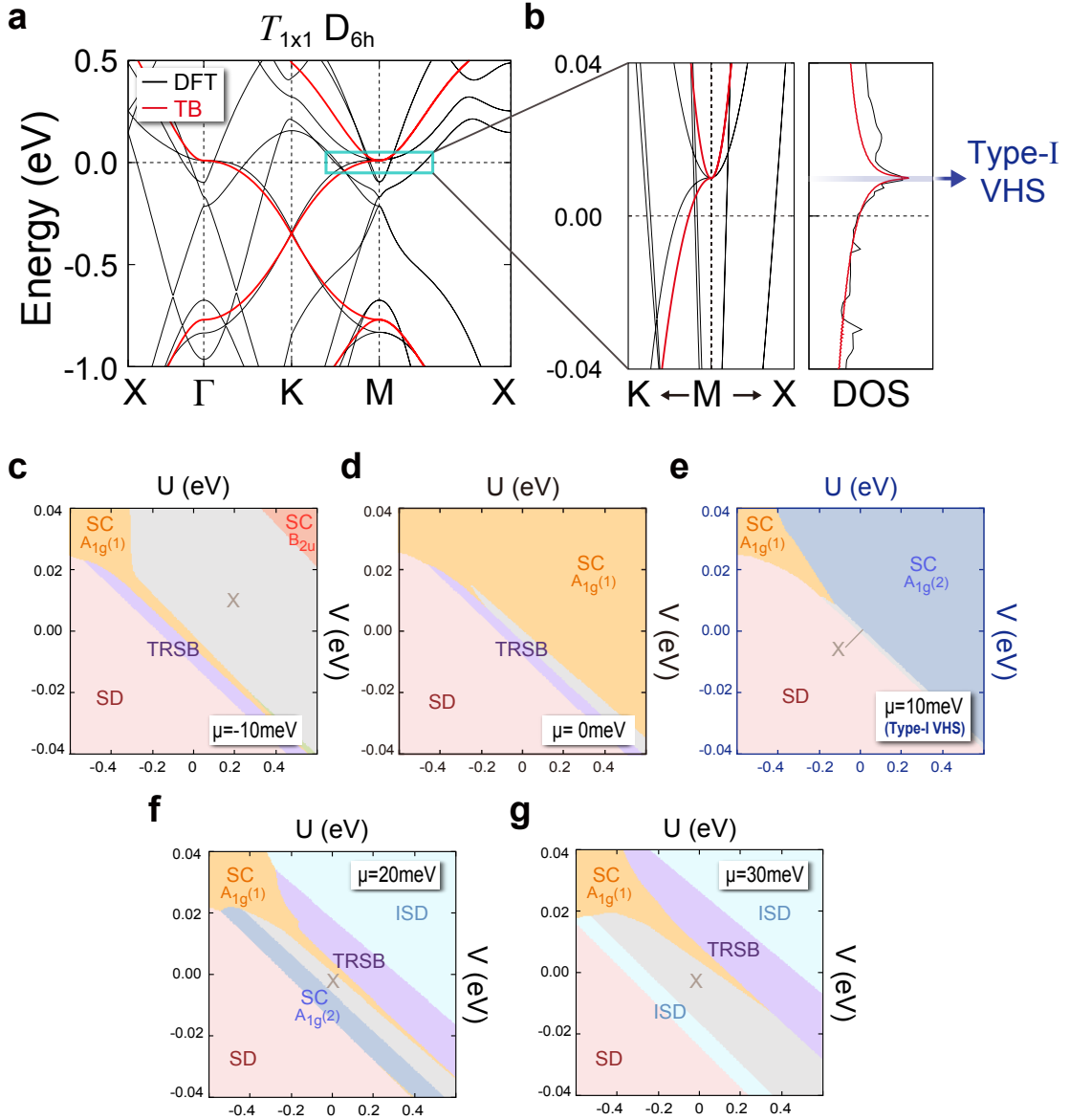

FIG. S19: **Electronic structures and chemical potential dependence of phase diagram of $A$V$_3$Sb$_5$ with $\mathcal{T}_{1\times1}$ and $D_{6h}$ symmetries.** **a-b** DFT and TB band structures of KV$_3$Sb$_5$ are illustrated by black and red colors. Here, although the structure with $D_{6h}$ point group symmetry has $\mathcal{T}_{1\times1}$ translational symmetry, we plot the band structures along the BZ of the $\sqrt{3}\times1$ unit cell to compare with the structure having $\mathcal{T}_{\sqrt{3}\times1}$ and $D_{2h}$ symmetries in Fig. S18. The density of states and the position of type-I VHS are expressed in **b**. **c-g** Mean field phase diagram at the chemical potentials $\mu = -10, 0, 10, 20,$ and $30$ meV. We set the tight-binding parameters $(\epsilon, t, t_2, \delta\epsilon, \delta t) = (0.01, 0.42, 0.03, 0, 0)$. Emergence and competition of various CDW and superconducting phases are highly contingent upon the fillings. The CDW and superconducting orders considered in the mean-field theory are tabulated in Tables S5 and S6. For numerical evaluations, we obtain 161 data points $\phi_i \in [-0.08, 0.08]$ of the ground state energy, with $80 \times 80$ $k$-points.

38

**Supplementary Note 10. Estimation of $U$ and $V$ values and their tunability**

We evaluate the $(U, V)$ values in monolayer by using the first-principles constrained random phase approximation (cRPA) method [15]. In cRPA method, we divide the full polarization function into two parts as $P = P_t + P_r$ in which $P_t$ and $P_r$ are calculated within the target and rest orbitals, respectively. The effective Coulomb interaction matrix $\mathcal{U}$ is computed via $\mathcal{U} = [1 - \mathcal{U}_{\text{bare}} P_r]^{-1} \mathcal{U}_{\text{bare}}$ where $\mathcal{U}_{\text{bare}}$ is a bare Coulomb interaction. We then average the matrix $\mathcal{U}$ to obtain $U$, $V$ parameters. We find that our screened Hubbard parameters $U$ and $V$ in the bulk are quite similar to those of the previous study [16], as shown in Table S7. The values of $U$ and $V$ in the monolayer are quite comparable to those of bulk. The effective $U$ and $V$ values depend on the target orbital models because the rest orbitals screen the target orbitals. Interestingly, due to the large screening effect, the $U$ and $V$ values are strongly renormalized as $U = 0.58$ and $V = 0.07$ eV for our $A_g$ orbital TB model used for mean-field calculations. The estimated $U$ and $V$ values are comparable to the $U$ and $V$ ranges considered in our mean-field phase diagram. We note that the effective $U$ and $V$ values obtained from the cRPA calculations can be changed and even become negative if the electron-phonon coupling is considered as we discussed in section 7.5.

TABLE S7: **On-site $U$ and nearest-neighbor $V$ Hubbard interaction parameters calculated by cRPA method [15] with the weighting approach [17].** Our $A_g$ orbital TB model used for mean-field calculations includes three V-$d$ target orbitals of $d_{3z^2-r^2}$, $d_{x^2-y^2}$, and $d_{xy}$ in global coordinates (Fig. S15), dubbed V-$d$3 model.

|  | Monolayer | | Bulk | |
|---|---|---|---|---|
| Target orbitals | $U(F_0)$ (eV) | $V$ (eV) | $U(F_0)$ (eV) | $V$ (eV) |
| All (bare) | 14.20 | 3.93 | 15.79 | 4.46 |
| V-$d$ Sb-$p$ | 5.87 | 1.96 | 5.34 | 1.48 |
| V-$d$ | 0.77 | 0.11 | 0.80 | 0.08 |
| V-$d$3 ($A_g$ orbital) | 0.58 | 0.07 | 0.60 | 0.06 |

We also show that the $U$ and $V$ values are tunable under mechanical strains, which allows us to access diverse phases. As shown in Fig. S20a, the calculations reveal that the $U$ value changes from 0.55 eV to 0.67 eV from 0 to 4 percents uniform tensile strains. Similarly, the $V$ value from 0.07 eV to 0.11 eV under the same modulation of tensile strain. Encouragingly,

within these changes of $U$ and $V$ values, we find that diverse phases are accessible. For example, for the chemical potential $\mu = 10$ meV, two different superconducting phases with $d$-wave ($B_{1g}$) and $p$-wave ($B_{2u}$) symmetries are accessible as shown in Fig. S20b. In addition, for the chemical potentials $\mu = 20$ and 30 meV, the SD-1 and TRSB-2 CDW and superconducting phases are observable within the ranges of the $U$ and $V$ values as shown in Figs. S20c-d. Therefore, we believe that diverse phases can be accessed in experiments via strain engineering.

We expect that another feasible route to tune the $U$ and $V$ values is to utilize different substrates, gate dielectrics, or gate configurations, which effectively change the screening of Coulomb interaction as in many two-dimensional systems [18–20].

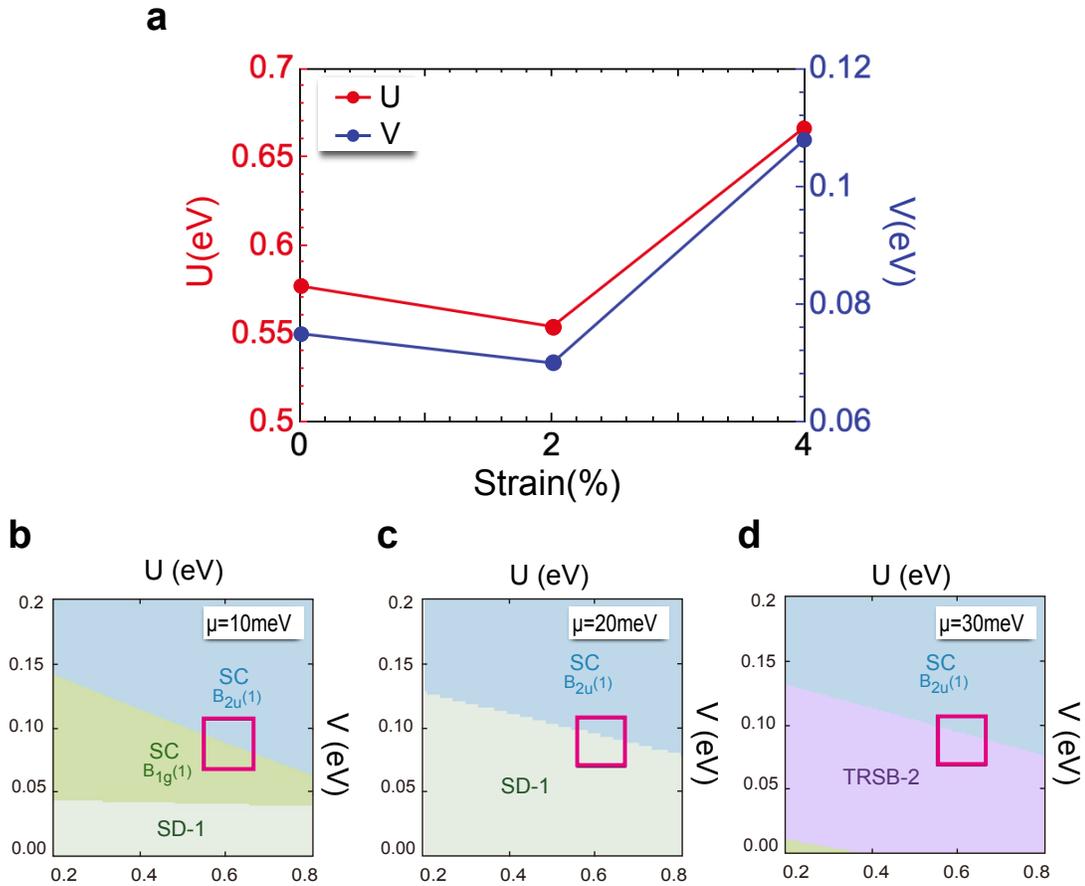

FIG. S20: **a** $U$ and $V$ values as a function of uniform tensile strain. **b-d** The phase diagrams in $U$ and $V$ space for various chemical potentials. **b** $\mu = 10$ meV **c** $\mu = 20$ meV, and **d** $\mu = 30$ meV. The red boxed regions show the accessible phases under strain engineering.

**Supplementary Note 11. Comments on Incommensurate CDW**

In this section, we provide some discussion about incommensurate CDW (IC-CDW) orders, which did not mainly cover in the main text. It is generally known that the incommensurate CDW (IC-CDW) orders are much more vulnerable under thermal fluctuations and impurities [21]. Within an IC-CDW state, all positions of charge density relative to the crystal lattice are equally likely, in stark contrast to the commensurate CDW phase. Applying this perspective to the monolayer $AV_3Sb_5$ system, the bond orders with any phase, $\phi_{\alpha\beta}$, at given incommensurate $Q$ vector, $\boldsymbol{Q}_{\alpha\beta}$, are basically degenerate, where

$$\Phi_{\alpha\beta} = \sum_{\boldsymbol{R}} \left[ \cos(\boldsymbol{Q}_{\alpha\beta}^{IC} \cdot \boldsymbol{R} + \phi_{\alpha\beta})\langle \alpha_{\boldsymbol{R}}^{\dagger} \beta_{\boldsymbol{R}}\rangle + \text{C.C.} \right], \tag{29}$$

with sublattice index $(\alpha, \beta) \in \text{AB, BC, CA}$. We note that the gapless goldstone, namely phason, can be carried by the phase of complex CDW order, $\phi_{\alpha\beta}$. Then, the dimensionality plays a key role since the continuous symmetry associated with IC-CDW cannot be broken in a dimension lower than the critical dimension, $d_{l,c} = 2$, at finite temperature, due to the celebrated Mermin-Wagner theorem.

To resolve the IC-CDW issue more concretely, we additionally performed numerical calculations, employing one of the well-known numerical approximation methods to incorporate incommensurate CDW orders [22]. We obtain a better picture of the clean monolayer kagome metals at zero temperature, finding the intriguing competing physics between IC-CDW and $2 \times 2$ CDW, as summarized in Fig.S21. Our new results can be considered as a demonstration of the new physics originating from type-II VHS points ($P_i$) which are off high-symmetry momenta ($M_i$). Below, we briefly sketch how the incommensurate CDW orders are considered.

In our improved mean-field ansatz, all IC-CDW configurations are included by extending the original ansatz which is restricted to access $2\times2$ CDW orders only. The possible $Q$ vectors of IC-CDW orders associated with (AB, BC, CA) bonds are constructed by connecting the rearranged VHS points,

$$\boldsymbol{Q}_{i,j}^{IC} = (\boldsymbol{P}_i^{(2)} - \boldsymbol{M}_1, \boldsymbol{P}_j^{(3)} - \boldsymbol{P}_i^{(2)}, \boldsymbol{M}_1 - \boldsymbol{P}_j^{(3)}), \tag{30}$$

where $P_i^{(n)}$ is a shifted type-II VHS point from $M_n$ by $d\boldsymbol{q}_i$ (See Fig. S22 i,ii of (a-d)). Here, $d\boldsymbol{q}_{1,3} = \mp[0.0498\boldsymbol{b}_1 - 0.0315\boldsymbol{b}_2]$, $d\boldsymbol{q}_{2,4} = \mp[0.315\boldsymbol{b}_1 - 0.0498\boldsymbol{b}_2]$ are defined with a $1 \times 1$ reciprocal vector $\boldsymbol{b}_{1,2}$. Depending on choosing different type-II VHSs, $(\boldsymbol{P}_i^{(2)}, \boldsymbol{P}_j^{(3)})$, there are 16 number of cases of IC-CDWs, distinguished by deviated $Q$ vectors, $d\boldsymbol{Q}_{i,j}^{IC}$, from those of $2 \times 2$ CDW orders $\boldsymbol{Q}_{2\times 2} \equiv (\boldsymbol{M}_2 - \boldsymbol{M}_1, \boldsymbol{M}_3 - \boldsymbol{M}_2, \boldsymbol{M}_1 - \boldsymbol{M}_3)$, whose explicit form is given by

$$d\boldsymbol{Q}_{i,j}^{IC} \equiv \boldsymbol{Q}_{i,j}^{IC} - \boldsymbol{Q}_{2\times 2} = (d\boldsymbol{q}_i, d\boldsymbol{q}_j - d\boldsymbol{q}_i, -d\boldsymbol{q}_j). \tag{31}$$

Here, the $D_{2h}$ symmetry simplifies the possible cases into 4 ones ($i = 3, j \in 1, 2, 3, 4$) and the associated possible $3Q$ and $1Q$ IC-CDW orders are taken into account (See Fig.S22). Note that $3Q$ orders mean that the relative amplitudes of $(Q_{AB}, Q_{BC}, Q_{CA})$ bonds are equivalent, while $1Q$ CDW order has only one finite amplitude of bond order among $(Q_{AB}, Q_{BC}, Q_{CA})$ bonds.

Technically, we have performed the supercell calculations with commensurate approximations, one of the well-established theoretical methods used for IC-CDW analysis. Following the previous literature [22], we could approximate the $Q$ vector of IC-CDW to the closest rational numbers of the $1 \times 1$ reciprocal vector, $\boldsymbol{b}_i$, which reproduce similar physical properties of the quasicrystal. For example, within $20 \times 32$ supercell approximation, IC-CDW wave vectors are approximated as

$$d\boldsymbol{q}_{1,3} = \mp[0.0498\boldsymbol{b}_1 - 0.0315\boldsymbol{b}_2] \to d\boldsymbol{q}_{1,3}^{\text{app}} \mp [\frac{1}{20}\boldsymbol{b}_1 - \frac{1}{32}\boldsymbol{b}_2], \tag{32}$$

$$d\boldsymbol{q}_{2,4} = \mp[0.0314\boldsymbol{b}_1 - 0.0498\boldsymbol{b}_2] \to d\boldsymbol{q}_{2,4}^{\text{app}} \mp [\frac{1}{32}\boldsymbol{b}_1 - \frac{1}{20}\boldsymbol{b}_2]. \tag{33}$$

The accuracy of these approximations increases as reducing the difference between actual values ($d\boldsymbol{q}_i$) and approximated ones ($d\boldsymbol{q}_i^{\text{app}}$) by enlarging the supercell size (say, $N_1 \times N_2$), as shown in Fig.S23, where approximation error, $(|d\boldsymbol{q}_i - d\boldsymbol{q}_i^{\text{app}}|/d\boldsymbol{q}_i)$, is plotted as a function of the supercell size, $(N_1, N_2)$. We note that our approximated IC-CDW Hamiltonian is already a $1920 \times 1920$ dimension, and increasing the matrix size is technically complicated since both the time and memory cost scale grows quadratically, $\propto (N_1 N_2)^2$ [23].

Based on our $20 \times 32$ supercell calculations, we obtained the new phase diagram at $U \neq 0, V = 0$, as depicted in Fig.S21. For numerical integration of $k$ points, we utilized the

61 × 61 meshes in the reduced Brillouin zone. We found that the incommensurate CDW replaces the 2×2 CDW in the provided phase diagram, but it cannot be an evidence of instability of the commensurate ISD, as the replacement only occurs, as we fine-tunned the chemical potential to the type-II VHS. By contrast, when the chemical potential is shifted, the 2×2 commensurate ISD phase in a wide range of repulsive interactions, suggesting the stability of the 2×2 commensurate ISD phase. To demonstrate this, we showed the phase diagrams at two different fillings $\mu = -5$ meV and $\mu = -7$ meV, respectively. As shown in Fig. S21, the commensurate ISD phase survives the incommensurate CDW at $\mu = -7$ meV in stark contrast to the $\mu = -5$ meV case, where the IC-CDW are energetically favored in the broad region of $U > 0$. We would like to highlight that the 2×2 commensurate CDW phase is as much as accessible as the IC-CDW phase due to the marked tunability of competing orders in the monolayered platform. This feature is also manifested in our susceptibility calculations (Fig. 3, main text), since the position of the peak is highly affected by chemical potentials. The competition between IC-CDW and $2 \times 2$ CDW phases is very versatile under varying interaction strength due to the tiny energy differences and more sophisticated analysis with advanced numerical methods is required for investigating these competing physics in the broad ragnge of chemical potentials.

Our new finding shows that IC-CDW orders are relevant in monolayer $AV_3Sb_5$ due to the symmetry breaking. Interestingly, the competition between $2 \times 2$ CDW and IC-CDW is also recently observed in the hole-doped $CsV_3Sb_5$ bulk samples, which is attributed to the symmetry breaking [24]. In this respect, we believe that our revised manuscript offers timely new physics in the kagome metal community, calling for future sophisticated theoretical and experimental studies to explore the interplay of physics between IC-CDW and other complex orders.

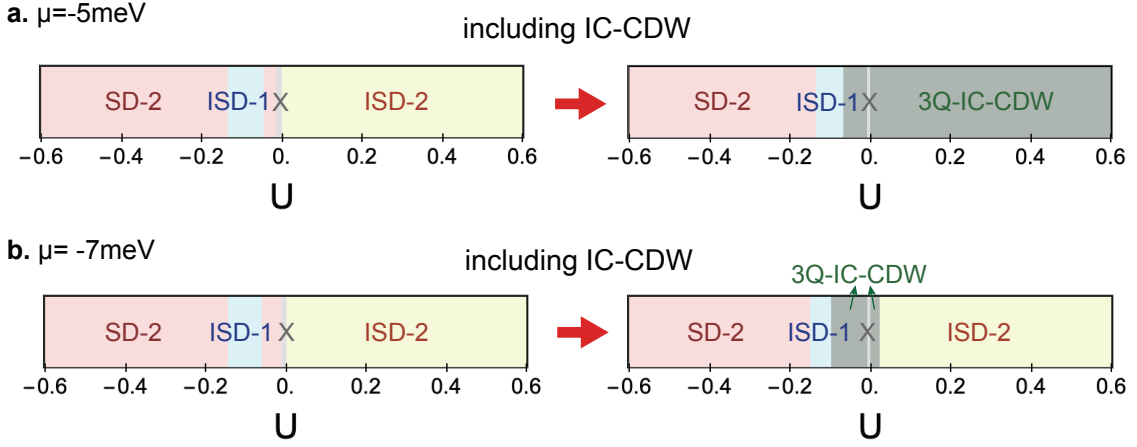

FIG. S21: **Chemical potential dependence of incommensurate CDW phase.** Phase diagrams are calculated at two different fillings (a) $\mu = -5$ meV and (b) $\mu = -7$ meV with (right) and without (left) the incommensurate CDW order. The order parameters of the incommensurate CDWs are given in Fig. S22. These are representative fillings near the type-II filling ($\mu = -6$ meV), which lead to a contrasting impact on the competition between the commensurate and incommensurate CDWs. For repulsive interactions ($U > 0$), the commensurate CDW (ISD-2) dominates over the incommensurate CDW for $\mu = -7$ meV, whereas the incommensurate CDW is dominant for $\mu = -5$ meV. The convergent results are obtained by using $20 \times 32$ supercell calculations with $61 \times 61$ $\mathbf{k}$ mesh points in the reduced Brillouin zone.

TABLE S8: **The shifted $Q$-vectors of IC-CDW from $2 \times 2$ CDW one, $d\mathbf{Q}^{IC}_{i,j}$ and their approximated values, $d\mathbf{Q}^{IC,\mathrm{app}}_{i,j}$, within $20 \times 32$ supercell for four different cases.** The three components in $d\mathbf{Q}^{IC}_{i,j}$, and $d\mathbf{Q}^{IC\mathrm{app}}_{i,j}$ correspond to the Q-vector of (AB,BC,CA) bonds. The four different cases are classified by choice of different VHS points ($P^{(2)}_i, P^{(3)}_j$), as illustrated in the Fig.S22.

| Case | $d\mathbf{Q}^{IC}_{i,j}$ | $d\mathbf{Q}^{IC,\mathrm{app}}_{i,j}$ within $20 \times 32$ supercell |
|---|---|---|
| Case 1 $(i,j)=(3,3)$ | $[0.0498\mathbf{b}_1 - 0.0315\mathbf{b}_2, \mathbf{0}, -0.0498\mathbf{b}_1 + 0.0315\mathbf{b}_2]$ | $\left[\frac{1}{20}\mathbf{b}_1 - \frac{1}{32}\mathbf{b}_2, \frac{0}{20}\mathbf{b}_1 + \frac{0}{32}\mathbf{b}_2, -\frac{1}{20}\mathbf{b}_1 + \frac{1}{32}\mathbf{b}_2\right]$ |
| Case 2 $(i,j)=(3,4)$ | $[0.0498\mathbf{b}_1 - 0.0315\mathbf{b}_2, -0.0183\mathbf{b}_1 - 0.0183\mathbf{b}_2, -0.0315\mathbf{b}_1 + 0.0498\mathbf{b}_2]$ | $\left[\frac{1}{20}\mathbf{b}_1 - \frac{1}{32}\mathbf{b}_2, \frac{0}{20}\mathbf{b}_1 + \frac{1}{32}\mathbf{b}_2, -\frac{1}{20}\mathbf{b}_1 + \frac{2}{32}\mathbf{b}_2\right]$ |
| Case 3 $(i,j)=(3,1)$ | $[0.0498\mathbf{b}_1 - 0.0315\mathbf{b}_2, -0.0996\mathbf{b}_1 + 0.063\mathbf{b}_2, 0.0498\mathbf{b}_1 - 0.0315\mathbf{b}_2]$ | $\left[\frac{1}{20}\mathbf{b}_1 - \frac{1}{32}\mathbf{b}_2, -\frac{2}{20}\mathbf{b}_1 + \frac{2}{32}\mathbf{b}_2, \frac{1}{20}\mathbf{b}_1 - \frac{1}{32}\mathbf{b}_2\right]$ |
| Case 4 $(i,j)=(3,2)$ | $[0.0498\mathbf{b}_1 - 0.0315\mathbf{b}_2, -0.0813\mathbf{b}_1 + 0.0813\mathbf{b}_2, 0.0315\mathbf{b}_1 - 0.0498\mathbf{b}_2]$ | $\left[\frac{1}{20}\mathbf{b}_1 - \frac{1}{32}\mathbf{b}_2, -\frac{2}{20}\mathbf{b}_1 + \frac{3}{32}\mathbf{b}_2, \frac{1}{20}\mathbf{b}_1 - \frac{2}{32}\mathbf{b}_2\right]$ |

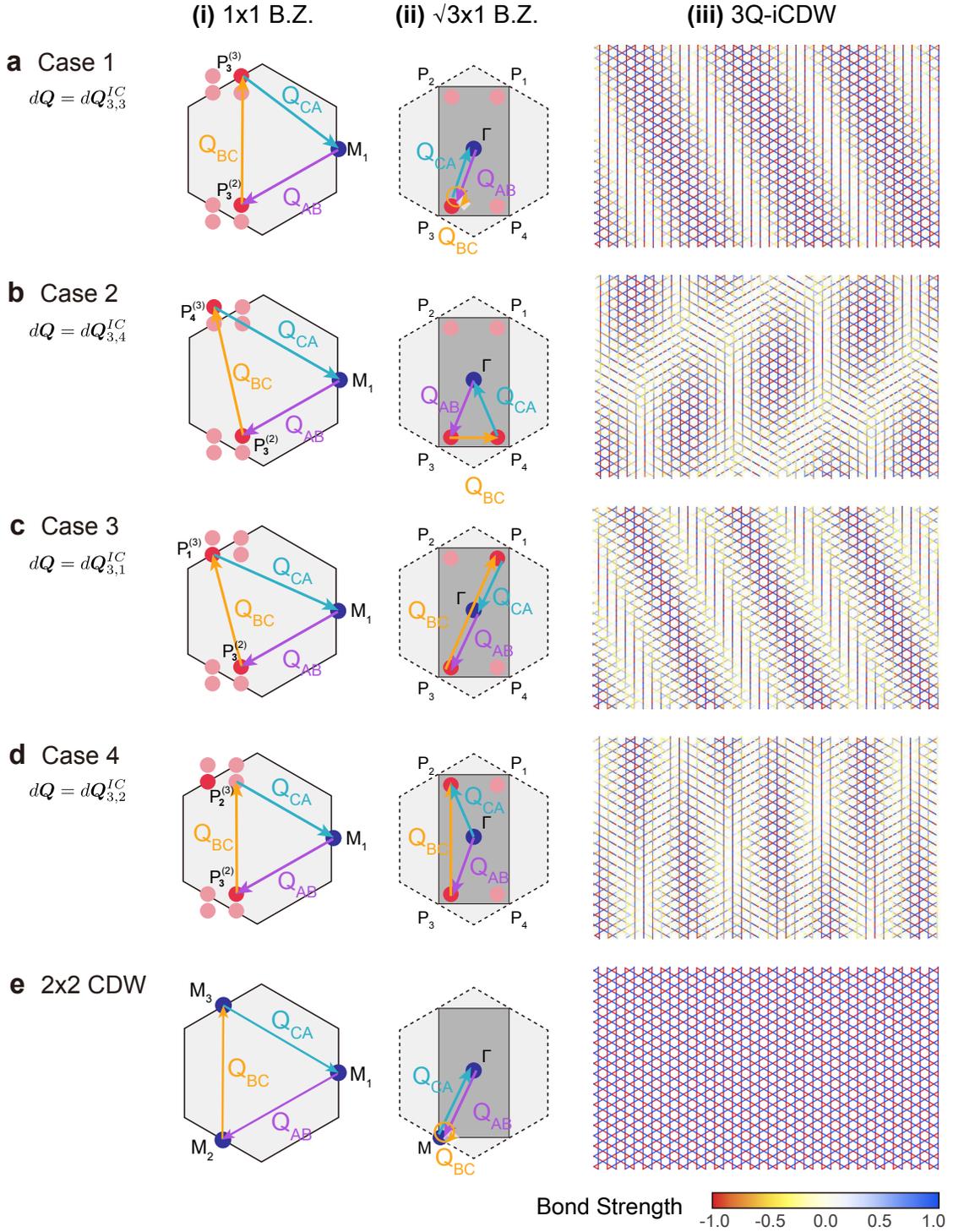

FIG. S22: **Schematic illustrations of incommensurate CDW (IC-CDW) order parameters.** (i) and (ii)-the columns show the wave-vectors of the CDW orders in the $1 \times 1$ and $\sqrt{3} \times 1$ Brillouin zones (BZs), respectively. (iii)-th column delineates the corresponding 3Q bond modulations in real space to exemplify the case. **(a-d)** IC-CDW orders. **(e)** Commensurate $2 \times 2$ CDW order. The type-I (type-II) VHSs are marked by blue (red) circles in the $1 \times 1$ and $\sqrt{3} \times 1$ BZs (i,ii of **a-e**). The order parameters are constructed, such that one of the three $Q$ vectors, $Q_{BC}$, connects two (equivalent or inequivalent) type-II $P_i$ out of four ($P_1, P_2, P_3,$ and $P_4$) and the rest two $Q$ vectors, ($Q_{AB}, Q_{AC}$), connect the chosen $P_i$ and the VHS at $\Gamma$, respectively. This construction results in four symmetry-inequivalent order parameters listed in **a-d**. The shifted bond Q vectors $dQ$ are chosen as $dQ_{3,3}^{IC}, dQ_{3,4}^{IC}, dQ_{3,1}^{IC},$ and $dQ_{3,2}^{IC}$ (i,ii of **a-d**). The $2 \times 2$ CDW corresponds to $dQ = 0$ (i,ii of **e**).

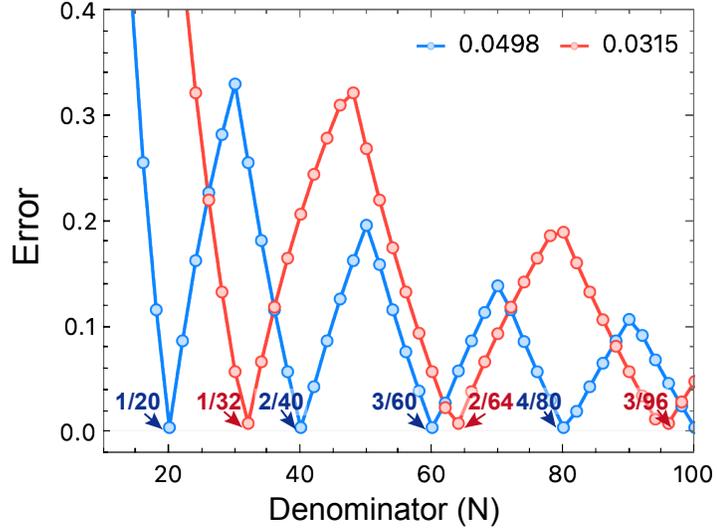

FIG. S23: **The error of commensurate approximation depending on the supercell size $N_1, N_2 < 100$.** To employ commensurate approximations, we approximate the two irrational values $(0.0498, 0.0315)$ to the closest rational numbers, expressed as a fraction $n/N$ of two integers $(n, N)$. Here, the numbers are inherited from $d\boldsymbol{q}_i$ (See Eqs.32,33), and the error is defined as |Exact − Approximated value|/Exact value. The approximated values $(1/20, 1/32)$ give the local minimum in the error with the actual values $(0.0498, 0.0315)$, thus we perform $20 \times 32$ supercell calculations.

**Supplementary Note 12. Engineering of VHS**

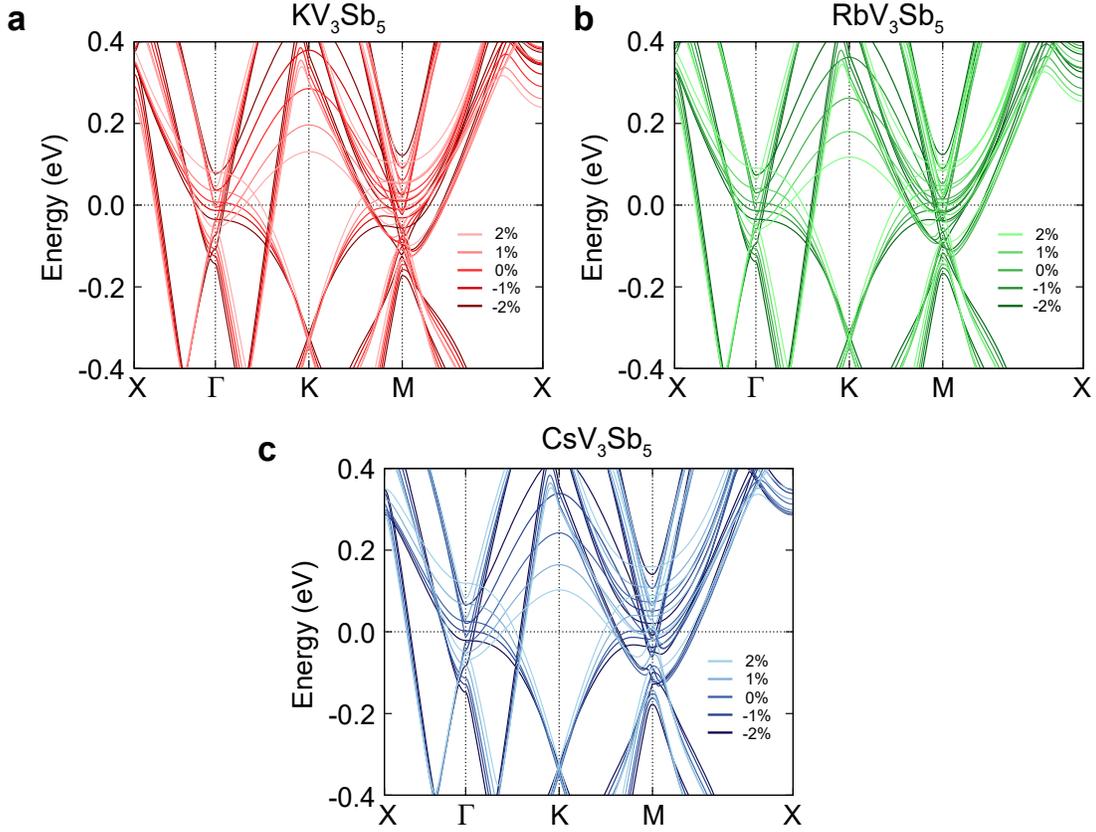

FIG. S24: **Strain engineering of VHSs in the $A\text{V}_3\text{Sb}_5$ monolayer. a-c** DFT Band structures of the $A\text{V}_3\text{Sb}_5$ monolayer family as a function of biaxial strain.

Figures S24**a-c** show DFT band structures as a function of biaxial strain for the $A\text{V}_3\text{Sb}_5$ monolayer family. Under the strain, we find that saddle point bands near $\Gamma$ and $M$ points drastically shift in energy compared with other bands for all three alkali atoms. The tunability of VHS energy is captured in the variations of chemical potential $\mu$ in the TB model under the strain, as displayed in Fig. 5**c** in the main text. It shows that small strain magnitudes of $\sim \pm 2\%$ can lead to large changes of $\mu$ ranging from about 50 meV to about $-100$ meV.

We also examine the doping effect on the location of VHSs. We simulate the doping effect by changing the number of alkali atom $A$ as $A_{1+x}\text{V}_3\text{Sb}_5$ using the $2\sqrt{3} \times 2$ supercell: $x > 0$ ($x < 0$) corresponds to the electron (hole) doping. Figures S25**a-c** show the unfolded band structures obtained at $x = 0.5$ and $-0.5$ for all three alkali atoms. The doping effect

leads to the tuning of locations of VHSs, which is captured in the variation of chemical potential $\mu$ of the TB model in Fig. 5d in the main text.

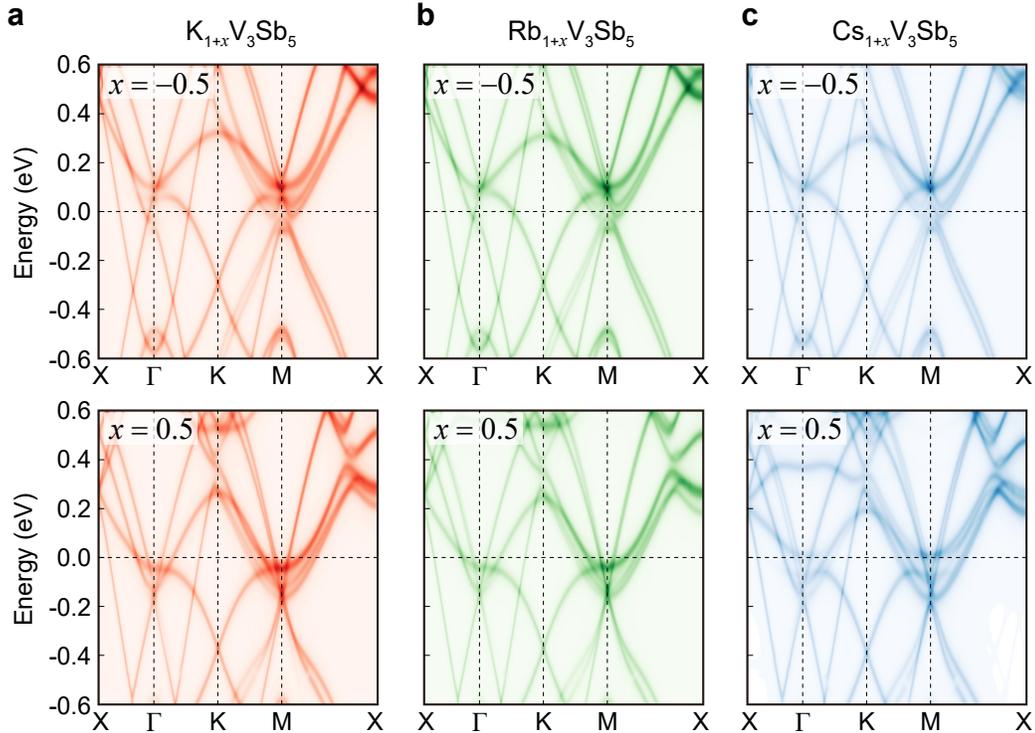

FIG. S25: **Doping engineering of VHSs in the $A$V$_3$Sb$_5$ monolayer. a-c** DFT band structures of the $A_{1+x}$V$_3$Sb$_5$ monolayer family at doping level $x = -0.5$ and $0.5$. Here, we use the $2\sqrt{3} \times 2$ supercell to simulate the doping effect and the band structures are unfolded into BZ of the $\sqrt{3} \times 1$ unit cell by using the method proposed by Popescu and Zunger [25] as implemented in the VaspBandUnfolding code [26].


\fi

\end{document}